\documentclass[usegraphicx,useAMS,usenatbib]{mn2e}
\usepackage{graphicx}
\usepackage{ulem}
\usepackage{amssymb}
\usepackage{amsmath}

\usepackage{color}

\def\fig{Fig.\,}

\def\sec{Section~}
\def\tab{Table\,}

\def\zsun{\rm{~Z_{\odot}}}
\def\Zsun{\rm{~Z_{\odot}}}

\def\Zcrit{\rm{~Z_{crit}}}
\def\msun{\rm{~M_{\odot}}}

\def\msunh{\rm{~M_{\odot}/{\it h}}}

\def\kpc{\rm{~kpc}}

\def\mpch{\rm{~Mpc}/{\it h}}

\def\Log{{\rm~Log}}

\title[Baryonic patterns in primordial haloes]{
Statistical properties of mass, star formation, chemical content and rotational patterns in early $z \gtrsim 9$ structures
}

\author[V. Biffi \& U. Maio]{
  V. Biffi$^{1,2}${\thanks{E-mail: biffi@sissa.it (VB)}},
  U. Maio$^{3,4}${\thanks{maio@oats.inaf.it (UM), ~Marie Curie Fellow}
  }
  \\
  $^{1}$ SISSA - Scuola Internazionale Superiore di Studi Avanzati, Via Bonomea 265, 34136 Trieste, Italy\\ 
  $^{2}$ Universidad Aut\'onoma de Madrid, Departamento de F\'isica Teorica, Ciudad Universitaria de Cantoblanco, 28049 Madrid, Spain\\
  $^{3}$ INAF -- Osservatorio Astronomico di Trieste, via Tiepolo 11, 34143 Trieste, Italy\\
  $^{4}$ Leibniz-Institut f\"ur Astrophysik, An der Sternwarte 16, 14482 Potsdam, Germany
}
 
\begin{document}

\pagerange{\pageref{firstpage}--\pageref{lastpage}} \pubyear{...}
\maketitle
\label{firstpage}


\begin{abstract}
We study the baryonic, chemical and dynamical properties of a
significantly large sample of early proto-galaxies in the first
500~Myr of the Universe (redshift $z\gtrsim 9$), obtained from
high-resolution numerical, N-body, hydrodynamical, chemistry
simulations including 
atomic and molecular networks,
gas cooling,
star formation, 
stellar evolution and metal spreading for population~III and
population~II-I regimes according to proper stellar yields 
and lifetimes.
We find that first star formation events take place in halos with
dark-matter mass $\rm M_{DM} > 2\times  10^6\rm\msun$.
Early star forming objects have:
molecular fractions from $x_{mol}\lesssim 10^{-4}$ in quiescent structures
up to $x_{mol}\gtrsim 0.1$ in active regions;
star formation rates $\rm SFR \sim 10^{-8}-10^{-3}\msun/yr$;
and metallicities in the range
$\sim 10^{-8}-10^{-2}\zsun$.
Roughly $\sim 10\%$ of high-$z$ haloes host population~II-I star formation and dominate the cosmic SFR density.
They usually are bursty objects with mean specific SFR around
$\rm \sim 10 \,Gyr^{-1}$ at $z\sim 9$ and increasing with redshift up to $\rm \sim 10^2\,\rm Gyr^{-1}$.
Stellar feedback effects alter the baryonic content of the haloes and
locally affect their chemical and thermo-dynamical properties, as
reflected by the broadening of various physical relations. 
The establishment of gaseous rotationally-supported cores is quite
uncommon, weakly related to the large-scale dark-matter behaviour and
evolving in an intermittent fashion.
The colder, molecular-rich phase tends to maintain any established
rotational motion longer with respect to the hotter, metal-rich
component, which is very sensitive to environmental processes.
While the fraction of haloes featuring a significant amount of
co-rotating, molecular-rich gas increases with cosmic time
(from a few per cent at $z\sim 20$ up to $\sim 5-15\%$
at $z \sim 9$), the chaotic nature of metal-enriched material
does not lead to particular trends. 
\end{abstract}




%
%
\section{Introduction}\label{sec:intro}
Cosmological structure formation at high redshift is a particularly interesting topic, because it probes the infancy of the very first galaxies, and the chemical and thermal conditions of the early Universe.
\\
Primordial, molecular-rich star forming regions are responsible for a variety of physical processes in the cosmic medium:
they are the main sources of cosmic metal and dust enrichment with consequent transition from pristine population III (popIII) stars to standard population II-I (popII-I) stars \cite[][]{Tornatore2007, Maio2010, Maio2011, Wise2012}; 
are relevant sites of production of photons that reionize the Universe \cite[][]{Abel2002, Sokasian2004, Robertson2010, Salvaterra2011, FernandezShull2011, PetkovaSpringel2011} or dissociate molecules \cite[][]{Ricotti2001, Whalen2008, PetkovaMaio2012}; 
are supposed to be at the origins of the cosmic near infra-red background \cite[][]{DwekArendt1998, Gorjian2000, Totani2001, Santos2002, Magliocchetti2003,Cooray2004, Kashlinsky2005, Kashlinsky2012, Fernandez2013arXiv};
are important places for black-hole seed formation \cite[][]{MadauRees2001, BrommLoeb2003, Alvarez2009, Johnson2012};
have observational counterparts in, e.g., Lyman-break galaxies or Lyman-$\alpha$ emitters \cite[][]{Bouwens2012arXiv, Akila2012};
are preferential hosts of long gamma-ray bursts (GRBs) \cite[][]{Campisi2011, MaioGRBs2012, Salvaterra2013};
reflect higher-order effects in the primordial velocity field \cite[][]{TH2010, MaioKoopmansCiardi2011, Stacy2011, Greif2011, McQuinn2012, Fialkov2012, Naoz2012, Naoz2013, Richardson2013arXiv};
and can be used as tracers of dark-matter non-Gaussianities \cite[][]{MaioIannuzzi2011, MaioNG2011, MaioKhochfar2012},
or even as dark-energy probes on large scales \cite[][]{Maio2006}.
\\
Although it is quite commonly accepted \cite[][]{GunnGott1972} that first star formation events take place in the growing dark-matter haloes, where gas can cool, fragment and acquire angular momentum \cite[][]{Toomre1977, WhiteRees1978}, there are still controversial debates about properties and degeneracies for the baryonic component in primordial mini-haloes \cite[][]{Bower2010}.
From low-redshift observations, it is rather well established that star formation activity is strictly linked to the gaseous content of cosmic structures \cite[][]{Schmidt1959, Kennicutt1998} and represents the main driver of metal pollution \cite[][]{Tremonti2004}.
Observational detections suggest the presence of early galaxies at $z\simeq 11$  \cite[e.g.][]{Zheng2012, Coe2013}, with corresponding formation redshift $z\sim 14$ \cite[][]{Bouwens2012arXiv}.
\\
Recent numerical works have pointed out that galaxies in the first billion years shape fairly well the observed luminosity function \cite[][]{Salvaterra2013}.
Never the less, uncertainties on the role of different feedback mechanisms on the first star formation sites, on their surrounding environments and on their local dynamical state still persist.
\\
Despite possible connections between the dynamical features of primeval dark-matter haloes and the hosted gas \cite[][]{deSouza2013}, $ad~hoc$ studies of individual cases \cite[e.g.][]{Stacy2013, Prieto2013} have claimed that the dynamical trends of collapsing gas can be completely decoupled from the large-scale hosting halo (i.e. the spins do not necessarily coincide).
Studies on dark-matter haloes have demonstrated that the total spin of the halo is only mildly correlating with mass, whereas the correlation with substructures, local environment and halo concentration is usually stronger \cite[see, e.g.,][where, however, disc formation in the gas component was not specifically addressed]{jeesondaniel2011,skibba2011}.
This is somehow consistent with additional analyses showing that gaseous disc and fragmentation properties result substantially influenced by gas cooling or feedback mechanisms \cite[e.g.][]{PiontekSteinmetz2011, Moster2012, Halle2013, Maio2013, Vogelsberger2013} and rotational patterns can experience enhancement or disruption according to the specific ambient medium \cite[mostly at high $z$;][]{Genzel2008}.
Moreover, the gaseous cold flows that are expected to give birth to baryonic stellar structures are invoked in literature both as responsible for the establishment of extended discs \cite[][]{Keres2005, Brooks2009, DekelNature2009} and as responsible of bulge growth \cite[][]{Sales2012}. Such contrasting conclusions testify that a unique and definitive understanding of this subject has not been reached, yet.
\\
At high $z$ our knowledge is particularly limited, because baryonic and dynamical properties in early objects could be only recently investigated with the help of suited numerical simulations.
In a few dedicated works \cite[][]{WiseAbel2007, Greif2008, RomanoDiaz2011, Stacy2013arXiv, Prieto2013, Xu2013arXiv} the thermodynamical state of the gas, the establishment of rotationally-supported cores and the morphologies of primordial haloes have been explored for some specific cases by means of hydrodynamical simulations including molecular cooling in pristine gas.
\\
However, additional, larger samples to draw firmer conclusions on the statistical trends at different times would be required.
Further studies need to be performed to unveil the effects of stellar evolution and consequent metal spreading according to proper yields and lifetimes on structure growth and disc formation.
Indeed, it is crucial to take into account hot, enriched gas ejected from type II and type Ia supernovae (SN) or during asymptotic-giant-branch (AGB) phases in the surrounding colder, molecular-rich environments to fairly assess the subsequent evolution of the surrounding gas.
More in detail, it would be interesting to know what the typical baryonic properties in primordial mini-haloes are; what their metal and chemical content is and how it changes with time; which fraction of haloes can host popIII or popII-I star formation; how thermal and dynamical properties are affected by supernova esplosions or feedback mechanisms; what connections between stellar growth and local environment can be established; whether it is possible to find relations between stellar mass and chemical composition (metallicity and/or molecular fraction); if the specific star formation rate has different trends at early and late cosmological times; how primordial, molecular-rich discs can form, host star formation and survive to stellar feedback; what rotational patterns can be expected in such high-redshift structures and how different components behave from a dynamical point of view.
\\
In this paper, we plan to address these issues by exploring both the typical features of early star forming gas and the implications for the resulting angular momentum and rotational patterns.
We will use high-resolution numerical simulations focusing on the first $\sim$500~Myr of the Universe and corresponding to $z \gtrsim 9$.
\\
The work is organized as follows: in \sec\ref{sec:sims} we describe the simulations and in \sec\ref{sect:selection} we present the data set selected for the analysis.
In \sec\ref{sec:results} we discuss the main results about: baryonic -- thermal and chemical -- properties of primordial haloes (\sec\ref{sub:basics}), their link to star formation activity (\sec\ref{sec:SFRandStars}) and the evolving dynamical features of the first molecular-rich star formation sites (\sec\ref{sub:L}).
We summarize and conclude in \sec\ref{sec:discussion}.

%
%
\section{Simulations of primordial haloes}\label{sec:sims}
We consider N-body, hydrodynamical, chemistry simulations \cite[see][]{Maio2010, Maio2011, MaioIannuzzi2011} in a standard $\Lambda$CDM model with geometrical parameters:
$\Omega_{0,\Lambda} = 0.7$,
$\Omega_{0,m} = 0.3$,
$\Omega_{0,b} = 0.04$,
for present-day $\Lambda$, total-matter, baryonic-matter density parameters,
and expansion parameter normalized to $100\,\rm km/s/Mpc$, $h=0.7$.
Spectral parameters are assumed to be $\sigma_8=0.9$, and $n=1$, for the power spectrum normalization via mass variance within $\rm 8~Mpc/\it h-$size sphere, and for the power spectrum slope, respectively.
The cosmological field is sampled at redshift $z=100$ with $2\times 320^3$ gas
and dark-matter particles in a cube of  $\sim 0.5\mpch$ a side, for a
resulting gas mass resolution of roughly $\sim 40\,\rm M_\odot/\it h$.
\\
The implementation includes non-equilibrium chemistry evolution for
$e^-$,
H,
H$^+$,
H$^-$,
He,
He$^{+}$,
He$^{++}$,
H$_2$,
H$_2^+$,
D,
D$^+$,
HD,
HeH$^+$ \cite[][]{Yoshida2003, Maio2007, PetkovaMaio2012}, gas cooling from resonant, fine-structure and molecular lines \cite[as in][]{Maio2007}, star formation, feedback \cite[][]{Springel2003}, stellar evolution and metal pollution from different stellar yields: He, C, O, Si, Fe, Mg, S, etc. \cite[][]{Tornatore2007, Tornatore2010}.
Stellar population transition with a switch on the initial mass function (IMF) takes place when a critical metallicity of $Z_{crit} = 10^{-4}Z_\odot$ is reached \cite[][]{Bromm2003,Schneider2003}: star forming regions with $Z < Z_{crit}$ (popIII) are assumed to have a top-heavy IMF in the range $\rm 100-500 \,M_\odot$ with a slope of $-2.35$, while star forming regions with $Z \ge Z_{crit}$ (popII-I) form stars according to a Salpeter IMF.
For further details we refer the interested reader to \cite{Maio2007, Maio2010, Maio2011, Maio2013z7arXiv}, and references therein.
We stress that this kind of implementation permits us to properly follow detailed chemical evolution of the medium both in pristine, popIII regimes and in enriched, popII-I ones, simultaneously taking into account the changes in the local stellar populations and the consequent energetics of feedback mechanisms.
\\
In the following we will clarify our selection criteria and show the main results about the baryonic and dynamical properties of primordial structures.
%
%
%
%
%
\begin{figure*}
\centering
\includegraphics[width=0.44\textwidth]{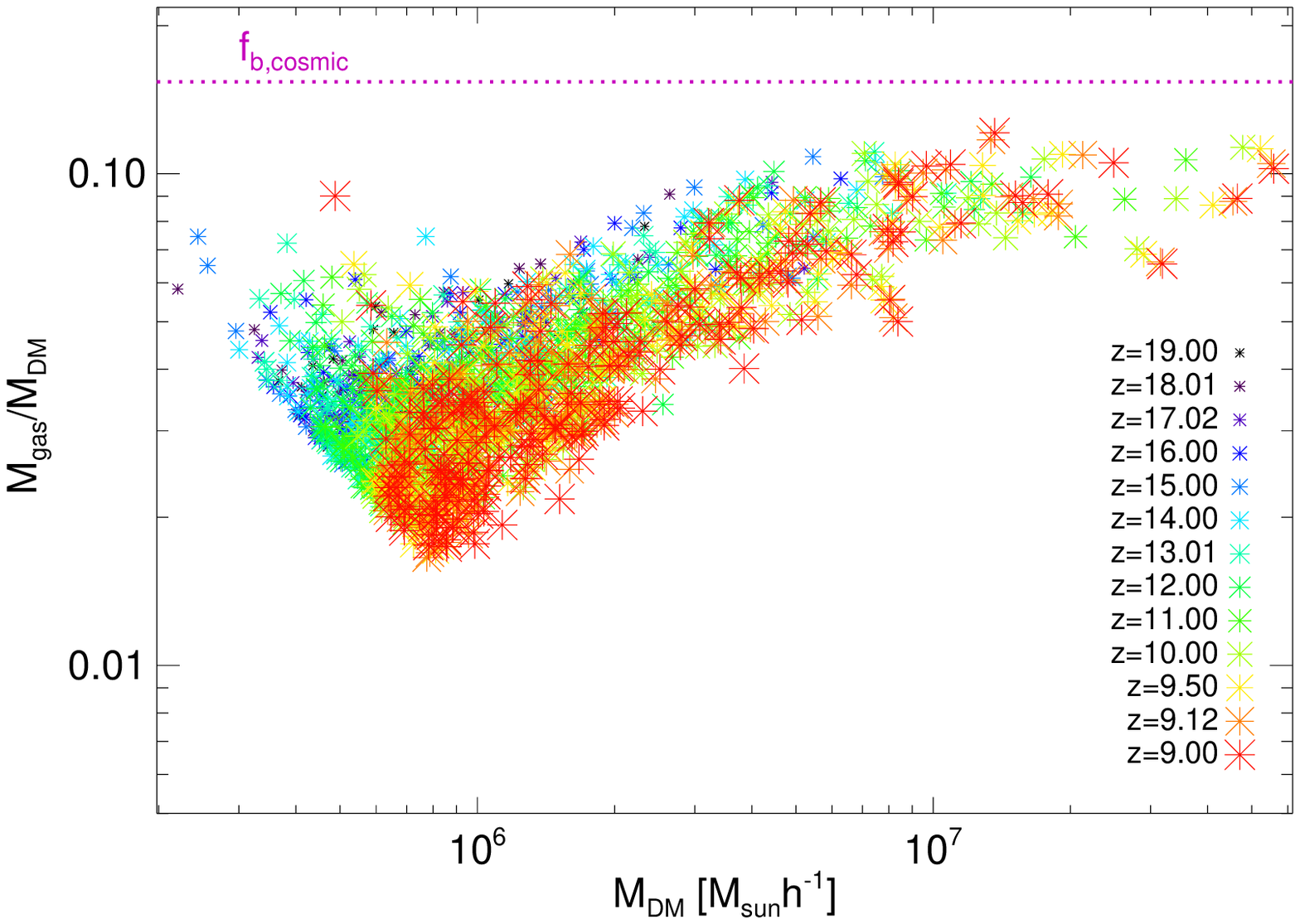}
\includegraphics[width=0.44\textwidth]{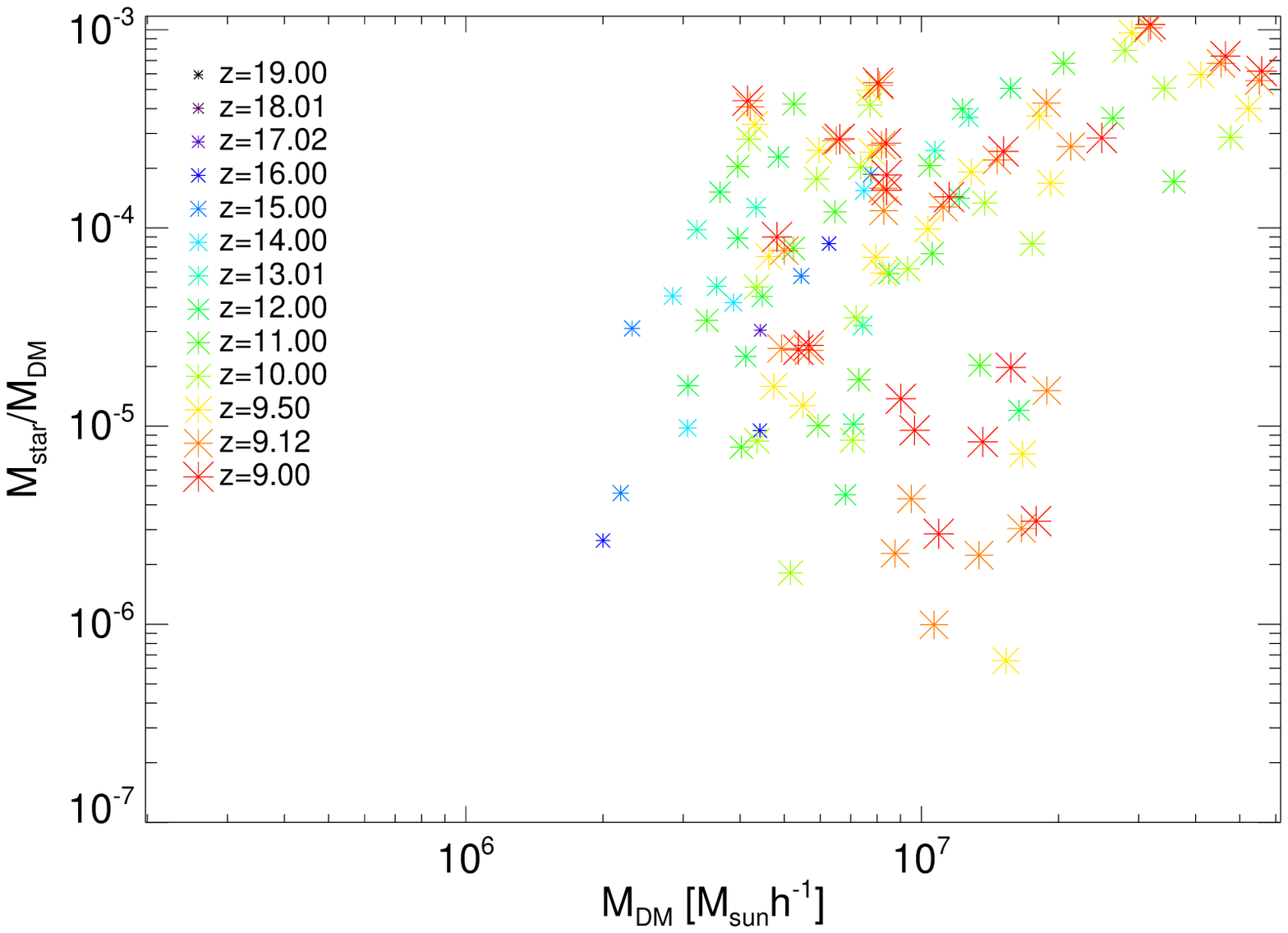}
\caption{
  Relations between the gas fraction (left) and stellar fraction (right) as function of dark-matter mass, $M_{\rm DM}$ at different redshifts (see legends).
The horizontal dotted line marks the cosmological baryon fraction, computed according to the simulation parameters.
}
\label{fig:masses}
\end{figure*}
%
%
\section{Selection and data set}\label{sect:selection}
We track cosmological structures with their dark-matter, gaseous, and stellar components by applying a friend-of-friend (FoF) algorithm \cite[][]{Dolag2009} with linking length equal to 20 per cent the mean inter-particle separation and by storing their properties (constituting particles, centers of mass, masses, positions, velocities, metal abundances, atomic and molecular fractions, star formation rates, temperatures, etc.) for each redshift.
\\
The total number of haloes usually ranges from a few tousands at very high redshift up to several tens of tousands at later times and reaches about $\sim 25000$ primordial objects at $z\simeq 9$.
In order to have a set of haloes whose matter components are properly
resolved, at each snapshot we select only objects with at least 300
gas particles. We notice that this constraint is quite conservative and
protects us from deviations due to numerical artifacts. Indeed,
artificial results would be obtained if a number of gas particles
lower than 4 times the number of neighbours were employed
\cite[][]{BateBurkert1997}. Since we use 32 neighbors in the density
estimation, with our choice we select only objects for which no
spurious fragmentation takes place.
We also stress that the selection is done on the number of gas
particles only, so the selected objects have a {\it total} number of
particles that is much larger, usually of at least one order of
magnitude, and results in a minimum of at least a few tousands
particles.
This also avoids unphysical two-body numerical heating
\cite[e.g.][]{vazza2011}.
The final data set resulting from our selection consists of 1680
haloes between $z=9$ and $z\simeq 20$, of which about 200 at 
redshift $z=9$.
Gas and stellar fractions as a function of dark-matter mass are
summarized in \fig\ref{fig:masses} for different redshifts (see
discussion in the next). 
\\
We highlight that we additionally checked our results by applying
different cuts in the number of gas particles in each halo (i.e. 200,
250, 300), and, then, by considering the (50, 100, 200) largest haloes
in the catalogue list. 
However, all the resulting trends did not show evident changes and the
statistical patterns were basically unaffected. 
%
%
%

\begin{figure*}
\centering
\includegraphics[width=0.33\textwidth]{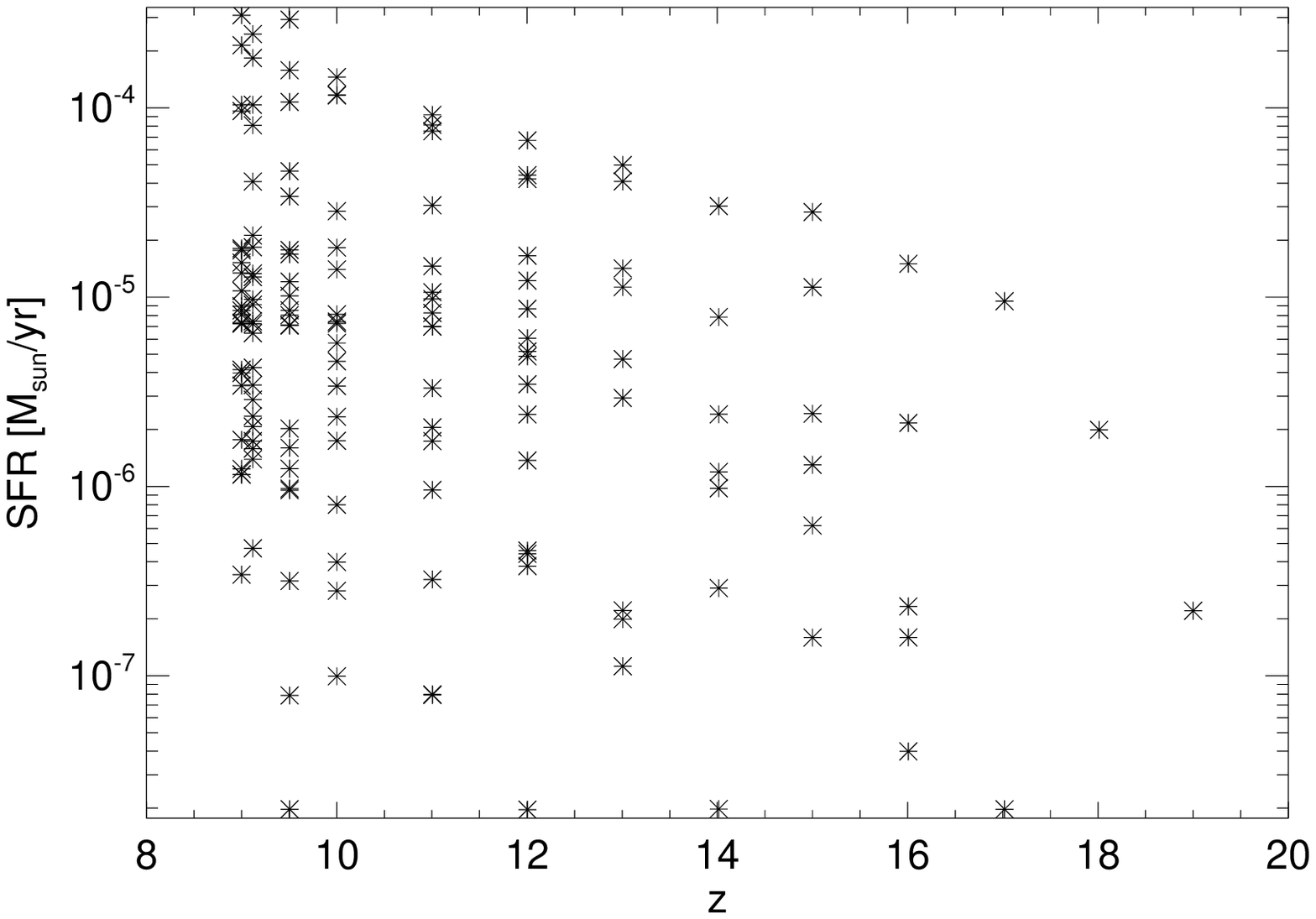}
\includegraphics[width=0.33\textwidth]{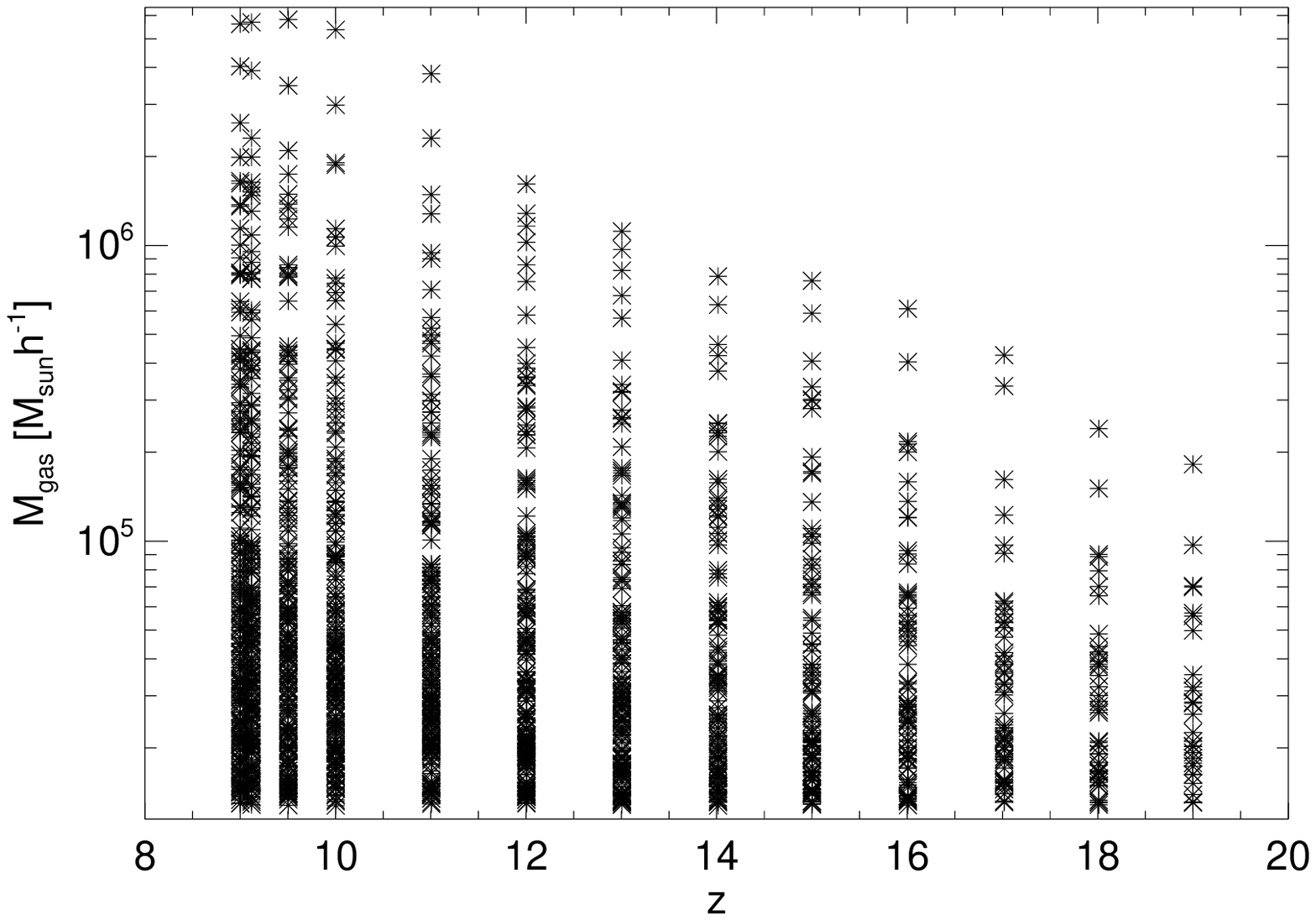}
\includegraphics[width=0.33\textwidth]{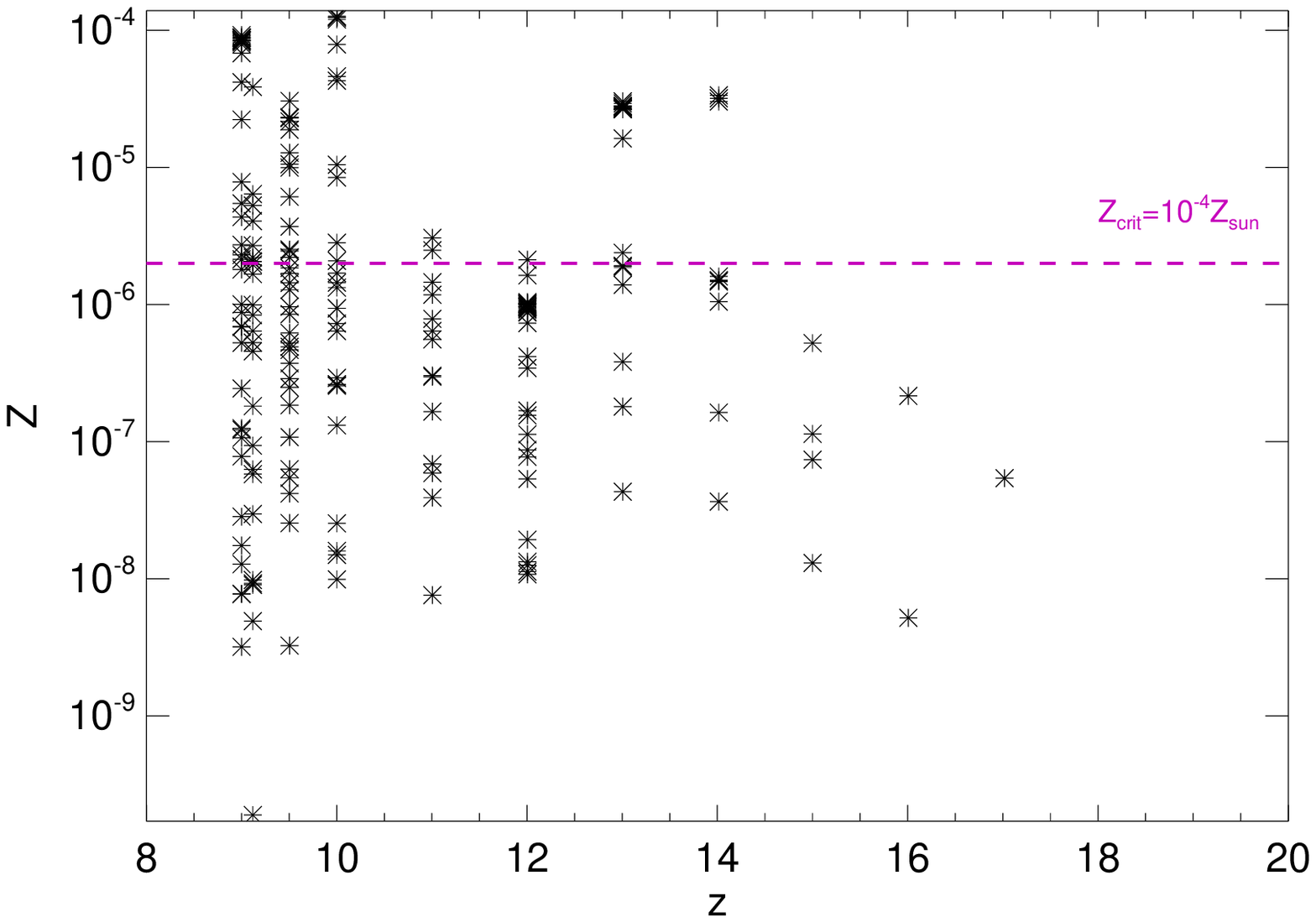}
\caption{
Redshift evolution of SFR (left), $M_{gas}$ (center) and total metallicity $Z$ (right) for our set of haloes. In the $redshift-Z$ panel, the purple dashed line marks the critical threshold of $\Zcrit=10^{-4}\Zsun$.
}
\label{fig:sfr-Mgas-Z}
\end{figure*}
\section{Results}\label{sec:results}
We will now show the principal features of the first galaxies at high redshift, and their baryonic and dynamical properties.\\
\fig\ref{fig:masses} shows gas (left-hand side panel) and stellar (right-hand side panel) fractions as a function of dark-matter mass, while the horizontal line marks the cosmological baryon fraction.
Both gas and stellar content can vary within haloes of different dark-matter masses, $M_{DM}$, due to the interplay of gas cold inflows and hot ejected material.
In particular, bigger haloes are able to retain the gas heated by star formation feedback, while smaller ones have shallower potential wells and can loose significant amounts of mass.
This is clearly demonstrated by the large spreads for low gas masses and the deep decrement around $\sim 10^5-10^6 \msun$.
Therefore, formation of stars in haloes with a dark mass below $\sim 2\times 10^6\,\rm \msun$ is strongly inhibited, as testified by the right panel and as also found by \cite{Wise2012}.
Stellar fractions are generally increasing with dark-matter
mass\footnote{Despite the scatter at low masses, the correlation of
  $M_{\star}/M_{DM}$ with halo mass is mildly positive, as quantified
  by the Pearson correlation coefficient, $\sim 0.74$, and by
  Spearman's rank correlation, $\sim 0.43$.} 
and span a range between $\sim 10^{-6}$ and $10^{-3}$ for hosting halo
masses of $\sim 10^6-10^8\msunh$. 
The basic reason why there is no relevant star formation below the minimum mass of $\sim 2\times 10^6\,\msun$ lies in the relatively low densities reached by the hosted gas.
As a consequence, in such small objects, H$_2$ and HD molecules either are not formed significantly or, in case, are not efficiently shielded, thus star formation feedback in close, more massive haloes can easily dissociate them.
\\
In the following sections, we will analyze in detail the properties of these objects: their baryonic features (\sec\ref{sub:basics}), the connections to star formation activity (\sec\ref{sec:SFRandStars}) and eventually their dynamical patterns (\sec\ref{sub:L}).
\subsection{Baryonic properties of early proto-galaxies}\label{sub:basics}
We start by discussing the basic physical properties of primordial popIII and popII-I galaxies (\sec\ref{sub:SFhaloes}), their gas, molecule, and metal content (\sec\ref{sub:chemicalproperties}) and their thermal state (\sec\ref{sub:Tproperties}). 
\\
We note that, since the number of well-resolved haloes is 1680 and the structures having a non-null star formation rate ($\rm SFR>0$) are 148, the fraction of cosmological objects effectively involved in high-redshift star formation is roughly $\sim 10\%$ per cent.
Pollution events from star formation episodes can influence a significant fraction of closeby regions, as 83 proto-galaxies are found to have $\rm SFR=0$ and $Z>0$, while the bulk of enriched ($Z>0$) objects consists of 206 early galaxies, i.e. about 2.5 times more.
\subsubsection{PopIII and popII-I star forming galaxies}\label{sub:SFhaloes}
As already discussed in different works in literature \cite[][]{Maio2010, MaioIannuzzi2011, Campisi2011, Wise2012, Salvaterra2013}, the popIII contribution to the total star formation process is important only during the very initial bursts of star formation, while the bulk of cosmic star formation is usually dominated by the popII-I regime.
Obviously, popIII star formation can still take place at later times, but only in pristine, unpolluted regions not yet affected by chemical feedback.
Never the less, the most star forming regions in the Universe are expected to be already significantly enriched by redshift $z\sim 10$.\\
\begin{figure}
\centering
\includegraphics[width=0.42\textwidth,height=0.20\textheight]{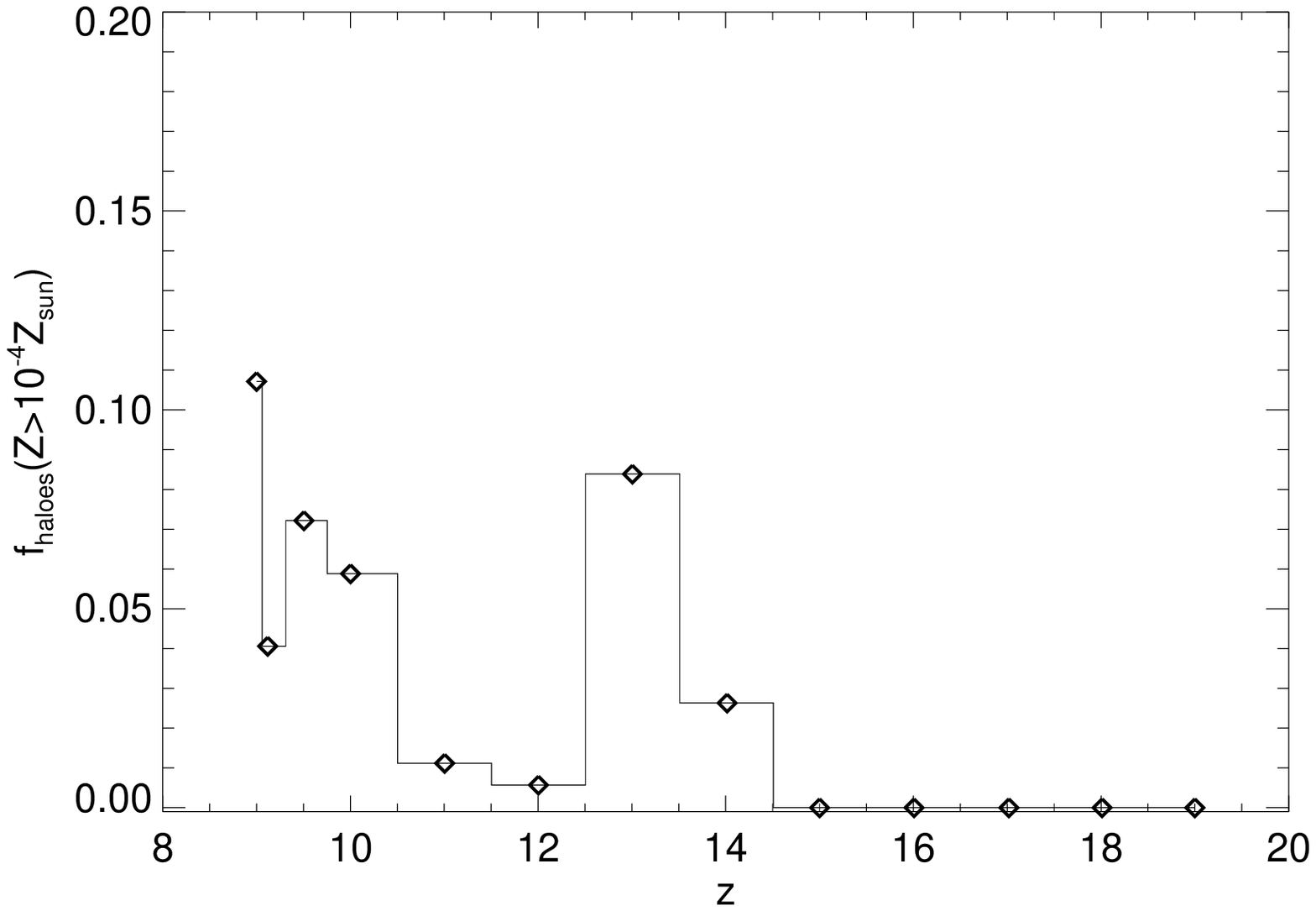}
\includegraphics[width=0.42\textwidth,height=0.20\textheight]{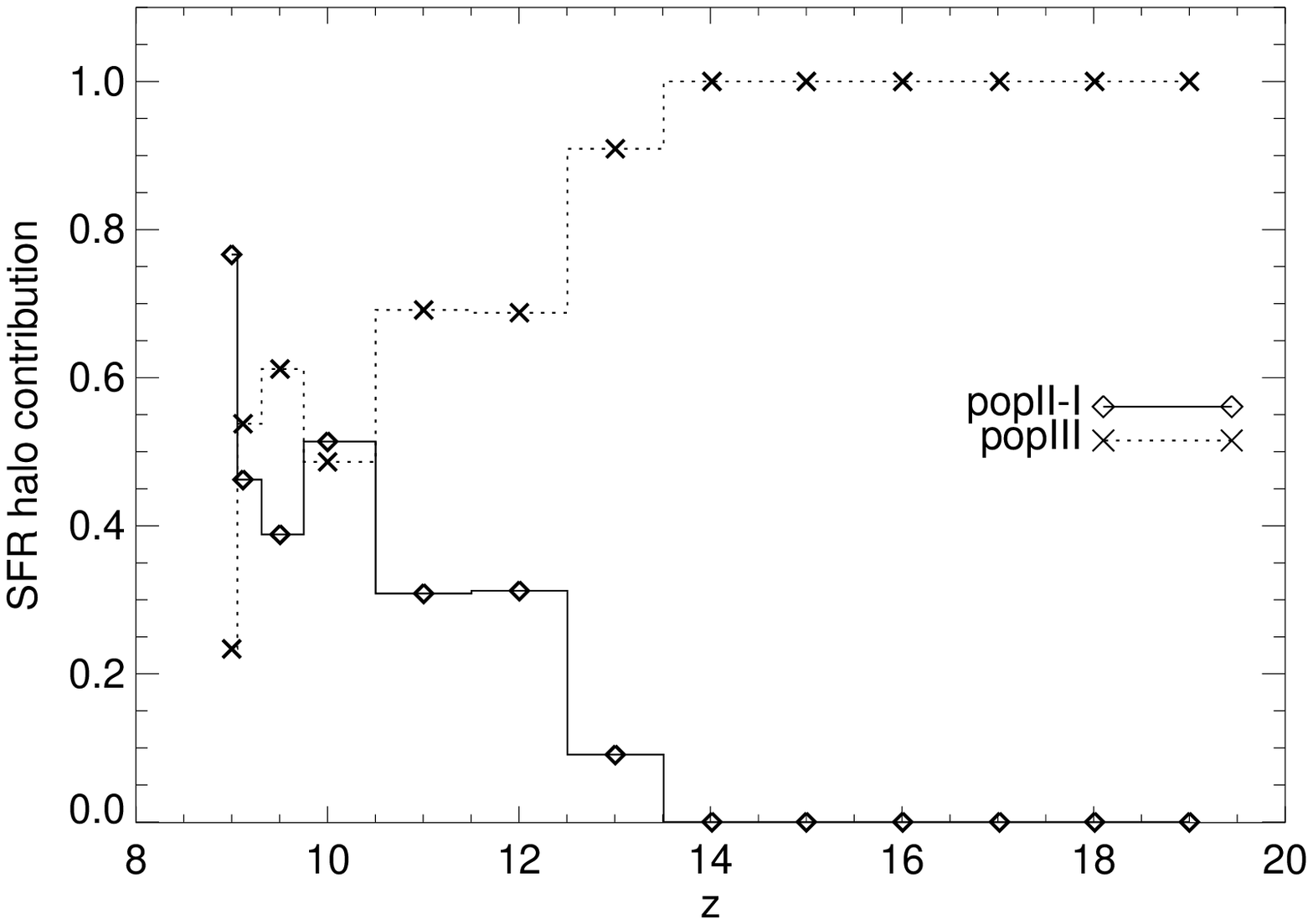}
\caption{{\it Upper panel:} redshift evolution of the fraction of
  haloes with gas metallicity above the critical threshold of
  $\Zcrit=10^{-4}\Zsun$ (popII-I haloes). 
  {\it Lower panel:} Redshift evolution of the relative
  contribution of popII-I (solid curve) and popIII (dotted) haloes to
  the total SFR.} 
\label{fig:red_f_enrich}
\end{figure}
To better quantify the contribution of the haloes involved in the enrichment process and their stellar regime properties, we show, in \fig\ref{fig:sfr-Mgas-Z}, the redshift evolution of the star formation rates, gas masses, and metallicities in our set of haloes.
Generally, these three quantities are increasing with time, as a
consequence of the growth of the cosmic perturbations that lead star
forming activity and consequent metal spreading. 
The objects considered here have typical gas masses of up to $\sim 10^7\,\rm M_\odot$ and star formation rates spanning almost 5 orders of magnitudes, from a few times $10^{-8}\,\rm M_\odot/yr$ up to several times $10^{-4}\,\rm M_\odot/yr$ at $z\simeq 9$.
Interestingly, some haloes are enriched well above $\Zcrit$ already at
$z\lesssim 14$. 
\\
A more quantitative result on the abundance of popIII and popII-I
haloes is plotted in \fig\ref{fig:red_f_enrich} (upper panel),  
where we show the redshift evolution of the fraction of enriched
haloes with gas metallicity above the critical threshold
$\Zcrit=10^{-4}\Zsun$. 
It emerges that usually only up to $\sim 10$ per cent of the enriched
haloes have $Z>\Zcrit$. 
Correspondingly, the lower panel of \fig\ref{fig:red_f_enrich} shows the relative contribution of different-population haloes to the cosmic star formation: pristine, molecular-cooling, popIII haloes (dotted line) and enriched metal-cooling, popII-I haloes (solid line).
The fraction of star formation processes hosted by popIII haloes
decreases rapidly with redshift, whereas the contribution by popII-I
haloes becomes more and more important. 
Basically, metal-rich haloes start contribute the cosmic star
formation activity and metal pollution at $z\lesssim 14$ and dominate
it below $z\sim 10$.
The initial rapid increment in $Z$ is
a consequence of efficient spreading from
the first star forming regions. 
The effects of metal pollution are particularly stronger at earlier
times, because of the more powerful and energetic feedback from
primordial popIII sources. 
This causes the peak of 
popII-I structures
at $z\sim 13$ 
(upper panel) with consequent onset of metal-driven star formation
(lower panel)
\footnote{We remind that among all the $Z>0$ haloes (206)
  between $z\sim 9$ and $z\sim 20$ about $40$ per cent (83) have null
  SFR.}. 
The non-very-regular trend in the upper panel of
\fig\ref{fig:red_f_enrich}, is due both to metals escaping from the
haloes, and to occurring merging events with low-metallicity objects
or pristine gas that can dilute metallicities. 
\\
We also note that \fig\ref{fig:red_f_enrich} supports the possibility
of forming substantial amounts of popIII galaxies at relatively lower
redshift, although passively evolving.
\begin{figure*}
\centering
\includegraphics[width=0.33\textwidth]{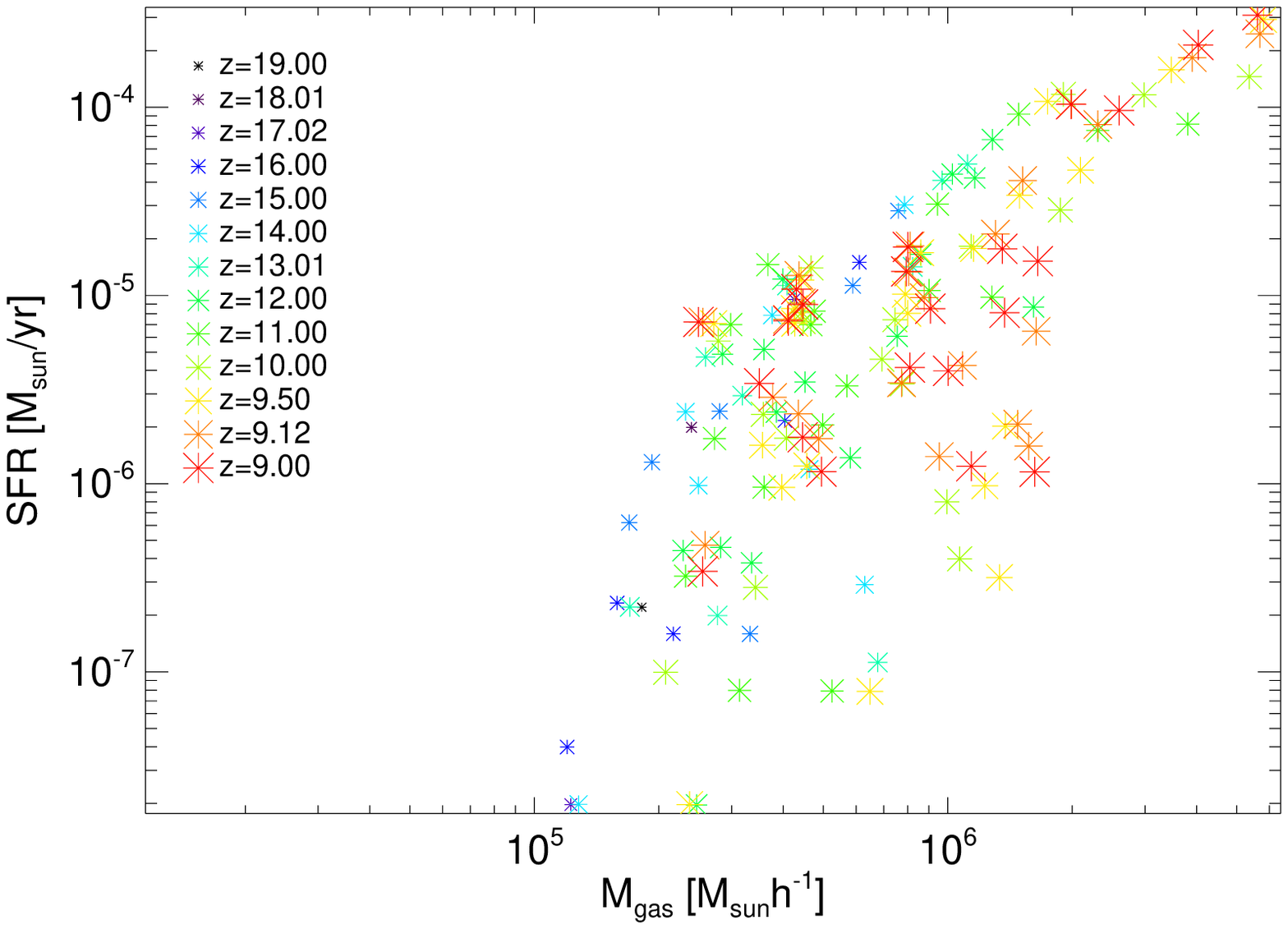}
\includegraphics[width=0.33\textwidth]{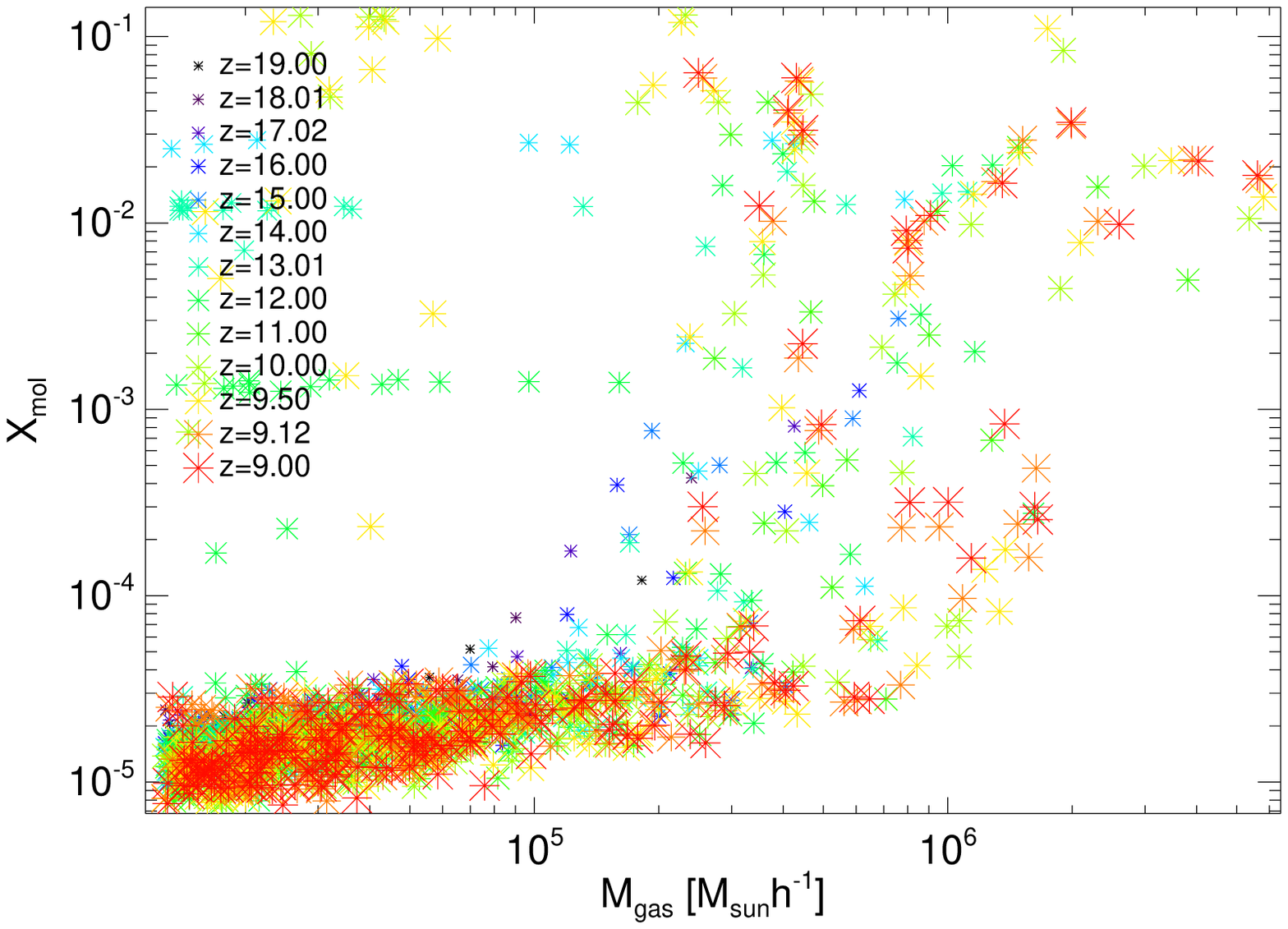}
\includegraphics[width=0.33\textwidth]{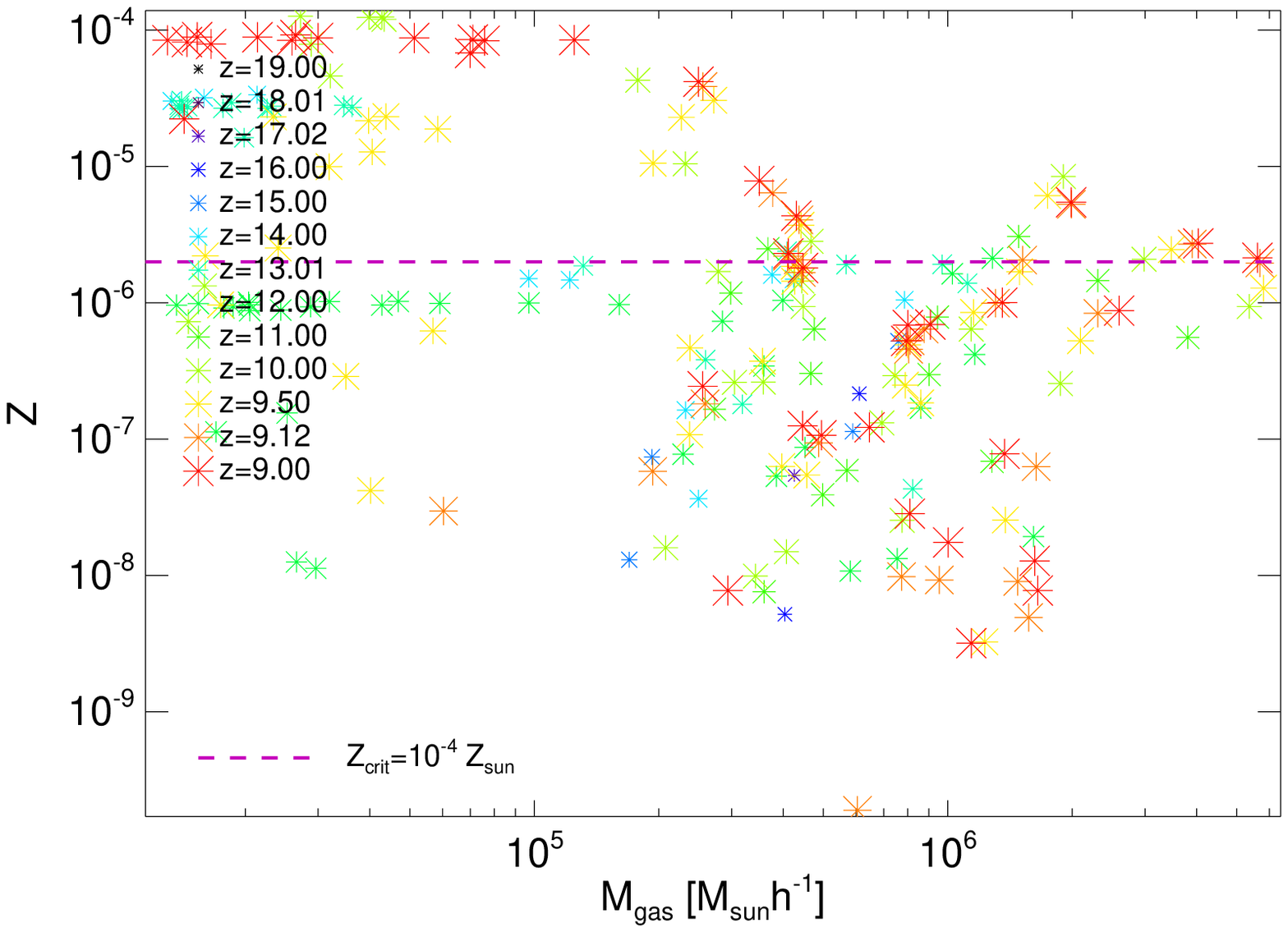}
\caption{
  Relations between gas mass, $M_{gas}$, and star-formation rate, SFR (left), molecular fraction, $x_{mol}$ (center), and metallicity, $Z$ (right). Color code and symbol size are given according to redshift, as in the legends. The purple horizontal dashed line in the right panel is drawn in correspondence of the critical metallicity $\Zcrit = 10^{-4}\Zsun$.
}
\label{fig:Mgasrelations}
\end{figure*}
%
%
\subsubsection{Chemical properties of the first galaxies}\label{sub:chemicalproperties}
In \fig\ref{fig:Mgasrelations}, we plot the dependencies of star formation rate (left panel), molecular fraction (central panel), and metallicity (right panel) from the gas mass of the various objects.
\\
The relation between star formation rate and gas mass results roughly linear, despite a wider spread at low masses, and consistent with \fig\ref{fig:masses} (right panel).
\\
In such regimes, proto-galaxies are experiencing the very first bursts of star formation (with a rapidly increasing star formation rate), but have not necessarily formed stars, yet.
In the whole redshift range considered, the smaller proto-galaxies
($\sim 10^5-10^6\,\msunh$ in gas) that are just undergoing their first
bursts ($\rm SFR > 0$, but $M_\star=0$)
correspond to roughly $\sim 20$ per cent of the total (27 objects out of 148) star
forming objects, while the remaining $\sim 80$ per cent (121 out of 148) has
already experienced star formation and feedback effects
($\rm SFR > 0$ and $M_\star>0$).
Thus, the spread at the low-mass end simply reflects these initial star formation stages, in contrast to those structures that have already an established star formation activity and lie on the roughly linear correlation individuated by the upper envelope of the distribution.
The general trend and the residual spread can be attributed to the competitive effects of different feedback mechanisms:
on one hand, the gas undergoing star formation is heated by SN explosions and inhibited to continue forming stars, mostly in smaller structures that suffer significantly gas evaporation processes (negative feedback);
on the other hand, shock compressions and spreading of metals in the medium enhance gas cooling capabilities and induce more star formation (positive feedback), mostly in bigger objects that can trap and re-process larger amounts of material.
The sensitivity to feedback is also well visible by comparing the trend of e.g. $z\sim 16-17$ (blue) points below $\sim 10^6\msunh$, that are quite aligned with the upper envelope of the dispersion plot and have not experienced feedback mechanisms, yet, and the corresponding (green) points at $z\sim 10-11$, in the same mass range, that are much more scattered due to the ongoing feedback effects in the local environments.
\\
When looking at the cold molecular gas in the different objects at various redshifts (central panel), we find that molecular fractions, $x_{mol}$, are rather low ($ x_{mol} < 10^{-4} $) mostly in small, quiescent, objects that are not star forming, like many of those with a gas mass $\lesssim 10^{5}\rm \msun$.
These galaxies populate significantly the bottom-left part of the diagram, where the corresponding star formation rates plotted on the left panel are null.
The regular trend of increasing molecular fraction with increasing mass is
visible up to $z\sim 15-16$, when initial collapse phases boost $x_{mol}$ up
to $\gtrsim 10^{-3}$, in galaxies of $\sim 10^5-10^6\,\msun$, that are forming
their first stars and are not altered by feedback mechanisms.
At $z \lesssim 15$, feedback effects from the first star forming galaxies become responsible for dispersing the values of molecular fractions of several orders of magnitude, either by dissociating molecules (hence, lowering their fraction), or by partially enhancing their formation (hence, increasing their fraction) by shocks and gas compression \cite[e.g.][]{Ricotti2001, Whalen2008, PetkovaMaio2012}.
For these reasons, molecular fractions in larger structures at later times have a scatter of a few orders of magnitudes, and some small mini-haloes have a somewhat more sustained molecular content at $z\sim 14 - 10$.
However, the total amount of cold gas in such small proto-galaxies is not enough to induce significant star formation episodes.
\\
Investigating the gas metallicity evolution in haloes of different mass (right panel) is useful to have hints about metal enrichment and to distinguish structures enriched by internal sources by those enriched by external ones.
Of course, we do not find points corresponding to low-mass and high-redshift haloes, where in fact there is only pristine, not enriched gas.
Gas starts to be enriched by metals at $z\sim 15-16$, (after the first bursts of star formation), when metallicities are initially $Z\sim 10^{-8} - 10^{-7}$ (i.e., $Z\sim 10^{-6} - 10^{-5} \Zsun$), and get enhanced up to higher values at lower redshift (see also \fig\ref{fig:sfr-Mgas-Z} and \fig\ref{fig:red_f_enrich}).
A comparison among the three panels of \fig\ref{fig:Mgasrelations} suggests that many low-mass galaxies with substantial enrichment (up to $Z\sim 10^{-4}$) at $z\lesssim 12$ have very low molecular fractions and no significant star formation:
they have been basically polluted by huge amounts of hot metals ejected by SN explosions or winds located in larger, closeby star forming haloes.
In particular, such enrichment process by external sources involves all the objects less massive than $\sim 10^5\,\rm\msun$ in gas, while larger ones can partially retain metals produced {\it in loco} and be self-enriched.
\\
As a practical example, in the $M_{gas}-{\rm Z}$ space, the (leftmost red point) halo at $z=9$ with the smallest mass ($M_{gas} = 1.25 \times 10^5 \msunh$), the highest metallicity (${\rm Z} = 8.46802\times10^{-5}\simeq 0.004\zsun$), but very small molecular fraction ($x_{mol}= 2.49226\times 10^{-5}$) has null star formation rate.
By tracing back its evolution, it comes out that the hosting halo has formed around $z\simeq 10$ and never underwent star formation episodes and consequent self-enrichment, since molecular fractions were to low to cool down the gas and trigger collapse and fragmentation.
This is actually a quiescent halo, formed in a pristine environment, and eventually enriched by hot metals coming from external sources (see next section).
%
%
\subsubsection{Thermal properties of the first galaxies}\label{sub:Tproperties}
\begin{figure}
\centering
\includegraphics[width=0.44\textwidth]{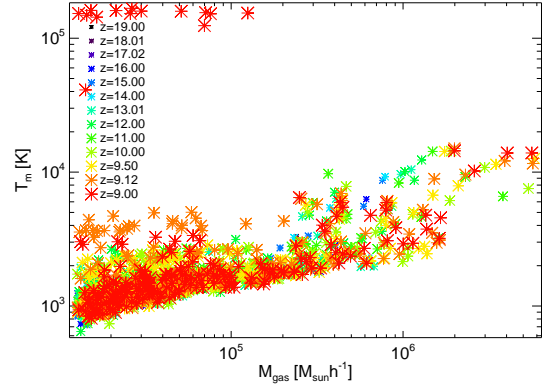}
\caption{
Relation between the gas masses, $M_{gas}$, and gas mass-weighted temperature, $T_m$. Color code and symbol size are given according to redshift, as in the legend.
}
\label{fig:tm_hottm}
\end{figure}
\begin{figure*}
\includegraphics[width=0.33\textwidth]{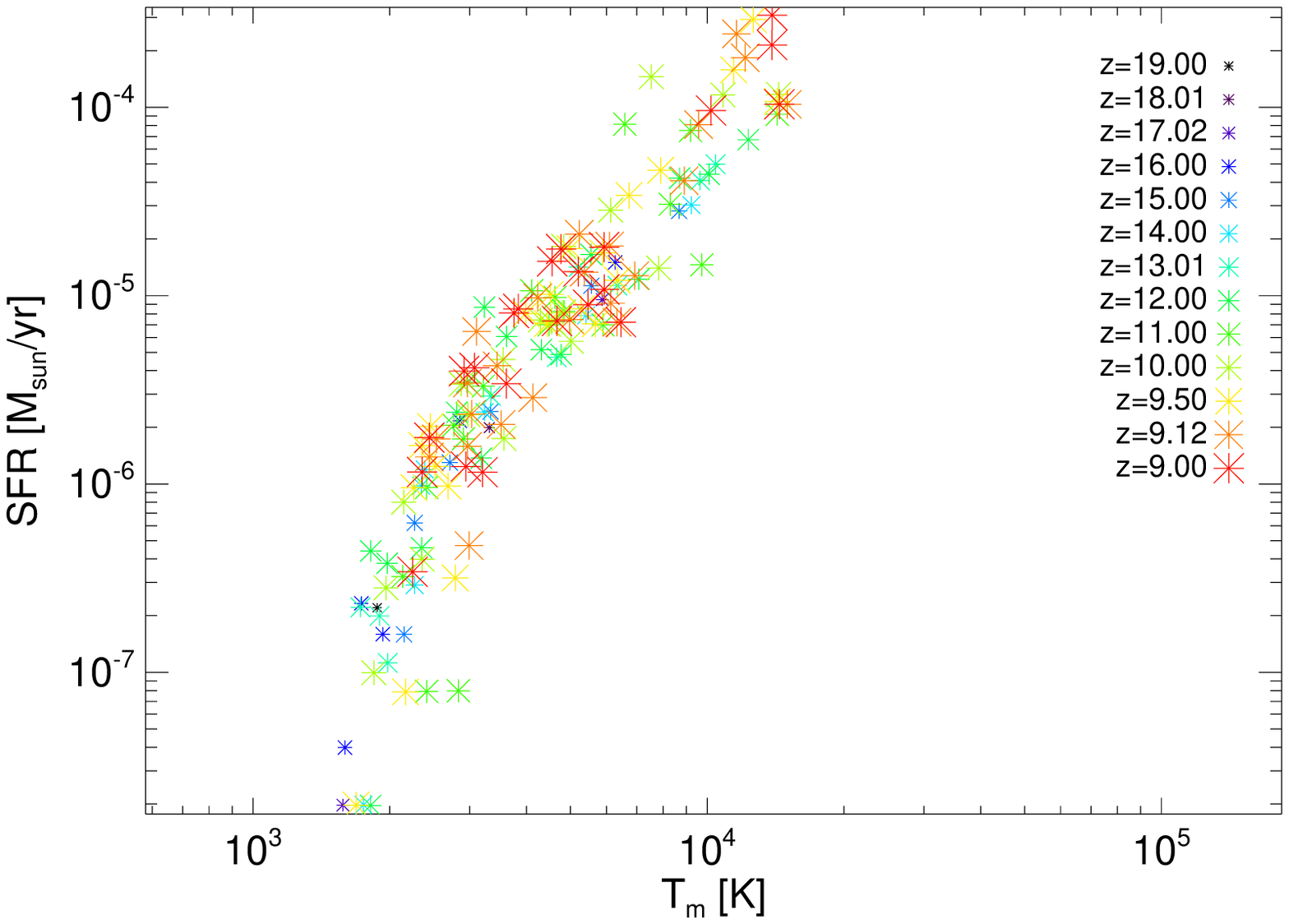}
\includegraphics[width=0.33\textwidth]{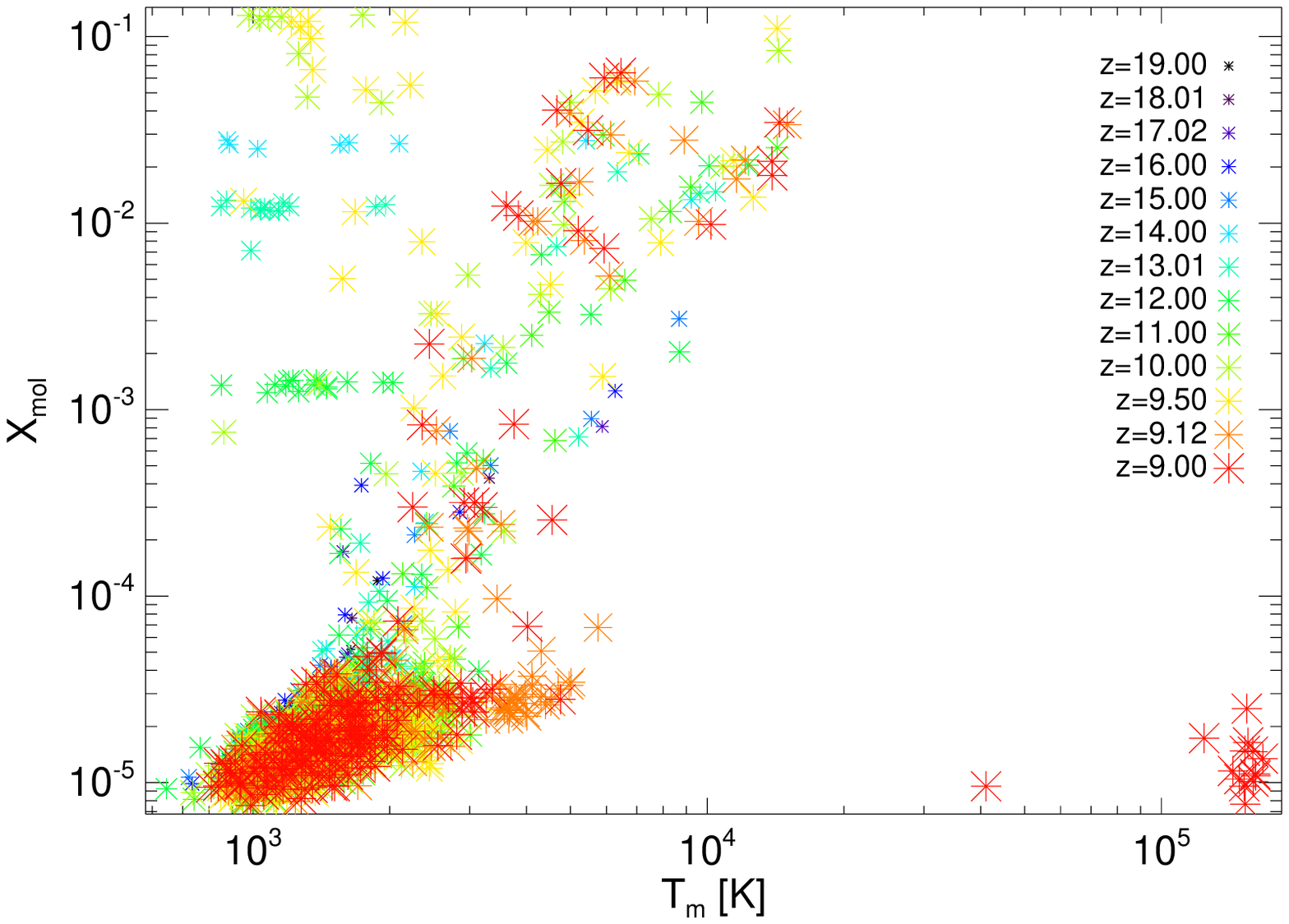}
\includegraphics[width=0.33\textwidth]{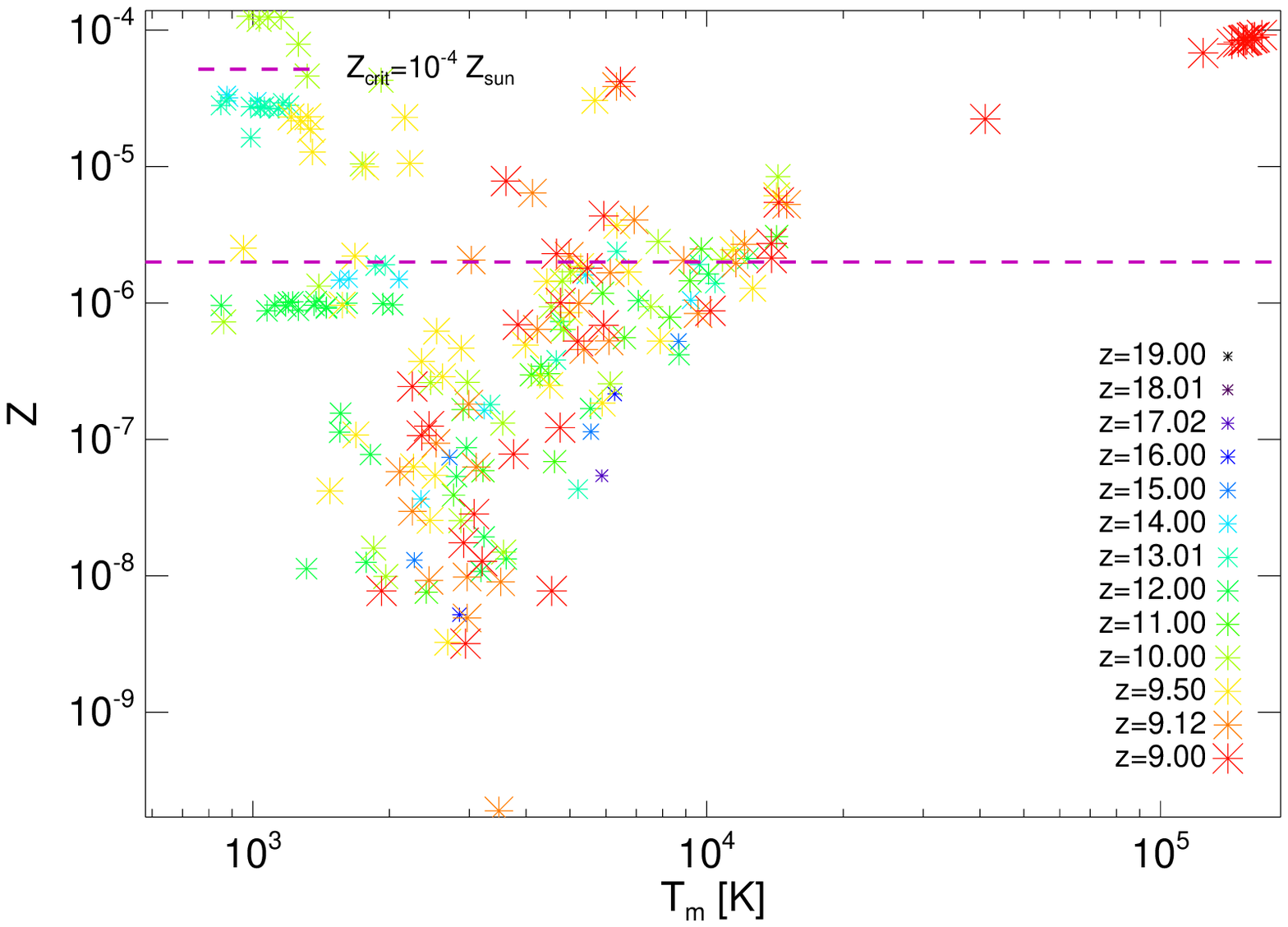}
\caption{
Relations between the gas mass-weighted temperature, $T_m$, and star formation rate, SFR (left), molecular fraction, $x_{mol}$ (center), and total metallicity, $Z$ (right). Color code and symbol size are given according to redshift, as in the legend.
The purple horizontal dashed line in the right panel is drawn in correspondence of the critical metallicity $\Zcrit = 10^{-4}\Zsun$.
}
\label{fig:Trelations}
\end{figure*}
\fig\ref{fig:tm_hottm} displays the mass-weighted gas temperatures, $T_m$, as a function of mass for the various redshifts considered.
As gas evolution is mainly led by molecular gas, typical temperatures fall in the range $\sim 500-10^4\,\rm K$, where H$_2$ shapes the thermal conditions of early objects.
Hotter temperatures are due to the thermal effects of SN explosions, whose shocks propagate in the primordial IGM, heat and enrich (see previous discussion) the gas in smaller objects (with masses $\lesssim 10^5\,\msun$) -- usually located within hundreds of parsecs \cite[e.g.][]{Creasey2012arXiv}.
The mass-weighted temperature increases with mass as a consequence of the virialization process in each halo, even though, the cold innermost regions will still be able to condense, collapse and fragment.
\\
In \fig\ref{fig:Trelations}, the basic relations of $T_m$ with star formation rate (left panel), molecular fraction (central panel), and metallicity (right panel) are presented.
It is interesting to stress that star formation is particularly
enhanced in the $\sim 10^3-10^4\,\rm K$ regime, where molecular
(H$_2$) cooling is mostly efficient and can determine gas collapse and
fragmentation. 
This underlines the importance of properly following chemistry
evolution to address primordial structure formation at proto-galactic
levels (in H$_2$-cooling haloes). 
Molecular and metallicity distributions better highlight:
\begin{itemize}
\item
  the actively star forming objects led by molecular-rich gas collapse at $T_m\sim 10^3-10^4\,\rm K$ (central panel) and responsible for metal spreading at $z\lesssim 17$ (right panel);
\item
  the many low-temperature, low-molecular-fraction ($x_{mol}\lesssim 10^{-4}$), pristine, quiescent objects (bottom left in the central panel, absent in the rigth panel);
\item
  some small (see \fig\ref{fig:tm_hottm}), high-temperature objects, whose properties are contaminated by hot enriched material at $T_m\gtrsim 10^5\,\rm K$ (upper right corner in the right panel) that enhances molecule dissociation in the local gas (bottom right corner in the central panel).
\end{itemize}
These objects (14, with metallicity $Z>10^{-4}\Zsun$) are all affected
by thermal feedback processes, but particularly smaller structures
(i.e. those with a gas mass $\lesssim 2\times 10^5\,\rm \msun$) tend
to be fully dominated by metal enrichement from external sources, due
to the lack of significant local star formation ($\rm SFR=0$) (see
also discussion about \fig\ref{fig:masses} and
\fig\ref{fig:Mgasrelations}). 
Roughly speaking, they represent $\sim 1$ per cent of the whole sample
at $z \gtrsim 9$ and correspond to $\sim 7$ per cent of the enriched
haloes. 
This means that $\sim 93$ per cent of polluted structures are
completely or partially self-enriched and, therefore, self-enrichment
is a significant pollution mechanisms of cosmic haloes. 
Finally, when considering only passive, non-star-forming haloes (83)
it turns out that purely externally enriched account for about 17 per
cent of the passive population.
\\

\noindent
A summary of early objects with given physical and/or chemical properties is presented in \tab\ref{tab:data}.
\begin{table}
\centering
\caption[Fractions]{Number of early haloes with given properties,
  classified in terms of: star formation rate ($SFR$), molecular
  fraction ($x_{mol}$), stellar mass ($M_\star$), metallicity
  ($Z$). The total number of considered objects is 1680.}
\label{tab:data}
\begin{tabular}{lccc}
\hline
           & $\rm SFR>0$   & $\rm SFR=0$     & $\rm SFR=0$       \\
           &               &                 & $x_{mol} < 10^{-4}$ \\
\hline
           &               &                 &\\
$M_\star=0$ & 27            &   1532          & 1470 \\
$M_\star>0$ & 121           &    0            & 0    \\
           &               &                 &\\
$Z=0 $     & 25            & 1449            & 1449 \\
$Z>0 $     & 123           & 83              & 21   \\
$Z>10^{-4}\Zsun$& 26        & 47              & 14   \\
           &               &                 &\\
\hline
\end{tabular}
\vspace{-0.4cm}
\end{table}
\subsection{Star formation and stellar activity}\label{sec:SFRandStars}
In the following we extend our investigations by 
presenting the stellar features of haloes with different 
dark and gaseous content (\sec\ref{sub:Mdm-Mstar-Mgas}),
by exploring the backreaction from star formation and stellar activity
(\sec\ref{sub:SFRproperties}) and by addressing
the physical relations with mass, metallicity and molecular content of
star forming  sites (\sec\ref{sub:SFRrelations}).
We conclude the section by discussing connections between theory and data 
(\sec\ref{sub:sSFRdiscussion}).
\begin{figure}
\begin{flushleft}
\includegraphics[width=0.44\textwidth]{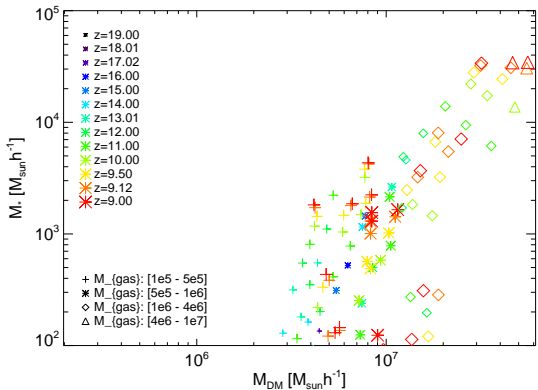}
\caption{
  Stellar mass as a function of dark-matter mass for the whole halo sample.
  Different symbols refer to different gas-mass bins:
  $\rm [1 \times 10^5,\, 5 \times 10^5]\msunh$ (crosses),
  $\rm [5 \times 10^5,\, 1 \times 10^6]\msunh$ (asteriscs),
  $\rm [1 \times 10^6,\, 4 \times 10^6]\msunh$ (diamonds),
  $\rm [4 \times 10^6,\, 1 \times 10^7]\msunh$ (triangles),
  as indicated by the legend on the bottom left.
}
\label{fig:MdmMstarMgas}
\end{flushleft}
\end{figure}
\subsubsection{Gas-to-star conversion efficiency}\label{sub:Mdm-Mstar-Mgas}
In the previous panels of \fig\ref{fig:masses} we have checked that larger dark-matter potential wells can retain larger amounts of gas and, hence, host more star formation.
The increase of the SFR with gas mass, as shown by e.g. \fig\ref{fig:Mgasrelations}, is obviously consistent with the fact that, if more gas is available, then higher SFRs and larger stellar masses are expected.
However, from our previous discussions it is still unclear whether the conversion of gas into stars will take place with the same {\it efficiency} or whether it will feature environmental dependencies.
\\
To this end, we combine the information about the dark, the gaseous and the stellar components and display in \fig\ref{fig:MdmMstarMgas} the trend of dark-matter mass versus stellar mass at given gas masses.
From the plot, it emerges that structures with
similar $M_{gas}$ are also characterised by similar $M_{DM}$, but
usually have very variable stellar masses.
Indeed, in the gas-mass bin $\rm [1 \times 10^5,\, 5 \times
10^5]\msunh$ (crosses), the spread in $M_\star$ is in excess of one 
dex, and this holds for the $\rm [5 \times 10^5,\, 1 \times
10^6]\msunh$ (asteriscs) bin, as well. 
For larger gas masses, as in the range $\rm [1 \times 10^6,\, 4 \times
10^6]\msunh$ (diamonds), expected values for the stellar component
have a scatter of more than two orders of magnitude. 
The most gas-rich objects, ranging within $\rm [4 \times 10^6,\, 1
\times 10^7]\msunh$ (triangles), show a 'smaller' scatter, of a factor
of a few, only because of statistical limitations. 
\\
For any fixed gas mass, haloes with very high stellar masses have
likely experienced significant boosts by merging events. 
In these cases, the role of their potential wells, mainly shaped by
the dark component, has been remarkable. 
On the other hand, more common haloes with intermediate or low stellar
fractions have either evolved in isolation or their stellar growth has
been (partially) suppressed by feedback mechanisms (see next
section). 
\\
Overall, this means that stellar-mass formation from gas collapse (a
process which is commonly parameterized by means of a gas-to-star
conversion efficiency) is very strongly dependent on the particular
environments and on the local ongoing physical processes. 
Indeed, \fig\ref{fig:MdmMstarMgas} suggests that same amounts of gas
mass can generate very different amounts of stellar mass, depending
on, e.g., host molecular content, heavy-element abundances, cooling
capabilities, SN heating, shocks, star formation feeback, gas
evacuation, photoevaporation, stripping, boosting by mergers, etc.. 
Interestingly, this conclusion is quite independent from the
particular redshift considered. 
\begin{figure*}
\centering
\includegraphics[width=0.33\textwidth]{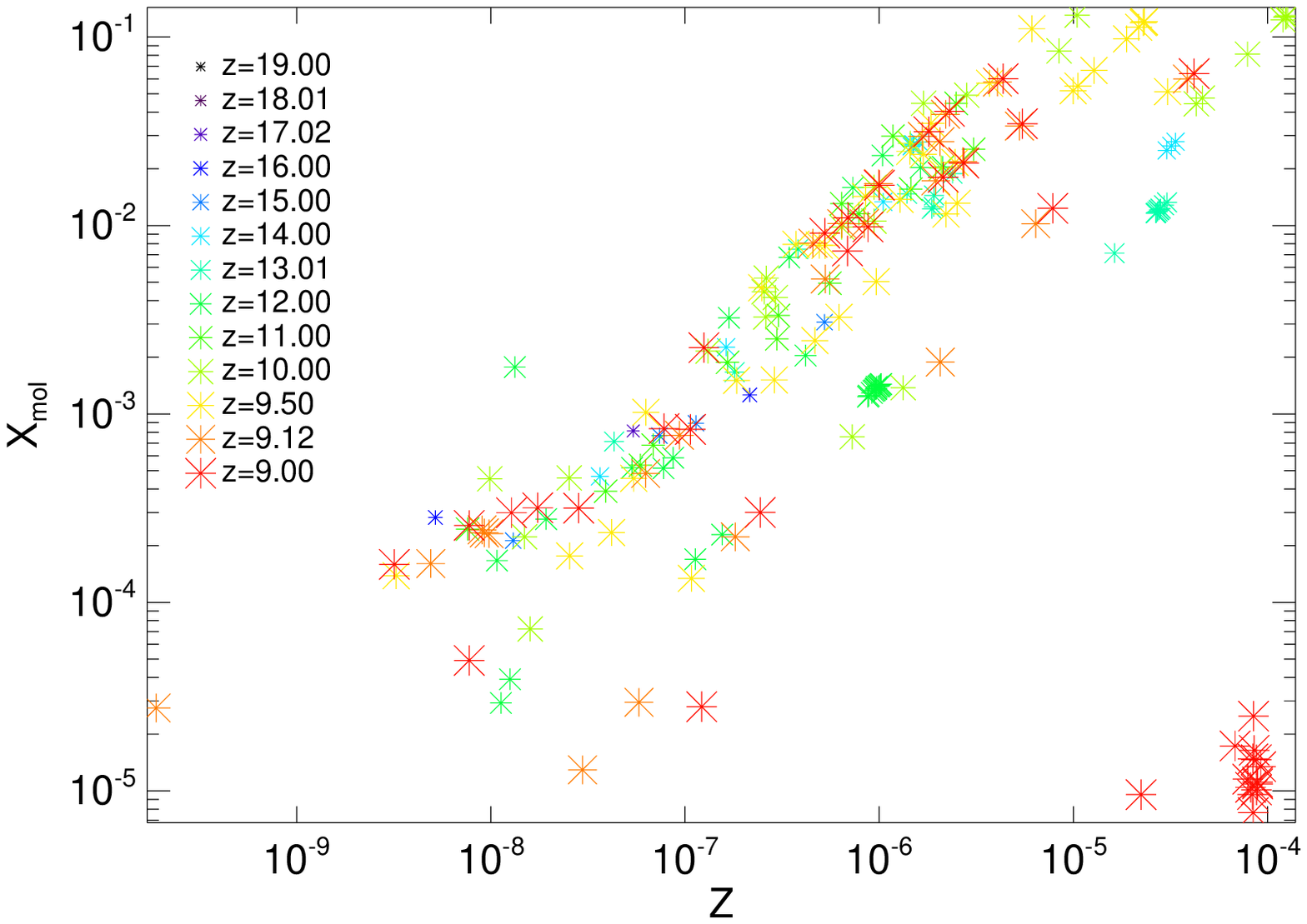}
\includegraphics[width=0.33\textwidth]{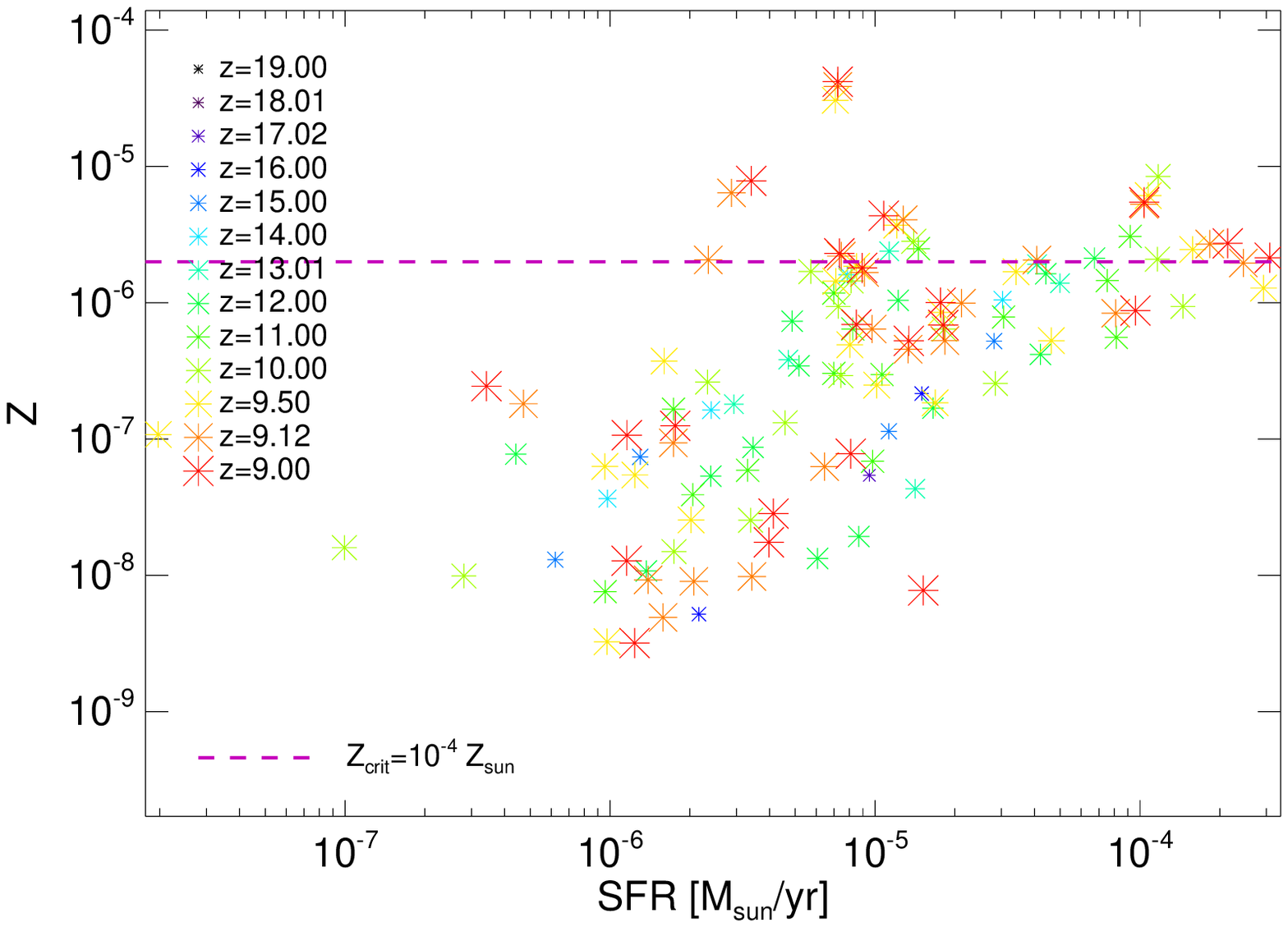}
\includegraphics[width=0.33\textwidth]{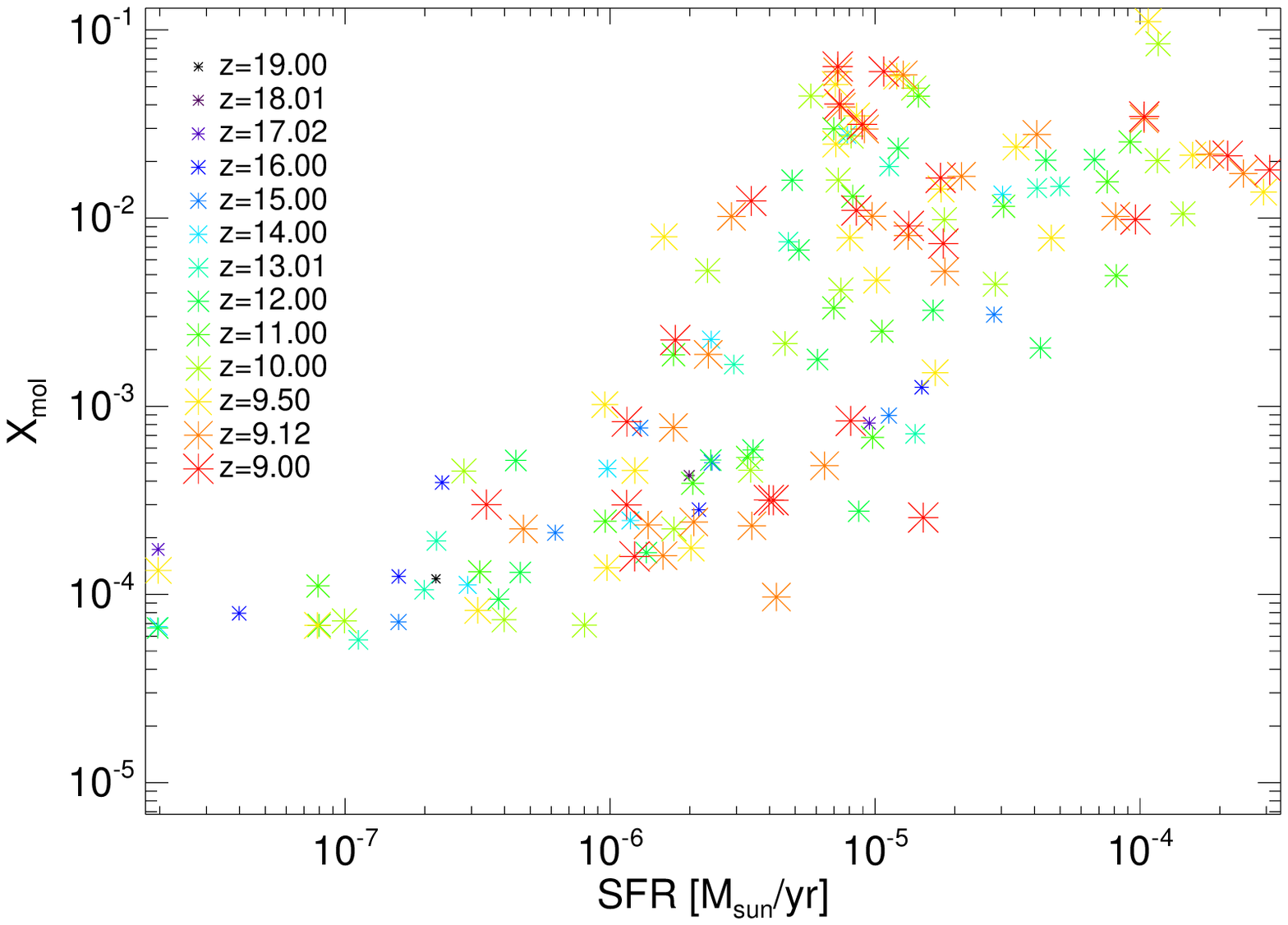}
\caption{
Relations between chemical properties and star formation. We plot the behaviour of molecular fraction, $x_{mol}$, and metallicity, $Z$ (left), metallicity and star formation rate, SFR (center), and molecular fraction, $x_{mol}$, and star formation rate, SFR (right).
Color code and symbol size are given according to redshift, as in the legend.
The purple horizontal dashed line in the central panel is drawn in correspondence of the critical metallicity $\Zcrit = 10^{-4}\,\Zsun$.
}
\label{fig:chemicalrelations}
\end{figure*}
\subsubsection{Effects from feedback mechanisms}\label{sub:SFRproperties}
\fig\ref{fig:chemicalrelations} explicitly highlights the connections between the gaseous chemical properties of proto-galaxies (i.e. molecules and metals) and their star formation activity.
\\
Molecule and metal distribution at different redshifts is plotted on the left panel for all the enriched galaxies in our sample.
Since metals are expected to be produced by short-lived stars in molecular-rich star forming sites, the larger the molecular fractions, the larger the metallicities.
A more interesting feature is the bottom right part, though: there the aforementioned objects with high metallicities and very low molecular fractions are placed. They are strongly displaced by the general trend, but this allows us to recognize the effects of metal enrichment of quiescent galaxies polluted by external sources.
The upper left corner is not populated because gas with large molecular fractions of $x_{mol} \sim 10^{-2}$ or higher would have very short cooling times, hence would form immediately stars polluting the surrounding medium\footnote{
This means that the probability of finding objects with $Z\simeq 0$ and $x_{mol}\sim 1$ is extremely low.
}.
Consequently, the bigger the deviation from the general increasing trend, the higher the effects of feedback mechanisms.
This is the case for the already mentioned smaller galaxies, while the larger ones, dominated by self-enrichment and not heavily affected by external enrichment, lie along the main trend for $x_{mol}$ and $Z$.
\\
In line with these conclusions, the central plot of \fig\ref{fig:chemicalrelations} shows the rather weak correlation between star formation rate and gas-phase metallicity.
Broadly speaking, there is an obvious tendency to have more enrichment for higher star formation rates, with the $\sim 10\%$ massive popII-I structures (see \sec\ref{sub:SFhaloes}) forming stars at rates $\gtrsim 10^{-5} - 10^{-4}\,\rm \msun/yr$.
However, feedback effects (more powerful for larger star formation rates) introduce a remarkable scatter that can be noted mostly at later times (i.e. $z \lesssim 10$), when such processes are at more advanced stages and additional metals have been ejected into a wider range of star forming environments.
\\
In addition, star formation feedback has further effects on molecular evolution (right panel).
Even though molecular fractions follow more tightly star formation rates (mostly
at early epochs, at $z \gtrsim 15$), the imprints of stellar evolution
become visible later on, as they influence directly the thermodynamical
properties of the ambient gas and hence the following formation or destruction of molecules within $\sim 2$ orders of magnitude.
\\
As a conclusion, feedback effects leave prominent traces in the baryonic growth of cosmological structures that are reflected in broader and more scattered trends at later times.
\\
Natural implications are that they can easily contaminate large-scale cosmological signatures, can be relevant sources of patchiness of cosmic chemical enrichment, can alter the gas thermal state at high $z$, and can affect the consequent HI or molecular emissions during the late dark ages.
\subsubsection{Relations between stellar and chemical properties}\label{sub:SFRrelations}
\begin{figure*}
\centering
\includegraphics[width=0.33\textwidth]{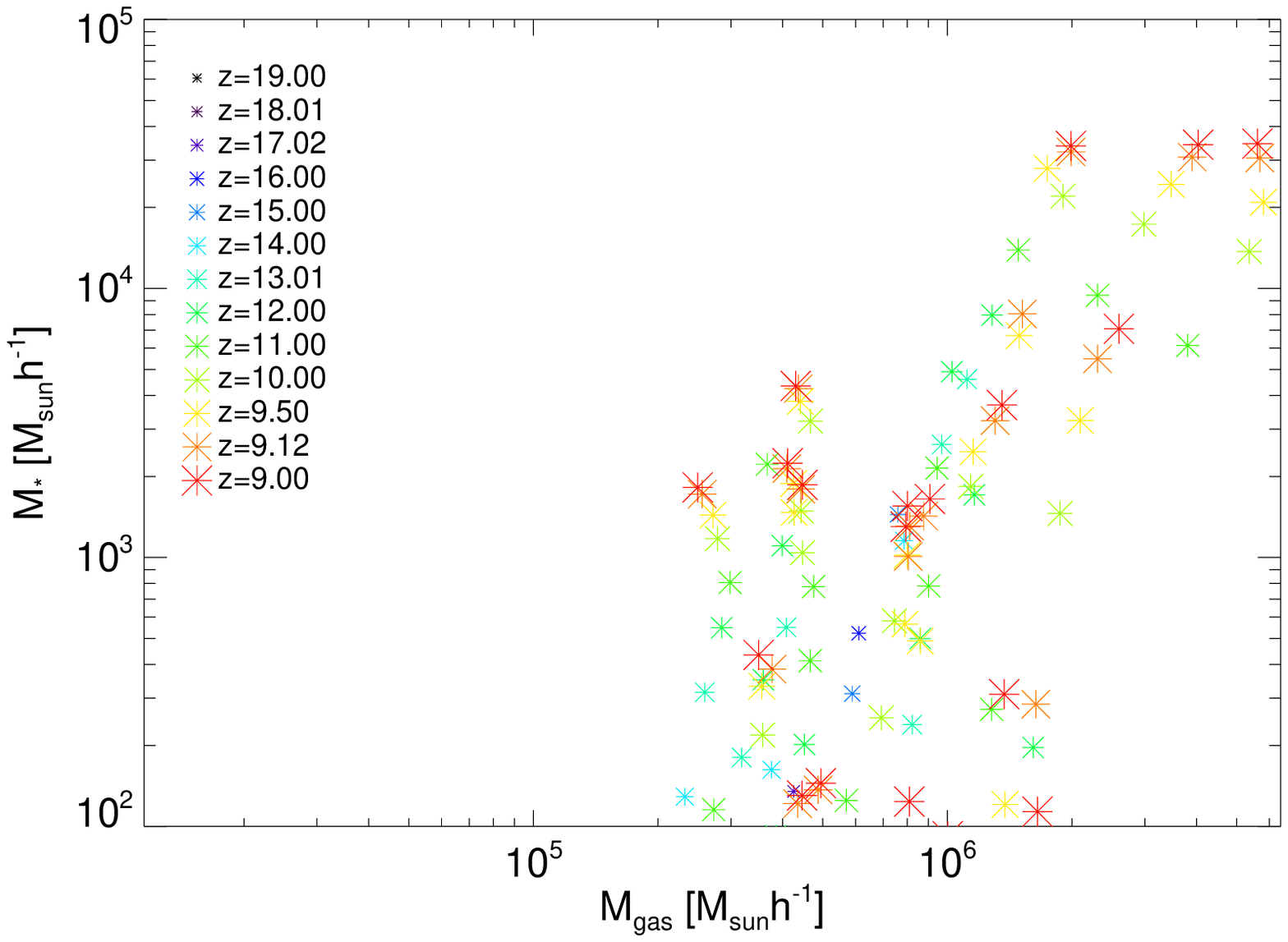}
\includegraphics[width=0.33\textwidth]{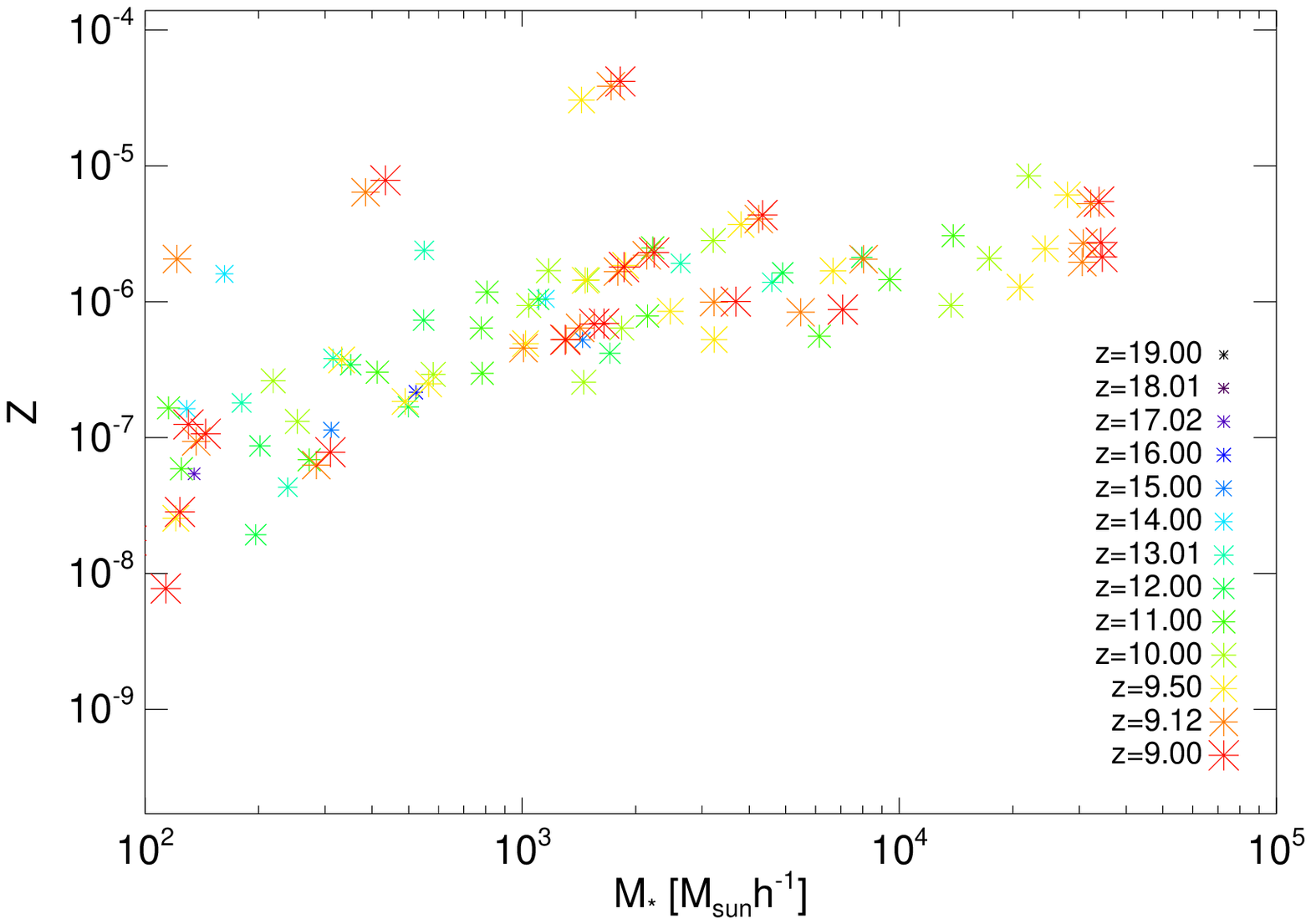}
\includegraphics[width=0.33\textwidth]{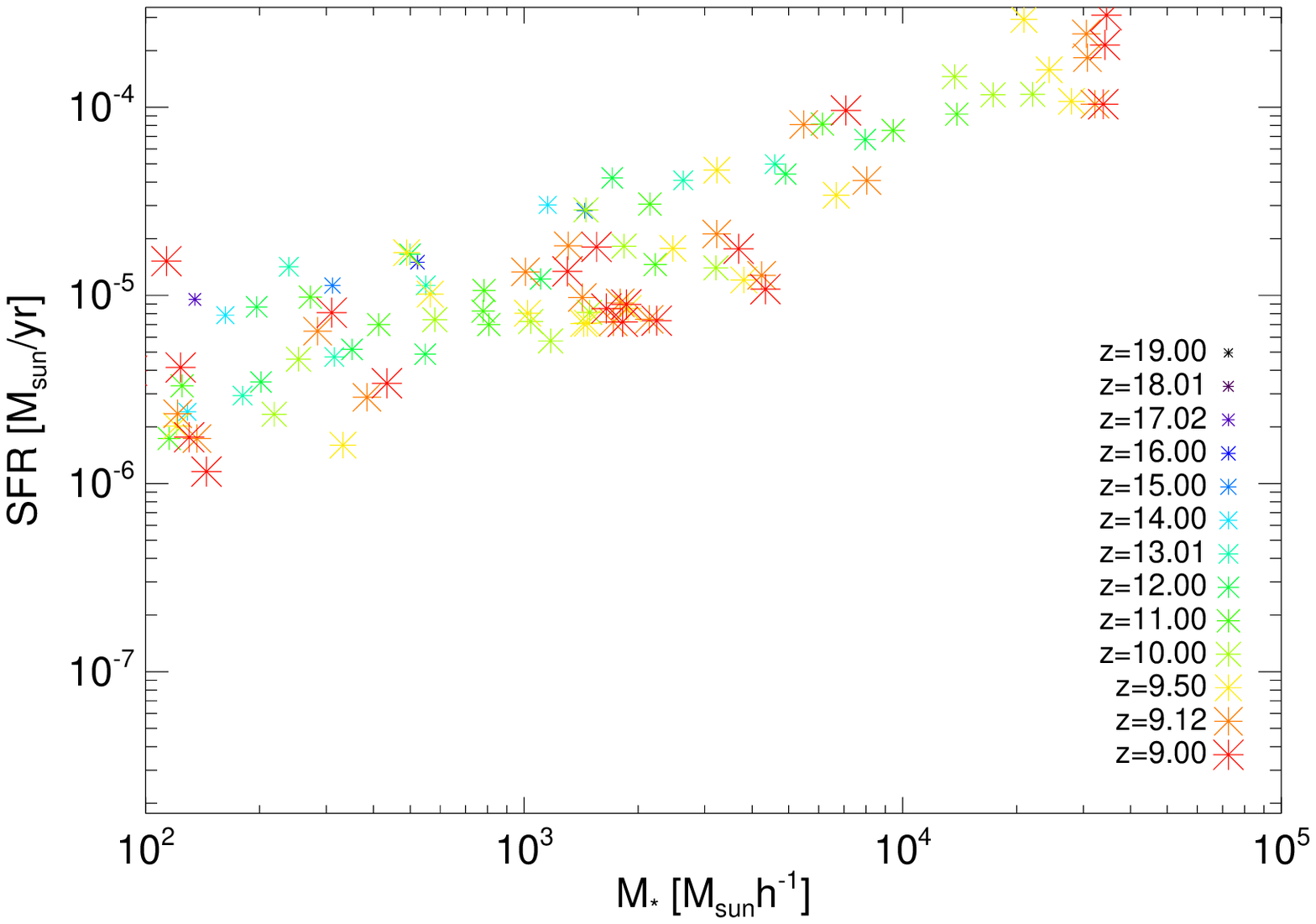}
\caption{
Relations between the stellar mass, $M_\star$, and gas mass, $M_{gas}$, (left), metallicity, $Z$ (center), and star formation rate, SFR (right). Isolated $Z$ values scattered around the bulk of the sample refer to structures subject to strong feedback and pollution processes.
}
\label{fig:Mstarproperties}
\end{figure*}
\begin{figure*}
\centering
\includegraphics[width=0.33\textwidth]{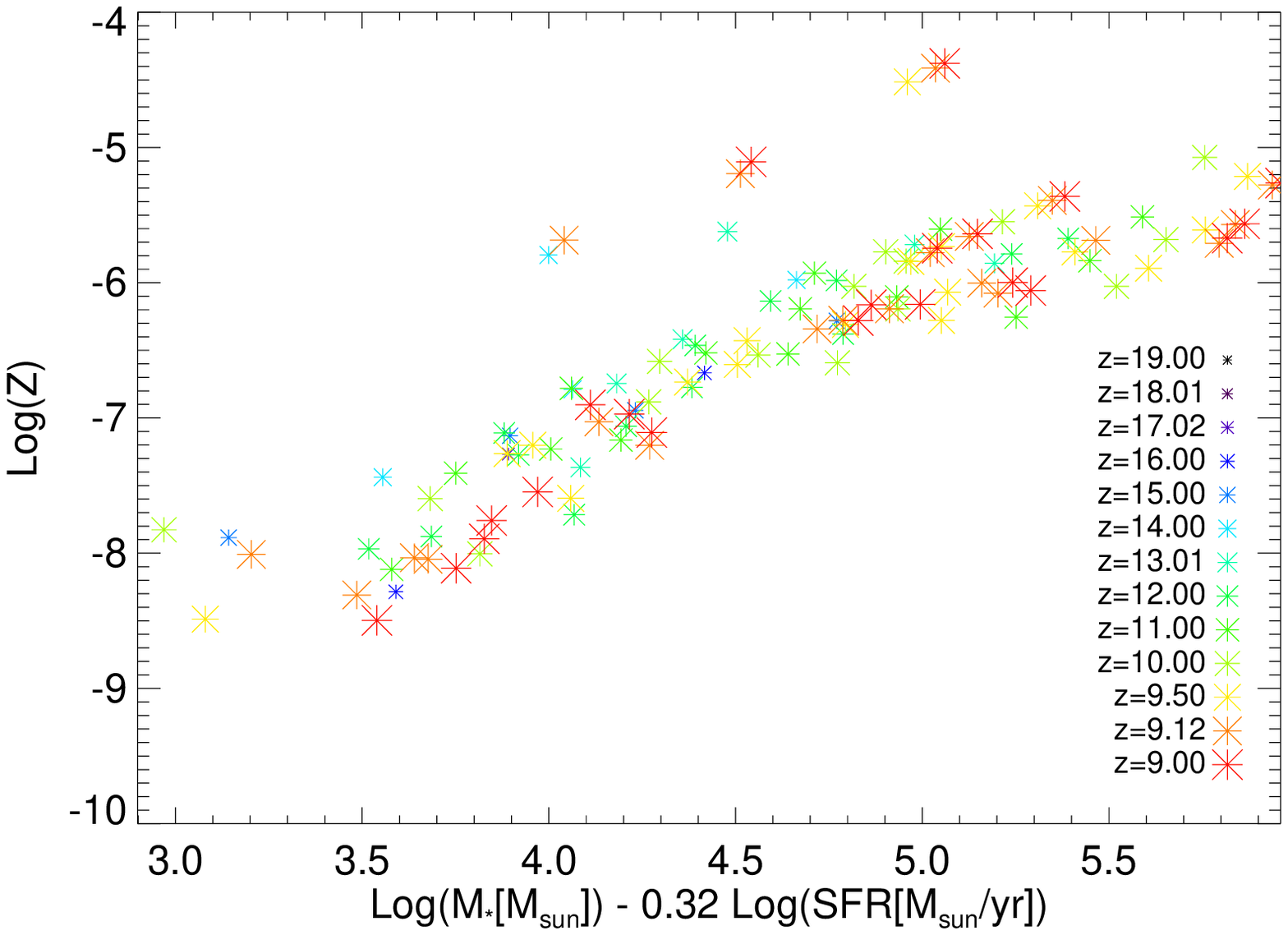}
\includegraphics[width=0.33\textwidth]{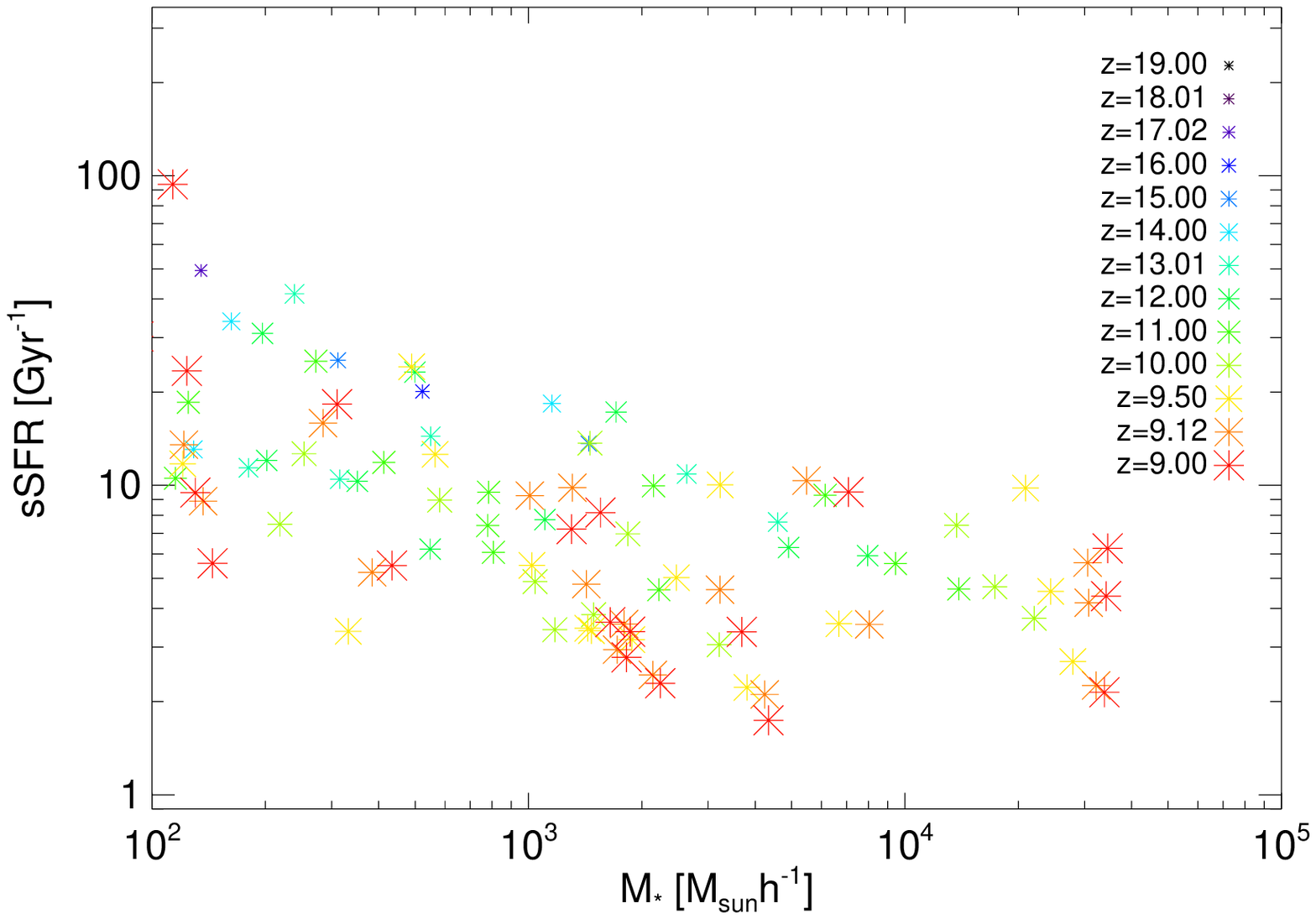}
\includegraphics[width=0.33\textwidth]{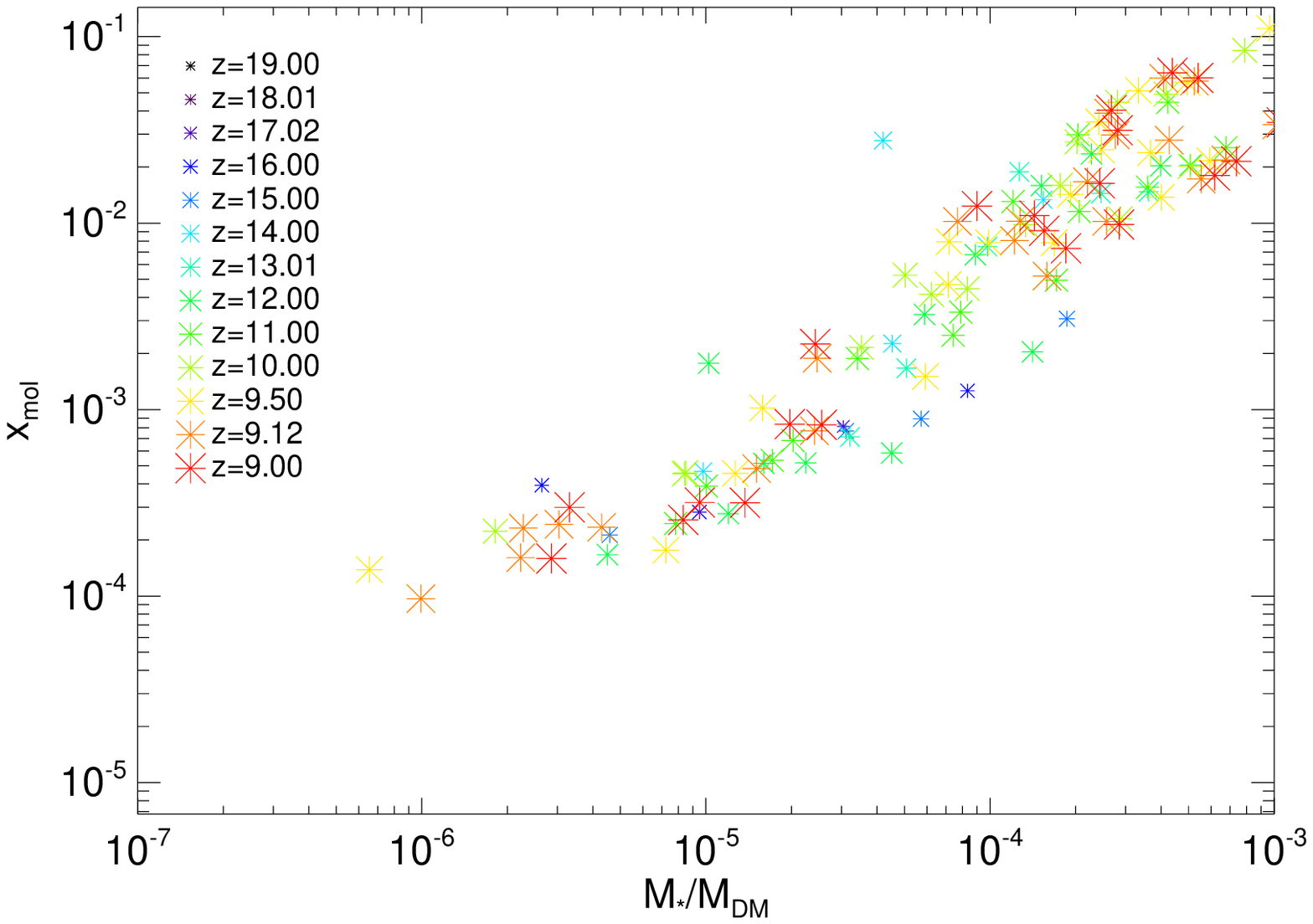}
\caption{
  $\rm SFR - M_\star - Z$ relation for our sample (left), discussed in the text.
  The relations sSFR-$M_\star$ (center) and $x_{mol}-f_\star$ (right) are also plotted for different redshifts.}
  \label{fig:FMR}
\end{figure*}
\begin{figure}
  \centering
  \includegraphics[width=0.44\textwidth]{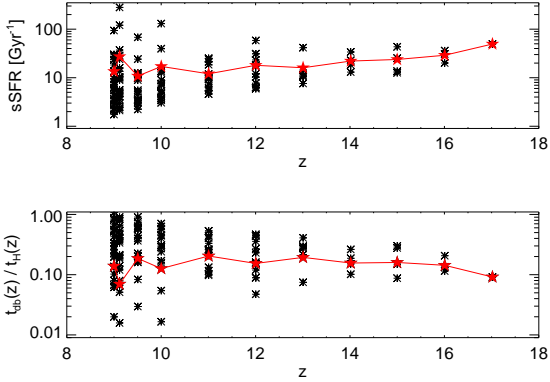}
  \caption{
    {\it Upper panel}: Redshift evolution of the specific star formation rate (sSFR) in Gyr$^{-1}$.
    {\it Lower panel}: Redshift evolution of the ratio between the inverse of the specific star formation rate (doubling time) and the age of the Universe (Hubble time), $t_{\rm H}$. The solid lines connect the average values (five-points stars) at each redshift.
  }
  \label{fig:sSFR}
\end{figure}

At this point, we study possible relations between stellar and chemical features in primordial environments.\\
In \fig\ref{fig:Mstarproperties} we plot the principal stellar properties for each object at each redshift.
Left panel displays the distribution of masses for gas and star component.
Although the stellar content is generally larger in objects with larger amounts of gas, it is not conceivable as a fixed stellar fraction parameter, $f_\star$, simply scaling with gas or dark-matter mass.
This because stars are expected to form in cold, dense, molecular-rich, collapsing gas, that is very sensitive to the typical, local environment.
The broadly increasing stellar content in objects with larger gas mass is related to the fact that bigger structures have a larger probability of hosting cold molecular gas and of forming stars.
On the other hand, feedback effects can (depending on the particular star formation regime, stellar evolution stages, SN rates, SFRs, halo interactions or mergers, etc.) inhibit or enhance stellar mass growth in very different ways, even in galaxies with comparable mass, causing the large spread in the data.
The mass-metallicity relation \cite[e.g.][]{Tremonti2004} in the central panel shows a much more clear trend that is understood in terms of more metal enrichment in larger and more stellar-rich structures. Also in this case the outliers in the mass-metallicity relation are small structures subject to strong feedback and metal enrichment from larger, closeby sources.
We note that the SFR is roughly proportional to $M_\star$, as demonstrated by the right panel, and this is consistent with local-Universe \cite[e.g.][]{Brinchmann2004, Salim2007, Peng2010} and intermediate-redshift \cite[e.g.][]{Noeske2007, Elbaz2007, Pannella2009, Rodighiero2011, Karim2011, Michalowski2010} data\footnote{
  We remind that observational data and stellar-mass reconstructions between $z\sim 0$ and $z\sim 3$ suggest a slope of $\sim 0.6-1$ and a normalization evolving with $(1+z)^{3.5}$ (see references in the text).
}.
\\
The left panel of \fig\ref{fig:FMR} summarizes the various dependencies with the help of a fit-constructed relation between metallicity, star formation rate and stellar mass, containing a free parameter, $\alpha$, tuned to minimize the scatter of $Z$.
More exactly, we plot $Z$ as a function of the adimensional combination  $\mu_\alpha \equiv \rm Log(M_\star[\msun]) - \alpha Log(SFR[\msun/yr])$, with $\alpha=0.32$ fixed by fitting observational SDSS data for field galaxies at low redshift \cite[][]{Mannucci2010}.
We note that the existence of the relation between $Z$ and $M_\star$, and the one between $M_\star$ and SFR (plotted in \fig\ref{fig:Mstarproperties}) would suggest by its-self a correlation between $Z$ and $\mu_\alpha$ (a function of $M_\star$ and SFR with an additional free parameter).
A more limited spread for $Z$ is obtained by tuning $\alpha$.
Our simulated sample shows a correlation even for early $z > 9$ proto-galaxies in primordial mini-haloes\footnote{
  This is in line with intermediate-redshift studies \cite[][]{Salvaterra2013} that reach similar conclusions for $z\sim 6 - 10$ galaxies.
},
even though a few outliers show up.
Indeed, some points have $Z$ values that lie significantly away from the bulk of the sample, almost 2 orders of magnitude (as in \fig\ref{fig:Mstarproperties}).
This is essentially due to feedback and pollution processes that are not properly accounted for by the $\mu_\alpha$ parameterization (since this relies on field galaxies only) and those can determine an evident spread of the data mostly in crowded regions.
It is worth stressing that, while the $\mu_\alpha$ parameterization might be a simple and observationally friendly way to fit data, more significative physical hints can be drawn by the following equivalent expression:
\begin{equation}
  \rm
  \mu_\alpha=(1-\alpha) \Log(SFR[\msun/yr]) - \Log(sSFR[yr^{-1}]),
\end{equation}
where $\rm sSFR \equiv SFR/M_\star$ is the specific star formation rate which, due to the $\rm SFR-M_\star$ relation of \fig\ref{fig:Mstarproperties}, correlates with stellar mass (central panel of \fig\ref{fig:FMR}), as well\footnote{
  The correlation has a negative power, as also expected from lower-redshift observations \cite[e.g.][]{Daddi2007, Rodighiero2011}.
}.
For any fixed mass bin, the sSFR features a larger spread with decreasing redshift, that appears as a consequence of the different environmental conditions developing during gas assembly and star formation episodes.
Larger sSFRs are typical for objects that are just forming their first stars, while smaller sSFRs are found in objects that have already consumed gas available for star formation and/or have experienced the backreaction from stellar evolution.
\\
A deeper insight into the star formation process is retrieved by means
of the right panel in \fig\ref{fig:FMR}, where the molecular content
of each object is plotted against its stellar fraction, $f_\star$.
The increasing behaviour of $x_{mol}$ for growing stellar component
shows how tightly star formation is linked to the chemical features of the collapsing medium.
From these considerations, molecular content results to be the main driver of 
stellar growth, with an approximate scaling
\begin{equation}
  f_\star \propto x_{mol}.
\end{equation}
In addition, this connotes $Z$ as strongly related to the ongoing star formation activity, i.e. SFR and sSFR, that are in turn driven by cooling and fragmentation processes in dense, cold, molecular-rich gas.
\\
Finally, we also plot the redshift evolution of the specific star formation rates (sSFR) in the upper panel of \fig\ref{fig:sSFR}, and the corresponding evolution of the doubling times, $t_{db}=\rm (sSFR)^{-1}$, in units of cosmic time, $t_{\rm H}$, in the lower panel, with their respective average values (solid lines).
The typical sSFR ranges between a few $\rm Gyr^{-1}$ and some tens of $\,\rm Gyr^{-1}$, with some very bursty objects reaching $\sim\rm 10^2\,Gyr^{-1}$.
Despite the evolution of the sSFR, the mean ratio between the doubling time and the corresponding Hubble time is always roughly constant in redshift, and equals: $t_{db}(z) / t_{H}(z) \simeq 0.1-0.2, $
in very good agreement with later-time expectations.
%
%
\subsubsection{Theory and data}\label{sub:sSFRdiscussion}
\begin{figure}
\begin{flushleft}
  \includegraphics[width=0.5\textwidth]{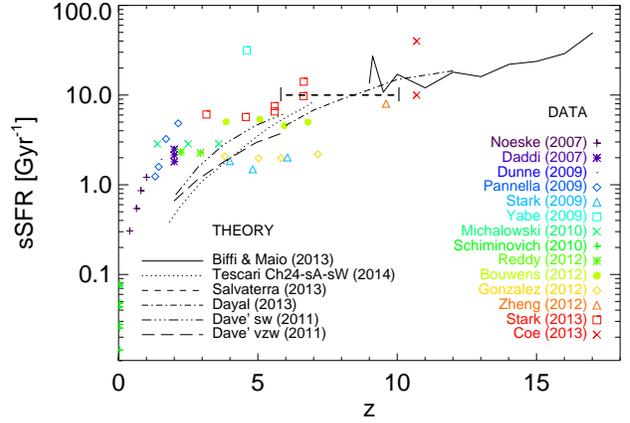}
  \caption{
    Redshift evolution of the specific SFR as inferred from different observational data samples and various numerical hydrodynamical simulations including gas cooling, star formation and feedback effects (see legends). The horizontal dotted line at $\rm sSFR\simeq 10\,Gyr^{-1}$ is the mean value at $z\sim 6-10$ expected by Salvaterra et al. (2013). The two points at $z\simeq 10.6$ are rough estimates for a stronlgy lensed image by Coe et al (2013).
  }
  \label{fig:sSFRcollection}
\end{flushleft}
\end{figure}

We stress that the sSFR values are larger than law- or
intermediate-redshift galaxies having typical $\rm sSFR \lesssim
0.3\,Gyr^{-1} $ at $z\lesssim 0.5$ 
\cite[][]{Noeske2007},
steeply increasing to $\rm sSFR \sim 2\,Gyr^{-1} $ at $z\gtrsim 2$
\cite[e.g.][]{Daddi2007, Pannella2009, Michalowski2010, Reddy2012}
and reaching $\rm sSFR \sim 2-10\,Gyr^{-1} $ at $z\sim 4-8$
\cite[][]{Gonzalez2012, Stark2013}.
Additionally, our findings are consistent with theoretical analyses of
star forming objects at $z\sim 6-10$ \cite[][]{Salvaterra2013} that
suggest $\rm sSFR \sim 10\,Gyr^{-1}$, and are in line with recent
observational estimates at $z\simeq 9.6$ \cite[][]{Zheng2012} and
$z\sim 10.6$ \cite[][their Sect. 7]{Coe2013}, as well, as summarized
in \fig\ref{fig:sSFRcollection}. 
We notice that theoretical calculations are fairly comparable with
data only at intermediate redshift, around $z\sim 5-9$. 
In those regimes, \cite{Salvaterra2013} could reproduce -- with the
help of simulations similar to the ones used here -- the observed UV
luminosity functions in the $\rm [-22, -16]$ magnitude range by
galaxies with stellar masses $\sim 10^4-10^9\,\msun$, and could
estimate typical SFRs (around $\sim 10^{-3} - 30\,\msun/yr$) and
metallicities ($\sim 10^{-3}-10^{-1}\zsun$) for high-$z$ GRB hosts. 
The expected values resulted to be consistent with recent
observational constraints \cite[for details, see][and references
therein]{Salvaterra2013}. 
\cite{Dayal2013tmp} reached reasonable agreement between the expected
stellar mass density and corresponding values inferred observationally
by e.g. \cite{Stark2013, Labbe2010}. 
Similarly, \cite{Dunlop2013} compared their observed UV slope $\beta
\simeq -2$ at $z\simeq 7-9$ with predictions by models of galaxy
formation within cosmological chemical evolution simulations
\cite[][]{Maio2010, Dayal2013tmp} and found consistent results. 
Very recently, theoretical predictions by \cite{Maio2013z7arXiv} have
demonstrated that cold collapsing gas or damped Ly-$\alpha$ candidates
might be hosted already at $z\simeq 7$ in molecular-rich objects with
gas masses $\sim 10^8\,\msun$ and metallicities spanning several
orders of magnitudes. 
At $z\lesssim 5$, when objects are usually bigger by a few orders of
magnitude, there are discrepancies appearing between theory and data. 
Models seem to be roughly consistent only down to $z\sim 3$, while
they fail to address star formation and stellar mass growth in massive
galaxies at $z< 3$, as found by e.g. \cite{Dave2011} and Tescari
(2014, private communication). 
On the other extreme, observational data at $z>9$ are highly
complicated to obtain. 
Stellar masses and SFRs for the $z=9.6$ and $z=10.6$ candidates in
\fig\ref{fig:sSFR} were inferred by SED fitting coupled to synthetic
stellar population modeling or to order-of-magnitude estimates
\cite[][]{Zheng2012, Coe2013}, for low-luminosity, strongly lensed
images. 
Considering the large number of unknowns, degeneracies and
assumptions, the resulting values carry relevant uncertainties and,
therefore, it is not easy to directly confront them with our
analysis. 
Never the less, the typical picture emerging from
\fig\ref{fig:sSFRcollection} is that the primordial Universe was made
of early small galaxies that were extremely bursty, with typical $\rm
sSFR \gtrsim 50\,\rm Gyr^{-1}$ at $z\sim 10-20$, decreasing down to
$\rm sSFR\sim 10\, Gyr^{-1}$ at $z\lesssim 10$, and $\rm sSFR <
1\,Gyr^{-1}$ at $z < 1$. 
Consistently, also metal pollution possibly led to an increase (by
several orders of magnitude over cosmological time) of the typical
galactic metallicities and of the expected cosmic metal filling factor
\cite[see also][]{MaioIannuzzi2011, Maio2011cqg, Maio2013z7arXiv}.
\subsection{Dynamical patterns in the first structures}\label{sub:L}
As a final task, we focus on the dynamical patterns of gas and dark
matter components and study the trends for angular momentum
calculations (\sec\ref{sub:Lcalculations}).
After showing a few explicative cases (\sec\ref{sub:cases}), we
will address the role of structure evolution in the formation and/or
destruction of primordial gaseous discs (\sec\ref{sub:evol}) and
in shaping their statistical rotational features
(\sec\ref{sub:statistics}). 
Then, we will check possible correlations between molecular discs and
hosting masses (\sec\ref{sub:rel_discs_mass}) and eventually the
occurrence of gas and dark-matter alignment
(\sec\ref{sub:global_alignment}) in primordial epochs. 
\subsubsection{Angular momentum calculations}\label{sub:Lcalculations}
Taking advantage of the full information accessible through our
numerical simulations, we are able to calculate the angular momentum
of any  particle in the haloes. This is done by using the
three-dimensional velocity associated to the particle, its
distance from the halo center and its mass. 
Then, as usually defined \cite[see, e.g.,][]{BinneyTremaine2008}, the
specific angular momentum, $\vec{j} $, of each particle is calculated
as 
\begin{equation}
\vec{j} = \vec{l}/m = \vec{r} \times \vec{v},
\end{equation}
where $\vec{l}$ is the angular momentum, $\vec{r}$ and $\vec{v}$ are the position and velocity of the particle with respect to the center, and $m$ the particle mass.
\\
In order to quantify the fraction of gas that might be found in rotating configurations, we calculate the mean angular momentum of the gas,  $\vec{J}_{mean}$, and we compare against the specific angular momentum of each gas element, considering the projection, $j_{z}$, along the direction defined by $\vec{J}_{mean}$.
Namely, the quantity
\begin{equation} 
\label{eq:eta}
\eta =j_z/|\vec{J}_{mean}|
\end{equation} 
provides a measure of the alignment of the angular momentum of every single gas element with the mean angular momentum.
\\
Alternatively, this alignment can be quantified by the cosine of the angle defined by $\vec{J}_{mean}$ and the angular momentum of the gas element, i.e.:
\begin{equation}
\label{eq:cos}
\cos\vartheta = \frac{\vec{j} \cdot\vec{ J}_{mean}}{|\vec{j}||\vec{ J}_{mean}|}.
\end{equation}
We note that, with the definition in equation (\ref{eq:eta}), the value of $\eta$ is not necessarily belonging to the range $[-1,1]$, while the cosine defined in equation (\ref{eq:cos}) is always in the range $[-1,1]$.
\\
To retrieve more information closely related to the circularity of the orbits, we calculate the commonly used quantity
\begin{equation}
\varepsilon=j_z/j_{circ},
\end{equation} 
where $j_z$ is the aforementioned component of the particle angular
momentum along the direction of $\vec{J}_{mean}$ and $j_{circ}$ is the
specific angular momentum expected for a circular orbit at the same
radial distance of the particle\footnote{We use 
\begin{equation}
j_{circ}=rv_{circ}(r), 
\end{equation}
where $r$ is the radial distance of the particle with respect to the
halo center, and $v_{circ}(r)=\sqrt{GM(r)/r}$ in spherical symmetry. 
}.
We stress that, if the particle is not bound to the halo but rather on an escaping or in-falling orbit (e.g. recently accreted), then $\varepsilon=j_z/j_{circ}$ can exceed unity.
We do not exclude explicitly such not-bound particles from the calculation, nor impose the requirement that $|\varepsilon| \leq 1$ 
(see also \sec\ref{sub:evol}).
\\
Given the definition, $\varepsilon$ is very sensitive to actual rotational orbits of the gas, so that values comprised between 0.5 and 1 ($-1$ and $-0.5$) more likely indicate co-rotating (contra-rotating) substructures in the velocity field of the gas.
\\
Combining the information provided by both $\eta$ (or equivalently
$\cos\vartheta$) and $\varepsilon$, one can put tighter constraints on
the presence of actual, ordered, rotational patterns in the gas for
all the 1680 haloes between $z=9$ and $z\simeq 20$ in our sample. 
\\
To study the trends of the different gas components, the calculations
are also done for the molecular and metal-rich gas phases separately. 
Specifically, we focus on:
\begin{itemize}
\item
{\it molecular-rich} gas with $x_{mol} > 5\times 10^{-4}$.
\item
{\it metal-rich} gas with metallicity $Z > 10^{-3}\Zsun$;
\item
{\it metal-rich} gas with metallicity $Z > 10^{-4}\Zsun$.
\end{itemize}
\subsubsection{Rotational patterns: representative cases}\label{sub:cases}
\begin{figure*}
\centering
\includegraphics[width=0.95\textwidth]{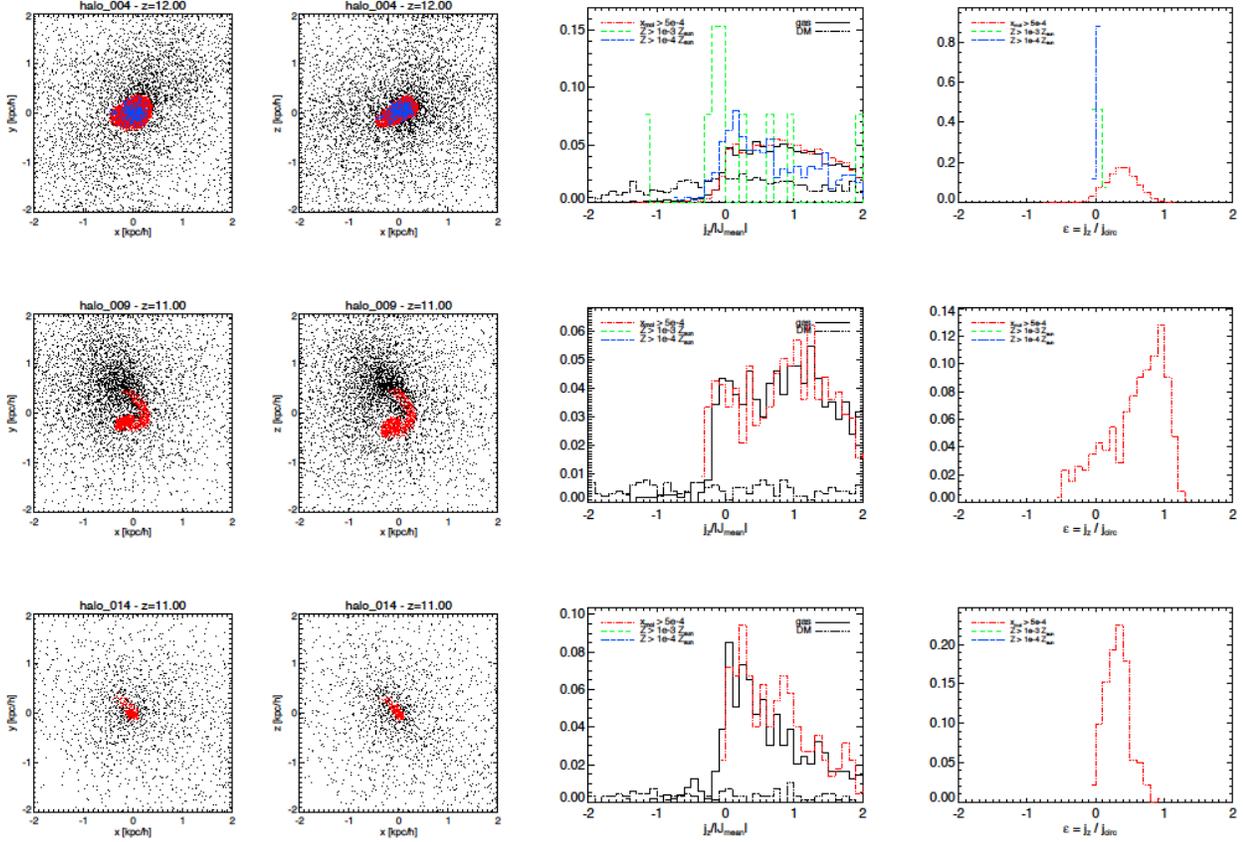}
%
\caption{
  Representative haloes at redshift $z\sim 12$ (top row) and $z\sim 11$ (middle and bottom rows), when the Universe was $0.361\,\rm Gyr$ and $0.407\,\rm Gyr$ old, respectively.
The three haloes show a variety of configurations, including both elongated molecular structures built in the contraction process of isolated haloes, and tail-like features where the rotational motion of the gas is clearly driven by environmental interactions (e.g. the in-falling and merging with a sub-halo.).
{\it First and second column}: spatial distribution of the gas in the $xy$ and $xz$ planes, respectively; different colours refer to the different gaseous components: 
all the gas (black points), 
molecular-rich gas with $x_{mol} > 5\times 10^{-4}$ (red points),
metal-rich gas with $Z > 10^{-3}\Zsun$ (green points),
metal-rich gas with $Z > 10^{-4}\Zsun$ (blue points).
{\it Third column}: $\eta$ distributions for gas (solid line) and dark-matter (dot-dot-dot-dashed line) components;
molecular-rich gas with $x_{mol} > 5\times 10^{-4}$ (red dotted line),
metal-rich gas with $Z > 10^{-3}\Zsun$ (green dashed line),
metal-rich gas with $Z > 10^{-4}\Zsun$ (blue dot-dashed line)
are also marked.
{\it Fourth column}: corresponding $\varepsilon$ distributions for the different gas phases.
}
\label{fig:represent_cases}
\end{figure*}

In \fig\ref{fig:represent_cases} a few haloes with different masses are chosen to represent possible dynamical configurations.
The spatial maps in the two projections give a visual impression of the gas structures in the halo central part, i.e. within $0.5\kpc/{h}$ comoving radius\footnote{
  We consider a typical region of $0.5\kpc/{\it h}$ comoving radius, which, for proto-galaxies at such high redshifts ($9 \lesssim z \lesssim 20$) represents a significant fraction of the virial radius.
}
(numerical studies and converges are discussed in Appendix~\ref{appA}).
Points in the maps are color-coded according to the gas phase, i.e. red for molecular-rich gas, green for metal-rich gas with metallicity $Z > 10^{-3}\Zsun$ and blue for metal-rich gas with metallicity $Z > 10^{-4}\Zsun$. 
Black points refer to all the gas particles in the plotted region.
The same color-code is applied to the distributions shown in the third and fourth columns.
For comparison, we overplot the $\eta$ distribution for all the gas in the selected region and for dark matter as well, marked with solid and dot-dot-dot-dashed black lines, respectively.
\\
As a first example, we select a case where both metal-rich (for both the two thresholds considered) and molecular-rich gas are present.
The gaseous components have very different behaviours.
Molecular-rich gas shows a more elongated morphology, which is confirmed by a broader distribution of $\eta$ at values $>0$.
The presence of rotational motions in the molecular-rich gas is also suggested by the values of $\varepsilon$, in the fourth panel, which are centered around $\varepsilon \sim 0.5$.
On the contrary, the $\eta$ distribution of very metal-rich ($Z>10^{-3}\zsun$) gas does not show any significant sign of coherent rotation, since for roughly $\sim 90\%$ of the gas $\eta \sim 0$ with scattered peaks at random values.
Metal-rich material with $Z>10^{-4}\zsun$ shows intermediate trends, being slightly affected by molecular evolution.
This halo is chosen explicitly to highlight the different behaviour of metal-rich and molecular-rich gas in an isolated halo, that is not subject to tidal interactions with closeby structures.
The dynamical differences in the various phases arise consequently to gas collapse, star formation and feedback mechanisms.
\\
The second halo (middle row in \fig\ref{fig:represent_cases}) shows instead the typical case where the circular motion of the molecular gas in the central core is mainly driven by interactions with merging sub-haloes occurring during the mass assembly of the main halo. 
This leads here to the formation of a tail-like feature in the molecular-rich gas, spinning around the halo center.
The peak around $\varepsilon \sim 1$ in the fourth-column is basically due to this motion.
The non-stable, non-ordered nature of this rotational pattern is reflected by the increasing $\eta$, which reaches values greater than 1 for a significant fraction of molecular-rich gas (middle-row, third-column panel).
\\
A very interesting example is provided by the isolated, low-mass halo in the last row of \fig\ref{fig:represent_cases}, which has no metal-rich gas, but presents an elongated, disc-like rotating structure in the molecular-rich phase.
The gas hosted in this halo has not hosted star formation episodes, yet, and its behaviour is led by molecular runaway collapse.
Similarly to the first example, also this small halo does not show
interacting features, leading to the conclusion that the establishment
of this rotational pattern is mainly driven by the fairly quiescent
contraction process, 
rather than 
by external forces.
\\
In general, the $\eta$ distributions for gas and dark matter show how
the two components in the central region of the halos respond differently to environmental effects.
In particular, the dark-matter distribution is generally flatter and
broader with respect to the gas one, meaning that ordered, rotational
patterns do not establish in the velocity field but rather preserve
random orbits. 
Regarding the total gas, its behaviour is mainly led by molecular-rich
or poorly enriched material, since these phases dominate the central
cooling region. 
\subsubsection{Evolutionary features: survival and destruction of primordial molecular discs}\label{sub:evol}
\begin{figure*}
\centering
\includegraphics[width=0.98\textwidth]{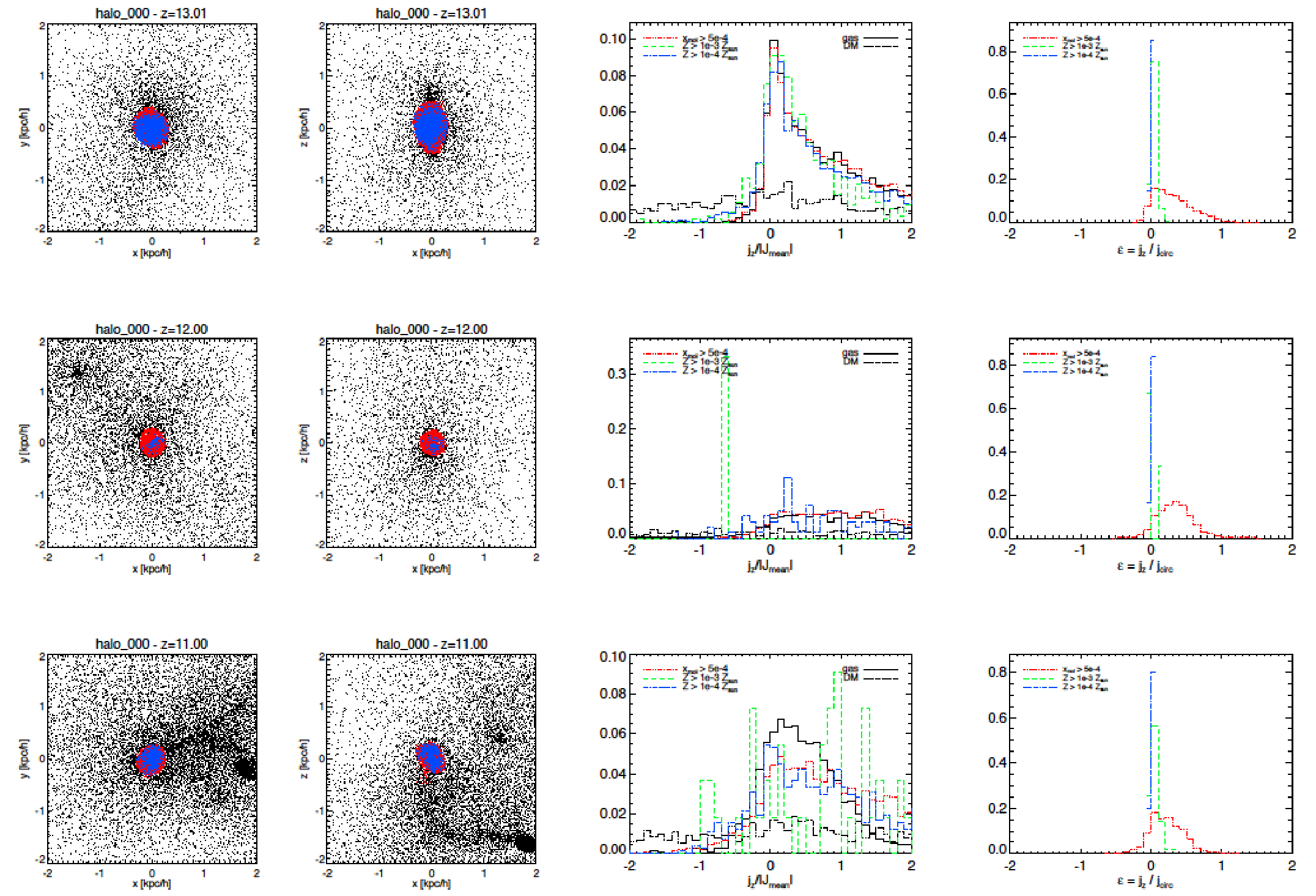}\\
\includegraphics[width=0.98\textwidth]{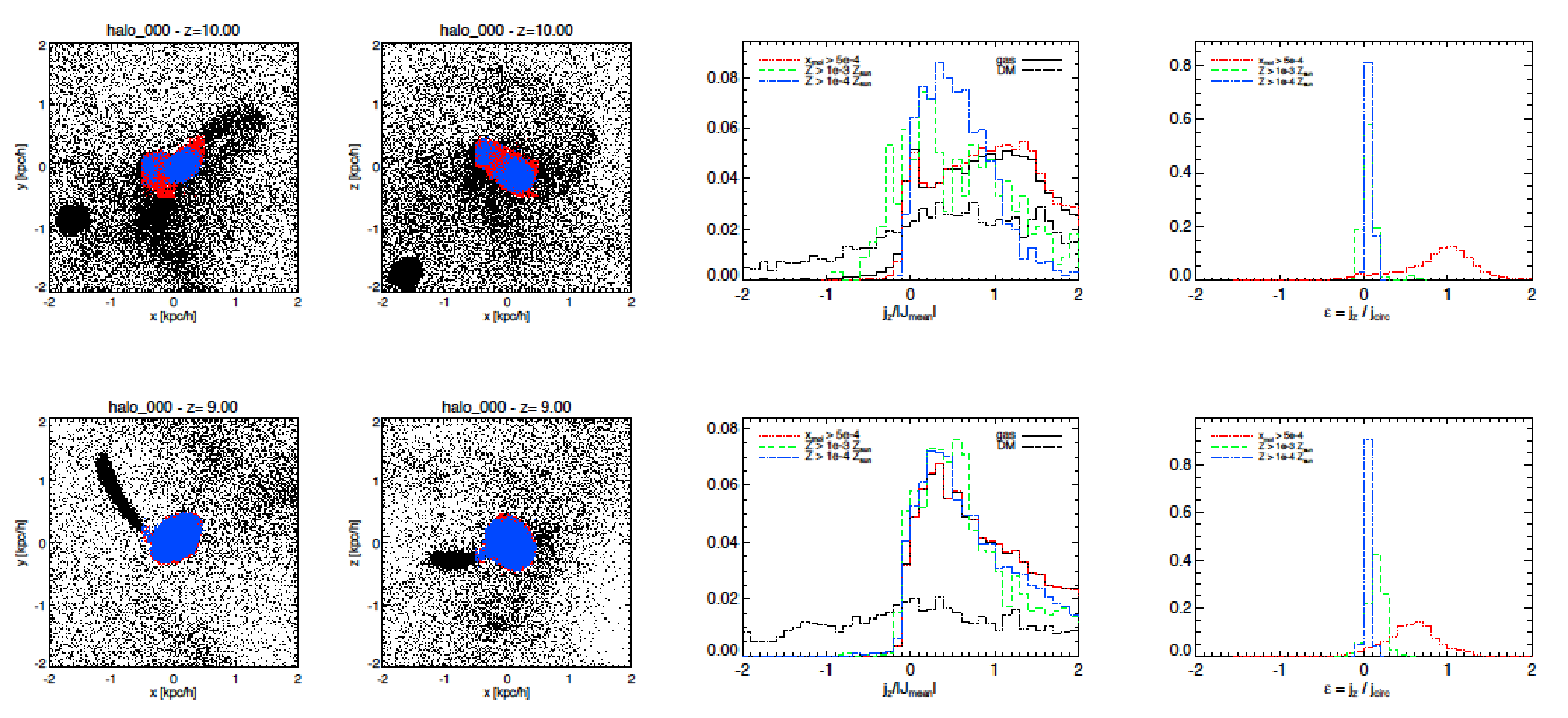}
%
\caption{
  Evolution with redshift of the biggest halo in the sample. The establishment and enhancement of rotational patterns in the molecular-rich gas of the halo is captured while the halo is merging with an in-falling sub-halo.
The snapshots refer to redshift $z = 13.01$ (first row), 12.00 (second row), 11.00 (third row), 10.00 (fourth row) and 9.00 (fifth row), when the corresponding age of the Universe was about 323, 361, 407, 464 and 535 Myr (in the standard $\Lambda$CDM model).
For the quantities plotted in the columns and color code, we refer to \fig\ref{fig:represent_cases}.
}
\label{fig:halo0_evol}
\end{figure*}
%
%
We analyse here the case of the most massive halo in the sample, for which the establishment of rotational substructures in the gaseous core region is captured during the subsequent phases of the merging process with an in-falling subhalo.
We present the results in \fig\ref{fig:halo0_evol} (colors and symbols are as in \fig\ref{fig:represent_cases}).
\\
This halo formed at redshift $z\sim 20$, when the Universe was $\sim 0.2\,\rm Gyr$ old, and is followed down to $z\simeq 8.5$.
As shown from the Figure, the central core of the halo goes from a more relaxed configuration around redshift $z \sim 13$ (when the age of the Universe was $\sim 0.3\,\rm Gyr$) to a strongly interacting phase due to the in-fall of a subhalo starting at around redshift $\sim 11$ and completing at $z\sim 9$ (when the Universe was about $0.5\,\rm Gyr$ old).
The sloshing of the core region has the effect of boosting the rotating motion of the molecular-rich gas especially, as one can see from the shift of the $\varepsilon$ distribution, towards values peaked around 1.
The establishment of a molecular rotating structure is particularly evident at $z\sim 10$, where the two halo cores are mixing.
\\
From the comparison of the maps and distributions at $z\sim 12$ and $z\sim 9$ we infer that the disc-like structure in the molecular-rich gas partially survives over time, since at lower redshifts (e.g. $z\sim 9$) the distribution of $\varepsilon$ is still biased towards non-zero values, which means that the molecular-rich gas in the very central region does show a mildly rotating component.
\\
In contrast to the disc-like structures that might arise in the star
component of galaxies, gaseous rotational patterns have essentially an
{\it intermittent} nature.
Given the collisional essence of gas, such rotational motions are
strongly influenced by dynamical interactions, which cause
them to establish as well as to be easily disrupted in less than a few
$ 10^7\,\rm yr$ timescales 
(see also the following \sec\ref{sub:statistics} for further discussion).
Basically, this 
demonstrates how the phenomenon is more
sensitive to close-by passages of sub-haloes accreted by the main
halo during its mass assembly, rather than to the passive evolution
during the collapse process 
\cite[a similar result was found, on larger scales, for the hot gas 
in simulated galaxy clusters by][]{Biffi2011}.
\\
This is well emphasized by the case study presented in
\fig\ref{fig:halo0_evol}:
the hotter, metal-rich gas, if present as in this case, is more likely
subject to environmental influences or feedback effects and no stable
rotational patterns form, while the dynamical properties of the
molecular-rich gas tend to survive longer. 
In fact, the blue and green $\eta$ distributions, corresponding to the
metal-rich gas for the two metallicity thresholds, show no significant
alignment. 
In particular, 
the high-metallicity ($Z > 10^{-3}\Zsun$) gas
has a 
noisy distribution (e.g. in the third column, third panel) 
that
reflects also the response of hot, enriched material to merging,
sloshing and interacting collisional forces.  
Indeed,
the quasi-totality of metal-rich gas is not found in
orbits close to circular, being the distribution peaked at
$\varepsilon \sim 0$ at all redshifts, 
independently of the dynamical status.
This is a direct consequence of the feedback from star formation,
because of which metals are spread randomly with no preferential
direction.
\\
As previously discussed in \sec\ref{sub:cases}, we remark here that
the dark-matter component does not show any strong effect of
the gas dynamical interactions in the halo core (see third-column
panels in \fig\ref{fig:halo0_evol}). 
\begin{figure*}
\centering
\includegraphics[width=0.33\textwidth]{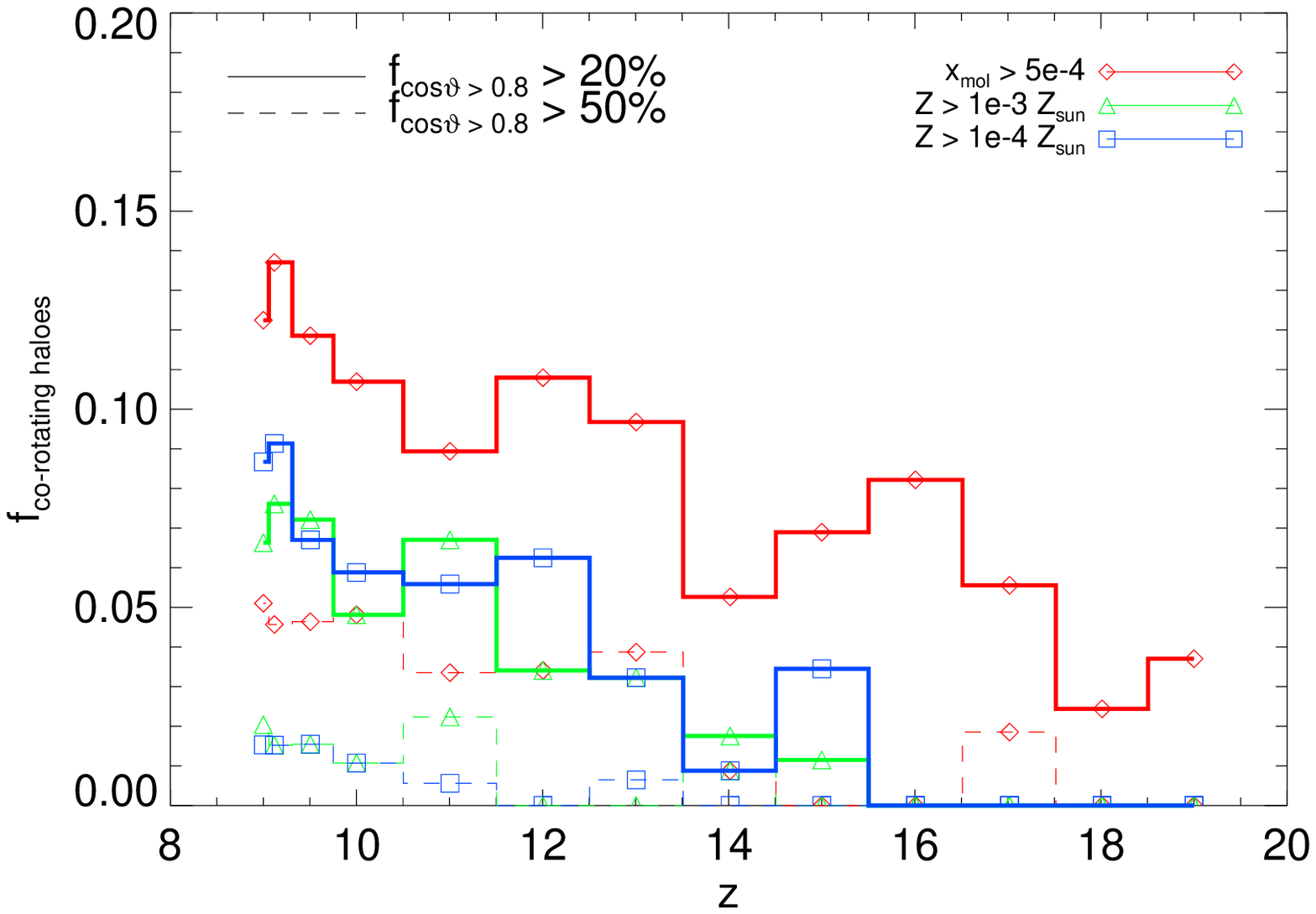}
\includegraphics[width=0.33\textwidth]{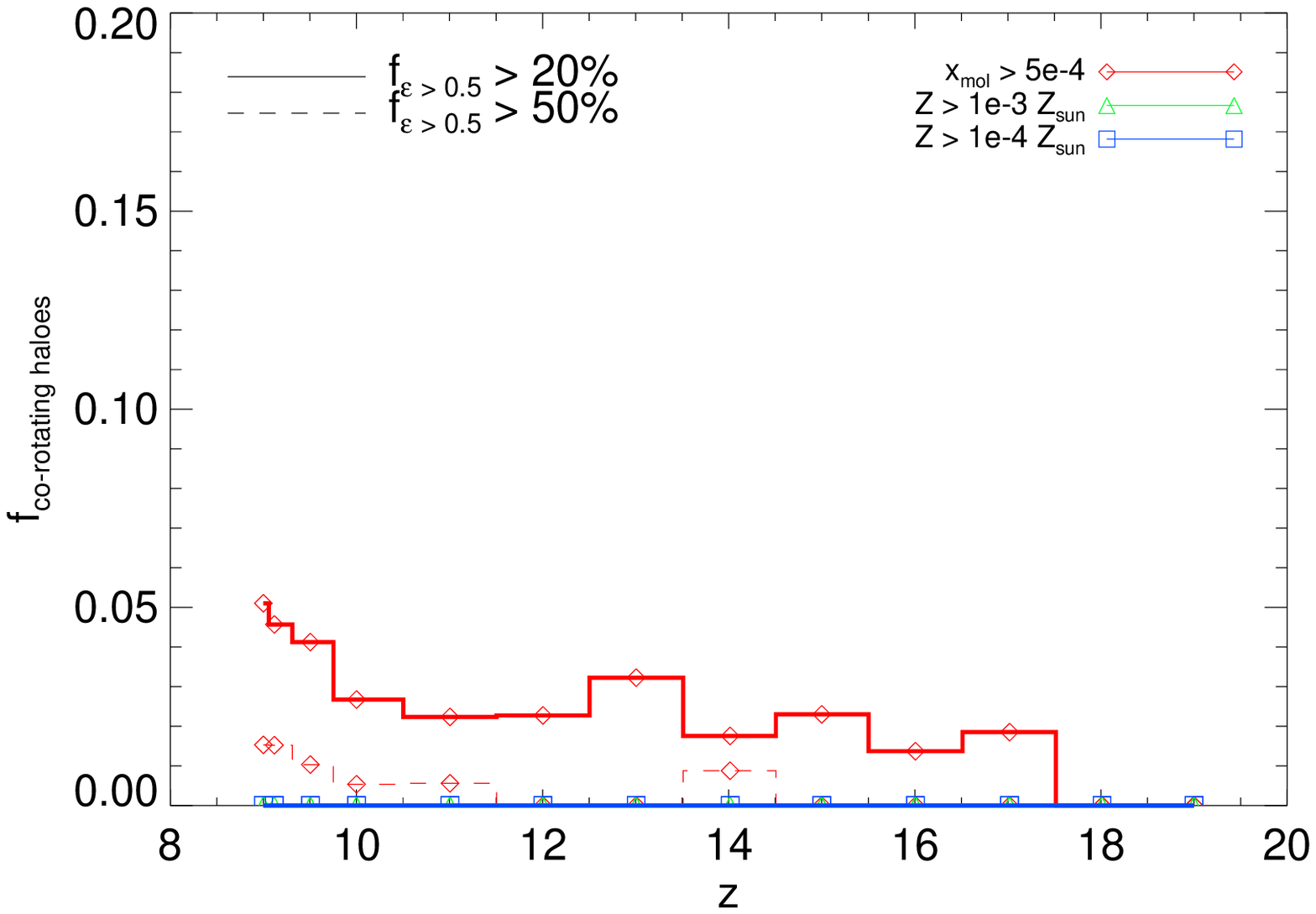}
\includegraphics[width=0.33\textwidth]{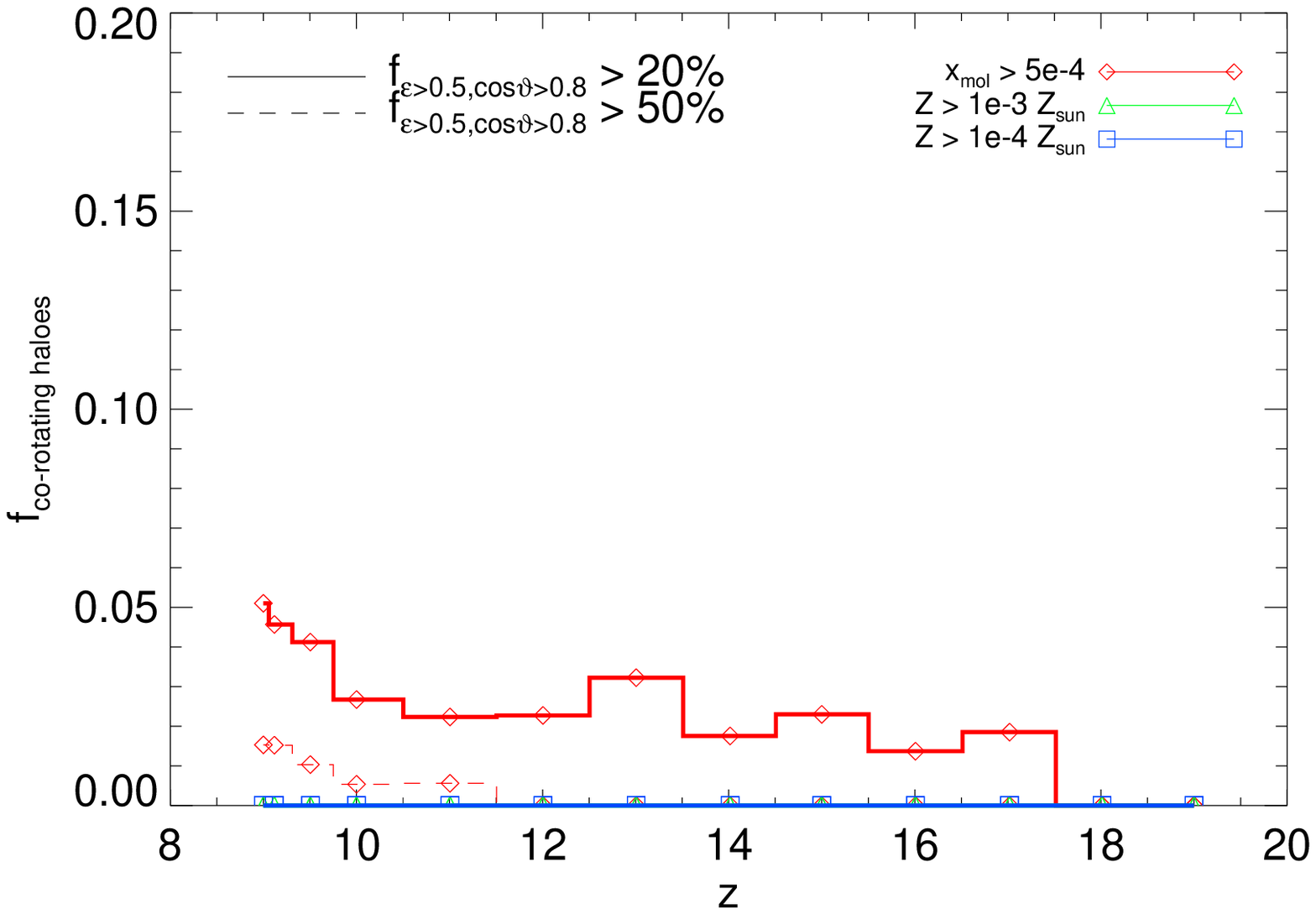}
\caption{
  Redshift evolution for the fraction of haloes hosting a significant
  percentage of co-rotating gas ($>20\%$, solid; $>50\%$, dashed),
  for the different gas phases considered, as in the legend (see also
  text and previous figures). 
  Different panels refer to the different diagnostics used to identify
  co-rotating gas: $\cos\vartheta > 0.8$ (left), $\varepsilon > 0.5$
  (center), and both indicators combined, i.e. $\cos\vartheta > 0.8$
  {\it and} $\varepsilon > 0.5$ (right).
}
\label{fig:summary_stat}
\end{figure*}
\subsubsection{Statistical properties of rotational patterns}\label{sub:statistics}
In order to explore the statistical occurrence of rotational gas patterns that establish in proto-galaxy haloes, we investigate the number of haloes featuring a significant amount of co-rotating gas fraction, as function of redshift.\\
More in detail,
we adopt the following procedure:\\
(i) we compute $\varepsilon$ and $\cos\vartheta$ for each particle in each object;\\
(ii) we calculate the fraction of gas particles with $\varepsilon$, or $\cos\vartheta$, above a significant threshold, $f_{\varepsilon > \varepsilon_{\rm th}}$ and $f_{\cos\vartheta > \cos\vartheta_{\rm th}};$\\
(iii) we require a halo to contain at least $20\%$ of co-rotating gas
particles -- as additional check, we repeat this last part of the analysis with the stricter
requirement of at least $50\%$ of co-rotating gas particles.
\\
For sake of clarity, we consider as {\it co-rotating gas particles}
those with either $\varepsilon > 0.5$ or $\cos\vartheta > 0.8$, 
and explore first the two cases separately.
In fact, the two thresholds chosen,
$\varepsilon_{\rm th}=0.5$ 
and
$\cos\vartheta_{\rm th}=0.8$ (i.e. $\theta_{\rm th}\simeq 37^o$),
represent reasonable values comprised among the cases investigated
in the parameter study (see also Appendix~\ref{appA})
\\
We stress that, by construction, the majority of the haloes, independently of the parameters used to quantify the rotating fraction of the gas (either $\varepsilon$ or $\cos\vartheta$) as well as of redshift and of molecular or metal content, show prevalently co-rotating features (namely, positive alignment).
Barely for a handful of haloes, the $\cos\vartheta$ diagnostics points
out a contra-rotating gas fraction ($f_{\cos\vartheta <
  -\cos\vartheta_{\rm th}}$) up to $\sim 10-25\%$ (see \fig\ref{fig:param_examples},
top-row panels).
For this reasons, we concentrate the following statistical analysis on co-rotating gas, only.
\\
In \fig\ref{fig:summary_stat} we report the fraction of haloes
featuring more than $20\%$ (solid curves; or $50\%$, dashed curves) of
co-rotating gas, as a function of redshift. 
In each panel we show the trends for molecular-rich gas (red
histograms) and metal-rich gas above the two metallicity thresholds 
(green for gas with $Z > 10^{-3}\Zsun $ and blue for gas with $Z > 10^{-4}\Zsun $), separately.
\\
In the left panel of \fig\ref{fig:summary_stat} we show the results of
the $\cos\vartheta$ diagnostics.
Since this is the weaker indicator of potential rotation
of the two analysed so far ($\cos\vartheta$ and $\varepsilon$),
it allows us to single out, on average, more candidates for
halos hosting gas rotational motions.
In fact, when the co-rotating gas is defined by requiring
$\cos\vartheta > 0.8$, we find a fairly large fraction of haloes with at
least $20\%$ co-rotating molecular-rich gas, ranging from $2\%$ at
$z\sim 19$ up to $\sim 12\%$ at $z\lesssim 10$.
A non-negligible fraction of haloes (from a few per cent at $z\sim15$ up
to $6-8\%$ at $z\sim 9$) also shows rotational features
when the metal-rich gas phase is considered.
Lower fractions (by a factor of roughly 2-3), but similar trends with
redshift, are found when we consider only haloes with a minimum of
$50\%$ co-rotating gas. 
\\
The $\varepsilon$ diagnostics (\fig\ref{fig:summary_stat}, middle
panel) provides more stringent, quantitative estimates of the number
of haloes containing a non-negligible fraction of gaseous rotational patterns. 
This presents a qualitatively similar, but quantitatively different
picture with respect to the previous case.
On average, in a few per cent of the haloes between $z \sim 9$ and $z
\sim 19$, at least $20\%$ of the molecular-rich gas is co-rotating. 
At $z\sim 18$ the initial fraction of haloes hosting co-rotating gas
is $\sim 2\%$ and at $z\sim 9$ this fraction reaches $\sim 5\%$. 
If one restricts further the definition of disc and considers only
haloes hosting at least $50\%$ of molecular-rich, co-rotating gas,
then it turns out that barely $1-2\%$ of early objects contain
primordial molecular discs at $z\lesssim 14$. 
\\
As anticipated from the previous single case studies
(\sec\ref{sub:cases} and \sec\ref{sub:evol}),
the results from the $\varepsilon$ diagnostics
in \fig\ref{fig:summary_stat} also confirm that, statistically,
metal-rich gas does not feature rotational patterns.
Indeed, the metal-rich gas is hotter than the molecular-rich gas,
which makes its collisional nature to be significantly more important
in preventing disc-like structures to establish and, especially, to
survive long enough against tidal disruption and environmental
effects.
Because of its colder temperatures, typically $\lesssim 10^4 {\rm K}$,
the molecular-rich component is instead less strongly affected.  
In fact, even in the case when the formation of a rotational pattern
in the molecular-rich gas is temporarily driven by dynamical
interactions, it still survives in time and does not dissolve entirely
as quickly as in the metal-rich gas case (see
\fig\ref{fig:halo0_evol}, and previous discussion in
\sec\ref{sub:evol}).   
\\ 
For a final, more robust verification, we make a more stringent
requirement by defining {\it co-rotating gas particles} those with
both $\varepsilon > 0.5$ {\it and} $\cos\vartheta > 0.8$.
The combination of the two methods (\fig\ref{fig:summary_stat}, right
panel) 
basically mirrors
the $\varepsilon$ diagnostics, which in fact points out rotational
patterns in a more reliable way. 
The only difference between middle and right panels consists of one
single halo at $z\sim 14$, for which more than $50\%$ of the
molecular-rich gas has $\varepsilon>0.5$ but the alignment angle
$\vartheta$ is larger than $\sim 37^o$ (i.e. $\cos\vartheta < 0.8$). 
When the required fraction of rotating gas is $20\%$, absolutely no difference is found. 
\subsubsection{Establishment of rotational patterns: correlation with the hosting mass}\label{sub:rel_discs_mass}
\begin{figure}
\centering
\includegraphics[width=0.44\textwidth]{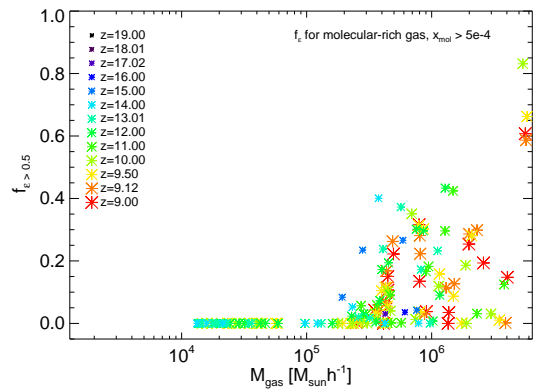}
\caption{
Fraction of the molecular-rich gas in each halo with $\varepsilon >
0.5$ as a function of the halo gas mass. Different colors and symbol
sizes refer to the redshifts analysed, as in the legend. The
correlation coefficient between $f_{\varepsilon >0.5}$ and $M_{gas}$
is $\sim 0.65$. 
When we consider $M_{DM}$ or $M_\star$, instead of $M_{gas}$, similar
behaviours are recovered with correlation coefficients of $\sim 0.59$
and $\sim 0.41$, respectively. 
}
\label{fig:discs_mass}
\end{figure}

The fraction of haloes for which the molecular-rich gas shows
significant co-rotating patterns could also depend on the hosting
mass. 
This possibility is studied in \fig\ref{fig:discs_mass}, where the
fraction of molecular-rich gas with $\varepsilon > 0.5$,
$f_{\varepsilon > 0.5}$, is plotted as a function of the halo gas
mass.
We focus, to this purpose, on the $\varepsilon$ diagnostic solely,
being this the most reliable indicator of rotational motions.
We note that on average $f_{\varepsilon > 0.5}$ increases with the
mass of the gas in the halo, quite independently of the redshift
considered, but with a large spread. 
Objects with low masses ($M_{gas}\lesssim 2\times 10^5 \msunh$)
present a fraction of co-rotating, molecular-rich gas which is
extremely low, basically zero, because they have not formed enough
molecules to cool and condense in a disc-like structure, yet. 
Such mass limit is consistent with the minimum dark-matter mass of
$M_{DM}\sim 2\times 10^6\,\rm \msunh$ needed to host runaway gas
collapse and star formation (as discussed in \sec\ref{sec:results} and
\ref{sect:selection}, and demonstrated in \fig\ref{fig:masses}). 
Slightly larger objects instead have a fraction of rotating,
molecular-rich gas, $f_{\varepsilon > 0.5}$, that ranges between $\sim
20\%$ and $\sim 60\%$ depending also on the local environments. 
For the larger and more evolved proto-galaxies with $M_{gas}\sim
10^7\,\rm \msunh$ and $M_{DM}\sim 10^8\,\rm \msunh$, $f_{\varepsilon >
  0.5}$ can easily exceed $60\%$ and reach $\sim 85\%$. 
The dependence of rotating, molecular gas on the hosting gas,
dark-matter or stellar mass is quantified by the Pearson correlation
coefficient between $f_{\varepsilon > 0.5}$ and $M_{gas}$, $M_{\rm
  DM}$ or $M_\star$. 
The resulting values turn out to be $\sim 0.65$, $\sim 0.59$ and $\sim 0.41$, respectively.
Thus, we conclude that the establishment of gaseous rotational
patterns in the innermost regions of early objects is mildly related
to the whole hosting mass.  
In particular, the null values at law masses and the higher values at
larger masses suggest the importance of overcoming the mass threshold
for gas cooling and collapse. 
The broad spread for $M_{gas}\gtrsim 2\times 10^5\,\msunh$
(i.e. $M_{\rm DM} \gtrsim 2\times 10^6\,\msunh$) is a consequence of
the backreaction of feedback mechanisms on the following gas
evolution. 
%
%
\begin{figure*}
\centering
\includegraphics[width=0.44\textwidth]{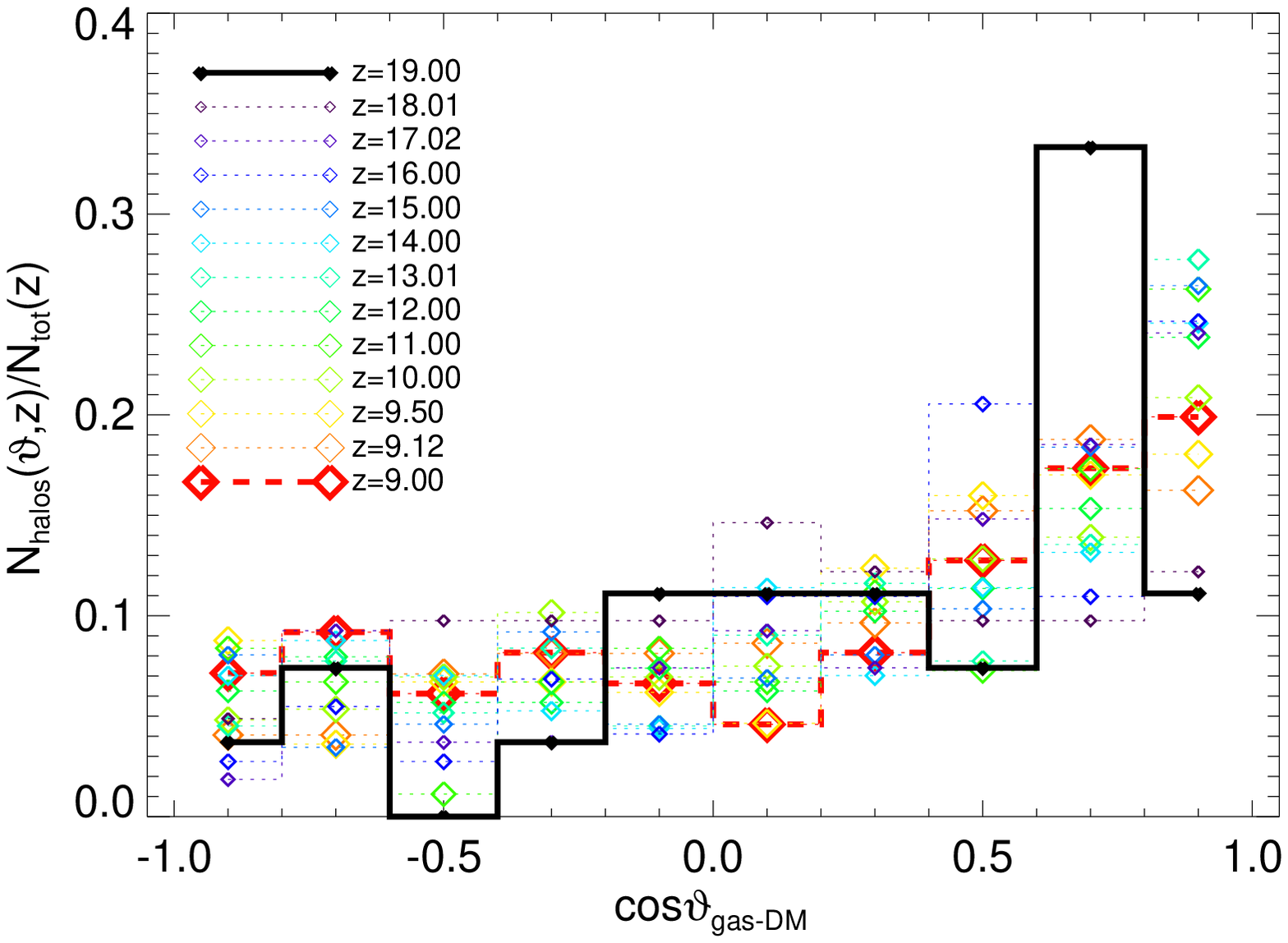}
\includegraphics[width=0.44\textwidth]{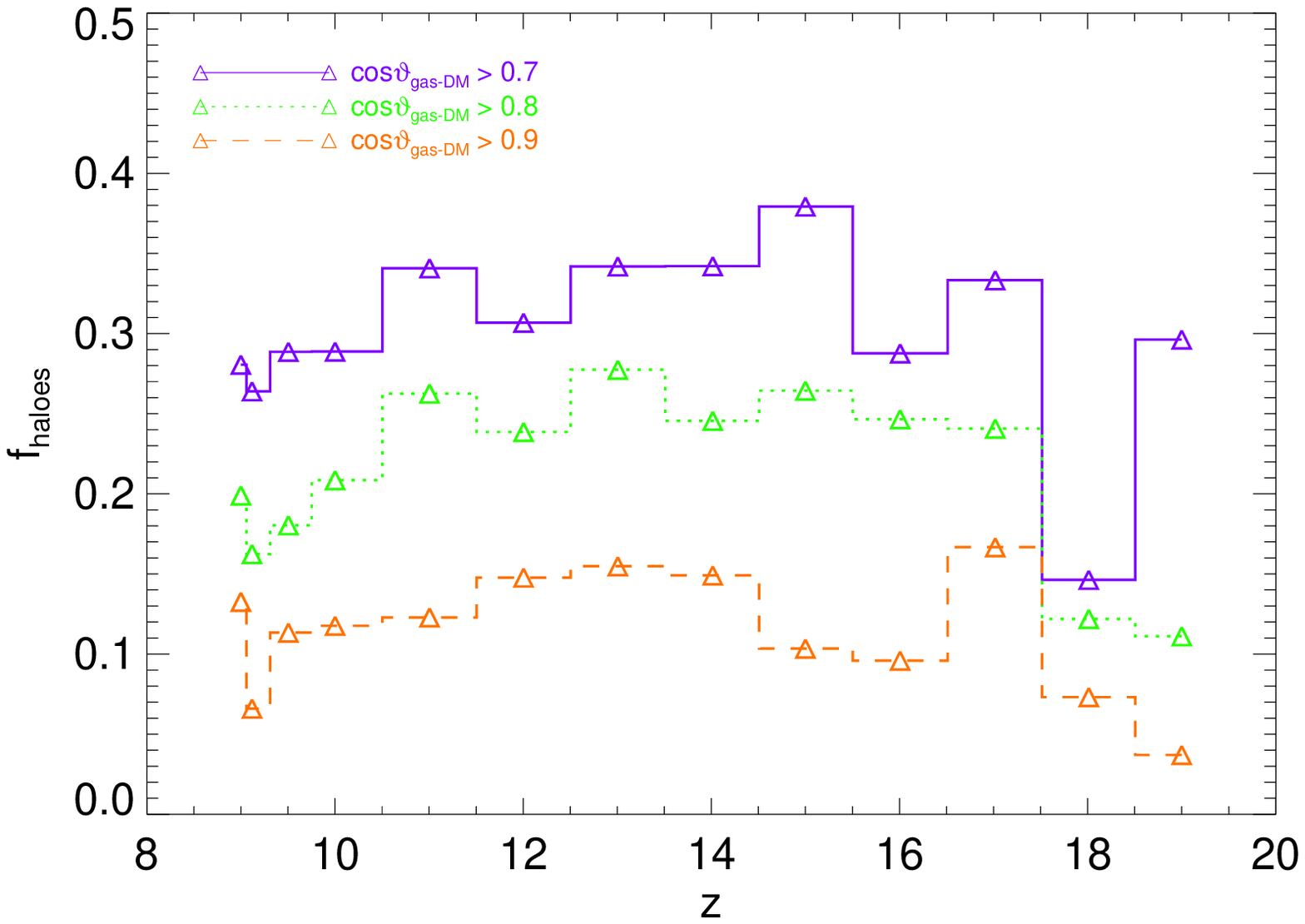}
\caption{
  {\it Left panel:} fraction of haloes, at each redshift, with a
  certain degree of alignment between the gaseous core
  and the large-scale dark-matter component, quantified by the cosine
  of the angle formed by their mean angular momenta, plotted on the $x$-axis; we mark with
  thicker lines the first ($z=19$) and last ($z=9$) redshift
  considered. 
  {\it Right panel:} distribution with redshift of the fraction of
  haloes for which $cos\vartheta_{\rm gas-DM}$ is greater than $0.7$
  (solid purple line), $0.8$ (dotted green line), $0.9$ (dashed orange line).
}
\label{fig:costh_mass_z}
\end{figure*}
\subsubsection{Angular momentum alignment of gas and dark-matter components}\label{sub:global_alignment}
\begin{figure}
\centering
\includegraphics[width=0.44\textwidth]{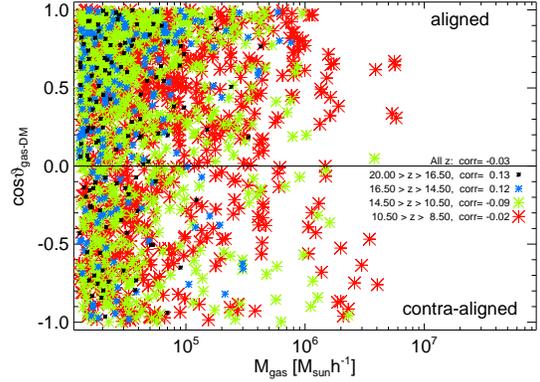}
\caption{
  Relation between the alignment of central gas and
  large-scale dark-matter angular momentum and the halo gas mass. As
  explained by the labels, four bins in redshift are considered
  separately. Values of $cos\vartheta_{\rm gas-DM} = 1,-1$ represent
  maximum alignment and contra-alignment, respectively. The
  correlation coefficient is close to zero, both for the individual
  redshift bins and considering all the points together. 
}
\label{fig:costh_mass}
\end{figure}

As a further issue, we explore the global alignment between the mean
angular momentum of the 
gas in the core region
and the one of the whole dark-matter distribution.
In this way we will gain hints on the dynamical effects of the
large-scale dark-matter component on the 
innermost gaseous core,
where the first events of baryonic structure formation
take place, the majority of molecular- and metal-rich gas resides
and the disc-like patterns are likely to establish.
\\
In particular, we investigate the statistical distributions for
$cos\vartheta_{\rm gas-DM}$ at different $z$
(\fig\ref{fig:costh_mass_z}) and the resulting correlation to the
hosting mass (\fig\ref{fig:costh_mass}).
\\
The left panel of \fig\ref{fig:costh_mass_z} shows a slightly increasing
trend of the distribution at $cos\vartheta_{\rm gas-DM} \gtrsim 0.5$
and therefore give support for a mild alignment of gas and dark-matter
components mostly at high redshifts. 
\\
The right panel of the Figure reports the 
fraction of haloes with prominent alignment (i.e. $cos\vartheta_{\rm gas-DM} >
0.9$) between the central 
gas and the large-scale
dark-matter halo, as a function of redshift. 
We note that this fraction increases from redshift $z\sim 20$ up to
$z\sim 14$, meaning that the large-scale dark-matter halo has a mild
effect on the central gas, building some alignment during the early
stages of the collapse process and mainly reflecting the ongoing star
formation activity, afterwards. 
At redshift $z\lesssim 14$, we find maximum alignment for about $\sim
10\%$ of the haloes, for which $cos\vartheta_{\rm gas-DM}>0.9$
(i.e. $\vartheta_{\rm gas-DM}\lesssim 26^o$). 
Similar results are recovered also when less stringent alignment
criteria are applied (namely, $\vartheta_{\rm gas-DM}\lesssim
37^o,45^o$), by which the fraction of haloes satisfying them naturally
increases at all redshifts, reaching $\sim 20-30\%$ at $z\sim 9$. 
The general, mild drop at later times is simply interpreted as a consequence
of the more chaotic environment established after first star formation
episodes and SN explosions. 
\\
To check whether the alignments are related to the hosting mass, in
\fig\ref{fig:costh_mass} we plot the cosine of the angle formed by the
angular momentum vectors of the 
(central) gas and dark-matter components as a function of mass, for all the
haloes and at all the redshifts considered.
As a result, we find no evidence for dependence on the hosting mass
of 
the alignment ($cos\vartheta_{\rm gas-DM}$). 
In a more quantitative way, the Pearson
correlation coefficient is very small, both by considering all the
redshifts together ($\sim -0.03$) and by considering individual
redshift bins (ranging from $\sim -0.09$ to $0.13$). 
The same conclusion is found when investigating correlation between
$cos\vartheta_{\rm gas-DM}$ and $M_{DM}$, for which the correlation
coefficient is only $\sim -0.04$. 
%
%
%
\section{Conclusions}\label{sec:discussion}
In the previous sections we have investigated baryonic and dynamical
features of the first proto-galaxies by analyses of numerical
hydrodynamical simulations including atomic and molecular chemistry
(following e$^-$, H, H$^+$, H$^-$, He, He$^+$, He$^{++}$ D, D$^+$,
H$_2$, H$_2^+$, HD, HeH$^+$), cooling, star formation and
self-consistent stellar evolution according to proper yields (for He,
C, O, Si, Fe, Mg, S, etc.) and lifetimes for popIII and popII-I
regimes (as outlined also in \sec\ref{sec:sims}). 
To this purpose, we have analysed a significantly large sample of early proto-haloes (1680
objects) all very well resolved with at least 300 gas particles
(\sec\ref{sect:selection}), at redshift $\gtrsim 9$. 
\\
First galaxies form in small H$_2$-cooling haloes with gas masses
$\sim 10^5-10^7\msun$ and stellar fractions $\lesssim 10^{-3}$
(\fig\ref{fig:masses}), broadly dispersed and with properties severely
affected by environmental effects (e.g. \fig\ref{fig:Mgasrelations}
). 
\\
The minimum mass for star formation is found to be 
$M_{DM} \sim  2\times  10^6\rm\msun$,
as gas in smaller structures does not reach densities high enough to
trigger significant molecule creation, cooling and fragmentation of
the medium (\fig\ref{fig:tm_hottm}).
\\
Early popIII galaxies are relevant in polluting the surrounding medium
up to $\sim10^{-2}\Zsun$ during the very first bursts
(\fig\ref{fig:sfr-Mgas-Z}).
In total, roughly $\lesssim 10$ per cent of the objects by $z\sim 9$
is enriched above the critical popII-I threshold,
but contribute to more than $\sim 80$ per cent of the halo SFR
(\fig\ref{fig:red_f_enrich}).
This suggests that pristine or weakly polluted (quiescent) 
proto-galaxies might well exist at lower $z$ 
\cite[as argued in][too]{Maio2013z7arXiv}.
\\
Furthermore, among the polluted structures, roughly $\sim 60\%$ 
are actively forming stars, the remaining quiescent $\sim 40\%$ 
being usually enriched by nearby sources (\sec\ref{sub:SFhaloes}). 
\\
PopIII and popII-I stellar evolution is responsible for enriching
completely closeby, low-mass, non-star-forming galaxies with hot
material that experienced feedback in relatively larger, more active
galaxies. 
This is the case for roughly $\sim 7$ per cent of the enriched-halo
sample at $z\gtrsim 9$. 
None the less, the majority of the polluted haloes (roughly $93\%$)
shows signature of both external and {\it in situ} enrichment 
(see \tab\ref{tab:data}).
\\
Feedback mechanisms alter their abundances and determine a significant
spread (of up to $\sim 2$ orders of magnitude) when looking at the
different correlations involving gas masses, temperatures, local star
formation rates, or chemical composition, as $x_{mol}$ and $Z$
(\fig\ref{fig:Trelations} and \fig\ref{fig:chemicalrelations}).
Their impact is less important for haloes that evolve in isolation or
far away from the main star formation sites, though.
\\
As a consequence of molecular-driven star formation and metal
pollution, relations between stellar mass, SFR and metallicity are
expected (\fig\ref{fig:Mstarproperties}), but outliers can persist due
to environmental dependencies.
\\
A fundamental link to understand early star formation lies in the
roughly linear $x_{mol}-f_\star$ relation (\fig\ref{fig:FMR}), 
while large scatter for $Z$ values in similar mass (or SFR) bins can be
simply determined by the effects of metal spreading in denser or rarer
regions.
\\
Typical star formation rates are around $\rm SFR \sim
10^{-7}-10^{-3}\msun/yr$, corresponding to mean specific values of
$\rm sSFR\sim 10 \,Gyr^{-1}$ at $z\sim 9$ and $\sim 50\,\rm Gyr^{-1}$
at $z\sim 10-20$. 
For individual primordial cases, sSFRs of the order of $\rm \sim
10^2\,\rm Gyr^{-1}$ at can be reached, as well (\fig\ref{fig:sSFR}). 
These values make primordial galaxies very bursty objects
(\fig\ref{fig:sSFRcollection}) . 
\\
Gaseous rotational features are quantified through the analysis of the
gas specific angular momentum. 
We find that the establishment of strong rotationally-supported gaseous cores
in early proto-galaxies is not very common, in agreement with recent
studies by \cite{Prieto2013}, but based on a larger sample (which
significantly increased the statistics) of numerical, hydro, chemistry
simulations, including star formation, stellar evolution, metal
production and feedback processes. 
\\
Additionally, we note that gaseous rotational patterns generally evolve 
in an intermittent fashion and show striking differences for 
different gas phases:
the molecular-rich component tends to maintain any established
rotational motion longer with respect to the hotter, metal-rich
gas phase, which is more sensitive to the environment (sub-haloes
in-fall, star formation bursts, feedback processes, mergers) and thus
shows a more chaotic nature (\fig\ref{fig:represent_cases} and
\fig\ref{fig:halo0_evol}). 
\\
In general, the establishment of such gas rotational motions in the halo innermost
region is found to be mildly dependent on the whole
hosting mass, with $\gtrsim 60 \%$ of molecular-rich gas featuring
rotation in haloes larger than $M_{gas}>10^6\msun/h$ and
$M_{DM}>10^7\msun/h$ (\fig\ref{fig:discs_mass}).
Moreover, the fraction of objects with a significant amount of
co-rotating molecular-rich gas increases in time from a few per cent
at $z\sim 18$ up to $\sim 5-15$ per cent at $z\lesssim 10$
(\fig\ref{fig:summary_stat}). 
\\
With our analysis we also observe that the (mis-)alignment between gas
and dark-matter angular momenta does not indicate a significant
dependence on the halo mass (\fig\ref{fig:costh_mass}).
On the other hand, a certain mild trend in redshift appears
(\fig\ref{fig:costh_mass_z}), with the fraction of haloes that show
maximum spin alignments 
increasing from $1-2$ per cent at $z\sim 19$ 
and reaching $\sim 10\%$ at redshift $z\sim 9$.
\\
We stress that we checked our conclusions by applying different resolution
requirements for each halo (as specified in \sec\ref{sect:selection}),
however, all the resulting trends did not show evident changes and the
statistical patterns were basically unaffected. 
%
%
\section*{Acknowledgments}
We acknowledge the anonymous referee for valuable comments that helped
to improve the presentation of the results.
We also would like to thank K.~Dolag for useful discussions and
E.~Tescari for kindly providing us with his numerical results (private
communication).
U.~M.'s research leading to these results has received funding from a
Marie Curie fellowship by the European Union Seventh Framework
Programme (FP7/2007-2013) under grant agreement n. 267251. 
The bibliographic research was possible thanks to the tools offered by
the NASA Astrophysics Data Systems.
%
%
\bibliographystyle{mn2e}
\bibliography{bibl.bib}

\begin{thebibliography}{}

\bibitem[\protect\citeauthoryear{{Abel}, {Bryan} \& {Norman}}{{Abel}
  et~al.}{2002}]{Abel2002}
{Abel} T.,  {Bryan} G.~L.,    {Norman} M.~L.,  2002, Science, 295, 93

\bibitem[\protect\citeauthoryear{{Alvarez}, {Wise} \& {Abel}}{{Alvarez}
  et~al.}{2009}]{Alvarez2009}
{Alvarez} M.~A.,  {Wise} J.~H.,    {Abel} T.,  2009, \apjl, 701, L133

\bibitem[\protect\citeauthoryear{{Bate} \& {Burkert}}{{Bate} \&
  {Burkert}}{1997}]{BateBurkert1997}
{Bate} M.~R.,  {Burkert} A.,  1997, \mnras, 288, 1060

\bibitem[\protect\citeauthoryear{{Biffi}, {Dolag} \& {B{\"o}hringer}}{{Biffi}
  et~al.}{2011}]{Biffi2011}
{Biffi} V.,  {Dolag} K.,    {B{\"o}hringer} H.,  2011, \mnras, 413, 573

\bibitem[\protect\citeauthoryear{{Binney} \& {Tremaine}}{{Binney} \&
  {Tremaine}}{2008}]{BinneyTremaine2008}
{Binney} J.,  {Tremaine} S.,  2008, {Galactic Dynamics: Second Edition}.
Princeton University Press

\bibitem[\protect\citeauthoryear{{Bouwens}, {Bradley}, {Zitrin} \& {and~35
  co-authors,}}{{Bouwens} et~al.}{2012}]{Bouwens2012arXiv}
{Bouwens} R.,  {Bradley} L.,  {Zitrin} A.,    {and~35 co-authors,} 2012, ArXiv
  e-prints

\bibitem[\protect\citeauthoryear{{Bower}, {Vernon}, {Goldstein}, {Benson},
  {Lacey}, {Baugh}, {Cole} \& {Frenk}}{{Bower} et~al.}{2010}]{Bower2010}
{Bower} R.~G.,  {Vernon} I.,  {Goldstein} M.,  {Benson} A.~J.,  {Lacey} C.~G.,
  {Baugh} C.~M.,  {Cole} S.,    {Frenk} C.~S.,  2010, \mnras, 407, 2017

\bibitem[\protect\citeauthoryear{{Brinchmann}, {Charlot}, {White}, {Tremonti},
  {Kauffmann}, {Heckman} \& {Brinkmann}}{{Brinchmann}
  et~al.}{2004}]{Brinchmann2004}
{Brinchmann} J.,  {Charlot} S.,  {White} S.~D.~M.,  {Tremonti} C.,  {Kauffmann}
  G.,  {Heckman} T.,    {Brinkmann} J.,  2004, \mnras, 351, 1151

\bibitem[\protect\citeauthoryear{{Bromm} \& {Loeb}}{{Bromm} \&
  {Loeb}}{2003a}]{BrommLoeb2003}
{Bromm} V.,  {Loeb} A.,  2003a, \apj, 596, 34

\bibitem[\protect\citeauthoryear{{Bromm} \& {Loeb}}{{Bromm} \&
  {Loeb}}{2003b}]{Bromm2003}
{Bromm} V.,  {Loeb} A.,  2003b, \nat, 425, 812

\bibitem[\protect\citeauthoryear{{Brooks}, {Governato}, {Quinn}, {Brook} \&
  {Wadsley}}{{Brooks} et~al.}{2009}]{Brooks2009}
{Brooks} A.~M.,  {Governato} F.,  {Quinn} T.,  {Brook} C.~B.,    {Wadsley} J.,
  2009, \apj, 694, 396

\bibitem[\protect\citeauthoryear{{Campisi}, {Maio}, {Salvaterra} \&
  {Ciardi}}{{Campisi} et~al.}{2011}]{Campisi2011}
{Campisi} M.~A.,  {Maio} U.,  {Salvaterra} R.,    {Ciardi} B.,  2011, \mnras,
  416, 2760

\bibitem[\protect\citeauthoryear{{Coe}, {Zitrin}, {Carrasco}, {Shu}, {Zheng} \&
  {18 co-authors}}{{Coe} et~al.}{2013}]{Coe2013}
{Coe} D.,  {Zitrin} A.,  {Carrasco} M.,  {Shu} X.,  {Zheng} W.,    {18
  co-authors} 2013, \apj, 762, 32

\bibitem[\protect\citeauthoryear{{Cooray}, {Bock}, {Keatin}, {Lange} \&
  {Matsumoto}}{{Cooray} et~al.}{2004}]{Cooray2004}
{Cooray} A.,  {Bock} J.~J.,  {Keatin} B.,  {Lange} A.~E.,    {Matsumoto} T.,
  2004, \apj, 606, 611

\bibitem[\protect\citeauthoryear{{Creasey}, {Theuns} \& {Bower}}{{Creasey}
  et~al.}{2012}]{Creasey2012arXiv}
{Creasey} P.,  {Theuns} T.,    {Bower} R.~G.,  2012, ArXiv e-prints

\bibitem[\protect\citeauthoryear{{Daddi}, {Dickinson}, {Morrison}, {Chary},
  {Cimatti}, {Elbaz}, {Frayer}, {Renzini}, {Pope}, {Alexander}, {Bauer},
  {Giavalisco}, {Huynh}, {Kurk} \& {Mignoli}}{{Daddi} et~al.}{2007}]{Daddi2007}
{Daddi} E.,  {Dickinson} M.,  {Morrison} G.,  {Chary} R.,  {Cimatti} A.,
  {Elbaz} D.,  {Frayer} D.,  {Renzini} A.,  {Pope} A.,  {Alexander} D.~M.,
  {Bauer} F.~E.,  {Giavalisco} M.,  {Huynh} M.,  {Kurk} J.,    {Mignoli} M.,
  2007, \apj, 670, 156

\bibitem[\protect\citeauthoryear{{Dav{\'e}}, {Oppenheimer} \&
  {Finlator}}{{Dav{\'e}} et~al.}{2011}]{Dave2011}
{Dav{\'e}} R.,  {Oppenheimer} B.~D.,    {Finlator} K.,  2011, \mnras, 415, 11

\bibitem[\protect\citeauthoryear{{Dayal}, {Dunlop}, {Maio} \& {Ciardi}}{{Dayal}
  et~al.}{2013}]{Dayal2013tmp}
{Dayal} P.,  {Dunlop} J.~S.,  {Maio} U.,    {Ciardi} B.,  2013, \mnras

\bibitem[\protect\citeauthoryear{{de Souza}, {Ciardi}, {Maio} \& {Ferrara}}{{de
  Souza} et~al.}{2013}]{deSouza2013}
{de Souza} R.~S.,  {Ciardi} B.,  {Maio} U.,    {Ferrara} A.,  2013, \mnras,
  428, 2109

\bibitem[\protect\citeauthoryear{{Dekel}, {Birnboim}, {Engel}, {Freundlich},
  {Goerdt}, {Mumcuoglu}, {Neistein}, {Pichon}, {Teyssier} \& {Zinger}}{{Dekel}
  et~al.}{2009}]{DekelNature2009}
{Dekel} A.,  {Birnboim} Y.,  {Engel} G.,  {Freundlich} J.,  {Goerdt} T.,
  {Mumcuoglu} M.,  {Neistein} E.,  {Pichon} C.,  {Teyssier} R.,    {Zinger} E.,
   2009, \nat, 457, 451

\bibitem[\protect\citeauthoryear{{Dolag}, {Borgani}, {Murante} \&
  {Springel}}{{Dolag} et~al.}{2009}]{Dolag2009}
{Dolag} K.,  {Borgani} S.,  {Murante} G.,    {Springel} V.,  2009, \mnras, 399,
  497

\bibitem[\protect\citeauthoryear{{Dunlop}, {Rogers}, {McLure} \& {16
  co-authors}}{{Dunlop} et~al.}{2013}]{Dunlop2013}
{Dunlop} J.~S.,  {Rogers} A.~B.,  {McLure} R.~J.,    {16 co-authors} 2013,
  \mnras, 432, 3520

\bibitem[\protect\citeauthoryear{{Dwek} \& {Arendt}}{{Dwek} \&
  {Arendt}}{1998}]{DwekArendt1998}
{Dwek} E.,  {Arendt} R.~G.,  1998, \apjl, 508, L9

\bibitem[\protect\citeauthoryear{{Elbaz}, {Daddi}, {Le Borgne}, {Dickinson},
  {Alexander}, {Chary}, {Starck}, {Brandt}, {Kitzbichler}, {MacDonald},
  {Nonino}, {Popesso}, {Stern} \& {Vanzella}}{{Elbaz} et~al.}{2007}]{Elbaz2007}
{Elbaz} D.,  {Daddi} E.,  {Le Borgne} D.,  {Dickinson} M.,  {Alexander} D.~M.,
  {Chary} R.-R.,  {Starck} J.-L.,  {Brandt} W.~N.,  {Kitzbichler} M.,
  {MacDonald} E.,  {Nonino} M.,  {Popesso} P.,  {Stern} D.,    {Vanzella} E.,
  2007, \aap, 468, 33

\bibitem[\protect\citeauthoryear{{Fernandez} \& {Shull}}{{Fernandez} \&
  {Shull}}{2011}]{FernandezShull2011}
{Fernandez} E.~R.,  {Shull} J.~M.,  2011, \apj, 731, 20

\bibitem[\protect\citeauthoryear{{Fernandez} \& {Zaroubi}}{{Fernandez} \&
  {Zaroubi}}{2013}]{Fernandez2013arXiv}
{Fernandez} E.~R.,  {Zaroubi} S.,  2013, ArXiv e-prints

\bibitem[\protect\citeauthoryear{{Fialkov}, {Barkana}, {Tseliakhovich} \&
  {Hirata}}{{Fialkov} et~al.}{2012}]{Fialkov2012}
{Fialkov} A.,  {Barkana} R.,  {Tseliakhovich} D.,    {Hirata} C.~M.,  2012,
  \mnras, 424, 1335

\bibitem[\protect\citeauthoryear{{Genzel}, {Burkert}, {Bouch{\'e}}, {Cresci},
  {F{\"o}rster Schreiber}, {Shapley}, {Shapiro}, {Tacconi} \&
  {17~co-authors}}{{Genzel} et~al.}{2008}]{Genzel2008}
{Genzel} R.,  {Burkert} A.,  {Bouch{\'e}} N.,  {Cresci} G.,  {F{\"o}rster
  Schreiber} N.~M.,  {Shapley} A.,  {Shapiro} K.,  {Tacconi} L.~J.,
  {17~co-authors} 2008, \apj, 687, 59

\bibitem[\protect\citeauthoryear{{Gonz{\'a}lez}, {Bouwens}, {Labb{\'e}},
  {Illingworth}, {Oesch}, {Franx} \& {Magee}}{{Gonz{\'a}lez}
  et~al.}{2012}]{Gonzalez2012}
{Gonz{\'a}lez} V.,  {Bouwens} R.~J.,  {Labb{\'e}} I.,  {Illingworth} G.,
  {Oesch} P.,  {Franx} M.,    {Magee} D.,  2012, \apj, 755, 148

\bibitem[\protect\citeauthoryear{{Gorjian}, {Wright} \& {Chary}}{{Gorjian}
  et~al.}{2000}]{Gorjian2000}
{Gorjian} V.,  {Wright} E.~L.,    {Chary} R.~R.,  2000, \apj, 536, 550

\bibitem[\protect\citeauthoryear{{Greif}, {Johnson}, {Klessen} \&
  {Bromm}}{{Greif} et~al.}{2008}]{Greif2008}
{Greif} T.~H.,  {Johnson} J.~L.,  {Klessen} R.~S.,    {Bromm} V.,  2008,
  \mnras, 387, 1021

\bibitem[\protect\citeauthoryear{{Greif}, {White}, {Klessen} \&
  {Springel}}{{Greif} et~al.}{2011}]{Greif2011}
{Greif} T.~H.,  {White} S.~D.~M.,  {Klessen} R.~S.,    {Springel} V.,  2011,
  \apj, 736, 147

\bibitem[\protect\citeauthoryear{{Gunn} \& {Gott} III}{{Gunn} \&
  {Gott}}{1972}]{GunnGott1972}
{Gunn} J.~E.,  {Gott} III J.~R.,  1972, \apj, 176, 1

\bibitem[\protect\citeauthoryear{{Halle} \& {Combes}}{{Halle} \&
  {Combes}}{2012}]{Halle2013}
{Halle} A.,  {Combes} F.,  2012, ArXiv e-prints

\bibitem[\protect\citeauthoryear{{Jeeson-Daniel}, {Ciardi}, {Maio},
  {Pierleoni}, {Dijkstra} \& {Maselli}}{{Jeeson-Daniel}
  et~al.}{2012}]{Akila2012}
{Jeeson-Daniel} A.,  {Ciardi} B.,  {Maio} U.,  {Pierleoni} M.,  {Dijkstra} M.,
    {Maselli} A.,  2012, \mnras, 424, 2193

\bibitem[\protect\citeauthoryear{{Jeeson-Daniel}, {Dalla Vecchia}, {Haas} \&
  {Schaye}}{{Jeeson-Daniel} et~al.}{2011}]{jeesondaniel2011}
{Jeeson-Daniel} A.,  {Dalla Vecchia} C.,  {Haas} M.~R.,    {Schaye} J.,  2011,
  \mnras, 415, L69

\bibitem[\protect\citeauthoryear{{Johnson}, {Whalen}, {Fryer} \&
  {Li}}{{Johnson} et~al.}{2012}]{Johnson2012}
{Johnson} J.~L.,  {Whalen} D.~J.,  {Fryer} C.~L.,    {Li} H.,  2012, \apj, 750,
  66

\bibitem[\protect\citeauthoryear{{Karim}, {Schinnerer},
  {Mart{\'{\i}}nez-Sansigre}, {Sargent}, {van der Wel}, {Rix}, {Ilbert},
  {Smol{\v c}i{\'c}}, {Carilli}, {Pannella}, {Koekemoer}, {Bell} \&
  {Salvato}}{{Karim} et~al.}{2011}]{Karim2011}
{Karim} A.,  {Schinnerer} E.,  {Mart{\'{\i}}nez-Sansigre} A.,  {Sargent} M.~T.,
   {van der Wel} A.,  {Rix} H.-W.,  {Ilbert} O.,  {Smol{\v c}i{\'c}} V.,
  {Carilli} C.,  {Pannella} M.,  {Koekemoer} A.~M.,  {Bell} E.~F.,    {Salvato}
  M.,  2011, \apj, 730, 61

\bibitem[\protect\citeauthoryear{{Kashlinsky}}{{Kashlinsky}}{2005}]{Kashlinsky%
2005}
{Kashlinsky} A.,  2005, \apjl, 633, L5

\bibitem[\protect\citeauthoryear{{Kashlinsky}, {Arendt}, {Ashby}, {Fazio},
  {Mather} \& {Moseley}}{{Kashlinsky} et~al.}{2012}]{Kashlinsky2012}
{Kashlinsky} A.,  {Arendt} R.~G.,  {Ashby} M.~L.~N.,  {Fazio} G.~G.,  {Mather}
  J.,    {Moseley} S.~H.,  2012, \apj, 753, 63

\bibitem[\protect\citeauthoryear{{Kennicutt}
  Jr.}{{Kennicutt}}{1998}]{Kennicutt1998}
{Kennicutt} Jr. R.~C.,  1998, \apj, 498, 541

\bibitem[\protect\citeauthoryear{{Kere{\v s}}, {Katz}, {Weinberg} \&
  {Dav{\'e}}}{{Kere{\v s}} et~al.}{2005}]{Keres2005}
{Kere{\v s}} D.,  {Katz} N.,  {Weinberg} D.~H.,    {Dav{\'e}} R.,  2005,
  \mnras, 363, 2

\bibitem[\protect\citeauthoryear{{Labb{\'e}}, {Gonz{\'a}lez}, {Bouwens},
  {Illingworth}, {Franx}, {Trenti}, {Oesch}, {van Dokkum}, {Stiavelli},
  {Carollo}, {Kriek} \& {Magee}}{{Labb{\'e}} et~al.}{2010}]{Labbe2010}
{Labb{\'e}} I.,  {Gonz{\'a}lez} V.,  {Bouwens} R.~J.,  {Illingworth} G.~D.,
  {Franx} M.,  {Trenti} M.,  {Oesch} P.~A.,  {van Dokkum} P.~G.,  {Stiavelli}
  M.,  {Carollo} C.~M.,  {Kriek} M.,    {Magee} D.,  2010, \apjl, 716, L103

\bibitem[\protect\citeauthoryear{{Madau} \& {Rees}}{{Madau} \&
  {Rees}}{2001}]{MadauRees2001}
{Madau} P.,  {Rees} M.~J.,  2001, \apjl, 551, L27

\bibitem[\protect\citeauthoryear{{Magliocchetti}, {Salvaterra} \&
  {Ferrara}}{{Magliocchetti} et~al.}{2003}]{Magliocchetti2003}
{Magliocchetti} M.,  {Salvaterra} R.,    {Ferrara} A.,  2003, \mnras, 342, L25

\bibitem[\protect\citeauthoryear{{Maio}}{{Maio}}{2011a}]{MaioNG2011}
{Maio} U.,  2011a, Classical and Quantum Gravity, 28, 225015

\bibitem[\protect\citeauthoryear{{Maio}}{{Maio}}{2011b}]{Maio2011cqg}
{Maio} U.,  2011b, Classical and Quantum Gravity, 28, 225015

\bibitem[\protect\citeauthoryear{{Maio}, {Ciardi}, {Dolag}, {Tornatore} \&
  {Khochfar}}{{Maio} et~al.}{2010}]{Maio2010}
{Maio} U.,  {Ciardi} B.,  {Dolag} K.,  {Tornatore} L.,    {Khochfar} S.,  2010,
  \mnras, 407, 1003

\bibitem[\protect\citeauthoryear{{Maio}, {Ciardi} \& {Mueller}}{{Maio}
  et~al.}{2013}]{Maio2013z7arXiv}
{Maio} U.,  {Ciardi} B.,    {Mueller} V.,  2013, ArXiv e-prints

\bibitem[\protect\citeauthoryear{{Maio}, {Dolag}, {Ciardi} \&
  {Tornatore}}{{Maio} et~al.}{2007}]{Maio2007}
{Maio} U.,  {Dolag} K.,  {Ciardi} B.,    {Tornatore} L.,  2007, \mnras, 379,
  963

\bibitem[\protect\citeauthoryear{{Maio}, {Dolag}, {Meneghetti}, {Moscardini},
  {Yoshida}, {Baccigalupi}, {Bartelmann} \& {Perrotta}}{{Maio}
  et~al.}{2006}]{Maio2006}
{Maio} U.,  {Dolag} K.,  {Meneghetti} M.,  {Moscardini} L.,  {Yoshida} N.,
  {Baccigalupi} C.,  {Bartelmann} M.,    {Perrotta} F.,  2006, \mnras, 373, 869

\bibitem[\protect\citeauthoryear{{Maio}, {Dotti}, {Petkova}, {Perego} \&
  {Volonteri}}{{Maio} et~al.}{2013}]{Maio2013}
{Maio} U.,  {Dotti} M.,  {Petkova} M.,  {Perego} A.,    {Volonteri} M.,  2013,
  \apj, 767, 37

\bibitem[\protect\citeauthoryear{{Maio} \& {Iannuzzi}}{{Maio} \&
  {Iannuzzi}}{2011}]{MaioIannuzzi2011}
{Maio} U.,  {Iannuzzi} F.,  2011, \mnras, 415, 3021

\bibitem[\protect\citeauthoryear{{Maio} \& {Khochfar}}{{Maio} \&
  {Khochfar}}{2012}]{MaioKhochfar2012}
{Maio} U.,  {Khochfar} S.,  2012, \mnras, 421, 1113

\bibitem[\protect\citeauthoryear{{Maio}, {Khochfar}, {Johnson} \&
  {Ciardi}}{{Maio} et~al.}{2011}]{Maio2011}
{Maio} U.,  {Khochfar} S.,  {Johnson} J.~L.,    {Ciardi} B.,  2011, \mnras,
  414, 1145

\bibitem[\protect\citeauthoryear{{Maio}, {Koopmans} \& {Ciardi}}{{Maio}
  et~al.}{2011}]{MaioKoopmansCiardi2011}
{Maio} U.,  {Koopmans} L.~V.~E.,    {Ciardi} B.,  2011, \mnras, 412, L40

\bibitem[\protect\citeauthoryear{{Maio}, {Salvaterra}, {Moscardini} \&
  {Ciardi}}{{Maio} et~al.}{2012}]{MaioGRBs2012}
{Maio} U.,  {Salvaterra} R.,  {Moscardini} L.,    {Ciardi} B.,  2012, \mnras,
  426, 2078

\bibitem[\protect\citeauthoryear{{Mannucci}, {Cresci}, {Maiolino}, {Marconi} \&
  {Gnerucci}}{{Mannucci} et~al.}{2010}]{Mannucci2010}
{Mannucci} F.,  {Cresci} G.,  {Maiolino} R.,  {Marconi} A.,    {Gnerucci} A.,
  2010, \mnras, 408, 2115

\bibitem[\protect\citeauthoryear{{McQuinn} \& {O'Leary}}{{McQuinn} \&
  {O'Leary}}{2012}]{McQuinn2012}
{McQuinn} M.,  {O'Leary} R.~M.,  2012, \apj, 760, 3

\bibitem[\protect\citeauthoryear{{Micha{\l}owski}, {Hjorth} \&
  {Watson}}{{Micha{\l}owski} et~al.}{2010}]{Michalowski2010}
{Micha{\l}owski} M.,  {Hjorth} J.,    {Watson} D.,  2010, \aap, 514, A67

\bibitem[\protect\citeauthoryear{{Moster}, {Macci{\`o}}, {Somerville}, {Naab}
  \& {Cox}}{{Moster} et~al.}{2012}]{Moster2012}
{Moster} B.~P.,  {Macci{\`o}} A.~V.,  {Somerville} R.~S.,  {Naab} T.,    {Cox}
  T.~J.,  2012, \mnras, 423, 2045

\bibitem[\protect\citeauthoryear{{Naoz}, {Yoshida} \& {Gnedin}}{{Naoz}
  et~al.}{2012}]{Naoz2012}
{Naoz} S.,  {Yoshida} N.,    {Gnedin} N.~Y.,  2012, \apj, 747, 128

\bibitem[\protect\citeauthoryear{{Naoz}, {Yoshida} \& {Gnedin}}{{Naoz}
  et~al.}{2013}]{Naoz2013}
{Naoz} S.,  {Yoshida} N.,    {Gnedin} N.~Y.,  2013, \apj, 763, 27

\bibitem[\protect\citeauthoryear{{Noeske}, {Weiner}, {Faber}, {Papovich},
  {Koo}, {Somerville} \& {23~co-authors}}{{Noeske} et~al.}{2007}]{Noeske2007}
{Noeske} K.~G.,  {Weiner} B.~J.,  {Faber} S.~M.,  {Papovich} C.,  {Koo} D.~C.,
  {Somerville} R.~S.,    {23~co-authors} 2007, \apjl, 660, L43

\bibitem[\protect\citeauthoryear{{Pannella}, {Carilli}, {Daddi}, {McCracken},
  {Owen}, {Renzini}, {Strazzullo} \& {13 co-authors}}{{Pannella}
  et~al.}{2009}]{Pannella2009}
{Pannella} M.,  {Carilli} C.~L.,  {Daddi} E.,  {McCracken} H.~J.,  {Owen}
  F.~N.,  {Renzini} A.,  {Strazzullo} V.,    {13 co-authors} 2009, \apjl, 698,
  L116

\bibitem[\protect\citeauthoryear{{Peng}, {Lilly}, {Kova{\v c}} \& {61
  co-authors}}{{Peng} et~al.}{2010}]{Peng2010}
{Peng} Y.-j.,  {Lilly} S.~J.,  {Kova{\v c}} K.,    {61 co-authors} 2010, \apj,
  721, 193

\bibitem[\protect\citeauthoryear{{Petkova} \& {Maio}}{{Petkova} \&
  {Maio}}{2012}]{PetkovaMaio2012}
{Petkova} M.,  {Maio} U.,  2012, \mnras, 422, 3067

\bibitem[\protect\citeauthoryear{{Petkova} \& {Springel}}{{Petkova} \&
  {Springel}}{2011}]{PetkovaSpringel2011}
{Petkova} M.,  {Springel} V.,  2011, \mnras, 412, 935

\bibitem[\protect\citeauthoryear{{Piontek} \& {Steinmetz}}{{Piontek} \&
  {Steinmetz}}{2011}]{PiontekSteinmetz2011}
{Piontek} F.,  {Steinmetz} M.,  2011, \mnras, 410, 2625

\bibitem[\protect\citeauthoryear{{Prieto}, {Jimenez} \& {Haiman}}{{Prieto}
  et~al.}{2013}]{Prieto2013}
{Prieto} J.,  {Jimenez} R.,    {Haiman} Z.,  2013, ArXiv e-prints

\bibitem[\protect\citeauthoryear{{Reddy}, {Pettini}, {Steidel}, {Shapley},
  {Erb} \& {Law}}{{Reddy} et~al.}{2012}]{Reddy2012}
{Reddy} N.~A.,  {Pettini} M.,  {Steidel} C.~C.,  {Shapley} A.~E.,  {Erb} D.~K.,
     {Law} D.~R.,  2012, \apj, 754, 25

\bibitem[\protect\citeauthoryear{{Richardson}, {Scannapieco} \&
  {Thacker}}{{Richardson} et~al.}{2013}]{Richardson2013arXiv}
{Richardson} M.~L.~A.,  {Scannapieco} E.,    {Thacker} R.~J.,  2013, ArXiv
  e-prints

\bibitem[\protect\citeauthoryear{{Ricotti}, {Gnedin} \& {Shull}}{{Ricotti}
  et~al.}{2001}]{Ricotti2001}
{Ricotti} M.,  {Gnedin} N.~Y.,    {Shull} J.~M.,  2001, \apj, 560, 580

\bibitem[\protect\citeauthoryear{{Robertson}, {Ellis}, {Dunlop}, {McLure} \&
  {Stark}}{{Robertson} et~al.}{2010}]{Robertson2010}
{Robertson} B.~E.,  {Ellis} R.~S.,  {Dunlop} J.~S.,  {McLure} R.~J.,    {Stark}
  D.~P.,  2010, \nat, 468, 49

\bibitem[\protect\citeauthoryear{{Rodighiero}, {Daddi} \& {32
  co-authors}}{{Rodighiero} et~al.}{2011}]{Rodighiero2011}
{Rodighiero} G.,  {Daddi} E.,    {32 co-authors} 2011, \apjl, 739, L40

\bibitem[\protect\citeauthoryear{{Romano-D{\'{\i}}az}, {Choi}, {Shlosman} \&
  {Trenti}}{{Romano-D{\'{\i}}az} et~al.}{2011}]{RomanoDiaz2011}
{Romano-D{\'{\i}}az} E.,  {Choi} J.-H.,  {Shlosman} I.,    {Trenti} M.,  2011,
  \apjl, 738, L19

\bibitem[\protect\citeauthoryear{{Sales}, {Navarro}, {Theuns}, {Schaye},
  {White}, {Frenk}, {Crain} \& {Dalla Vecchia}}{{Sales}
  et~al.}{2012}]{Sales2012}
{Sales} L.~V.,  {Navarro} J.~F.,  {Theuns} T.,  {Schaye} J.,  {White} S.~D.~M.,
   {Frenk} C.~S.,  {Crain} R.~A.,    {Dalla Vecchia} C.,  2012, \mnras, 423,
  1544

\bibitem[\protect\citeauthoryear{{Salim}, {Rich}, {Charlot} \& {21
  co-authors}}{{Salim} et~al.}{2007}]{Salim2007}
{Salim} S.,  {Rich} R.~M.,  {Charlot} S.,    {21 co-authors} 2007, \apjs, 173,
  267

\bibitem[\protect\citeauthoryear{{Salvaterra}, {Ferrara} \&
  {Dayal}}{{Salvaterra} et~al.}{2011}]{Salvaterra2011}
{Salvaterra} R.,  {Ferrara} A.,    {Dayal} P.,  2011, \mnras, 414, 847

\bibitem[\protect\citeauthoryear{{Salvaterra}, {Maio}, {Ciardi} \&
  {Campisi}}{{Salvaterra} et~al.}{2013}]{Salvaterra2013}
{Salvaterra} R.,  {Maio} U.,  {Ciardi} B.,    {Campisi} M.~A.,  2013, \mnras,
  429, 2718

\bibitem[\protect\citeauthoryear{{Santos}, {Bromm} \& {Kamionkowski}}{{Santos}
  et~al.}{2002}]{Santos2002}
{Santos} M.~R.,  {Bromm} V.,    {Kamionkowski} M.,  2002, \mnras, 336, 1082

\bibitem[\protect\citeauthoryear{{Schmidt}}{{Schmidt}}{1959}]{Schmidt1959}
{Schmidt} M.,  1959, \apj, 2129, 243

\bibitem[\protect\citeauthoryear{{Schneider}, {Ferrara}, {Salvaterra}, {Omukai}
  \& {Bromm}}{{Schneider} et~al.}{2003}]{Schneider2003}
{Schneider} R.,  {Ferrara} A.,  {Salvaterra} R.,  {Omukai} K.,    {Bromm} V.,
  2003, \nat, 422, 869

\bibitem[\protect\citeauthoryear{{Skibba} \& {Macci{\`o}}}{{Skibba} \&
  {Macci{\`o}}}{2011}]{skibba2011}
{Skibba} R.~A.,  {Macci{\`o}} A.~V.,  2011, \mnras, 416, 2388

\bibitem[\protect\citeauthoryear{{Sokasian}, {Yoshida}, {Abel}, {Hernquist} \&
  {Springel}}{{Sokasian} et~al.}{2004}]{Sokasian2004}
{Sokasian} A.,  {Yoshida} N.,  {Abel} T.,  {Hernquist} L.,    {Springel} V.,
  2004, \mnras, 350, 47

\bibitem[\protect\citeauthoryear{{Springel} \& {Hernquist}}{{Springel} \&
  {Hernquist}}{2003}]{Springel2003}
{Springel} V.,  {Hernquist} L.,  2003, \mnras, 339, 289

\bibitem[\protect\citeauthoryear{{Stacy} \& {Bromm}}{{Stacy} \&
  {Bromm}}{2013}]{Stacy2013arXiv}
{Stacy} A.,  {Bromm} V.,  2013, ArXiv e-prints

\bibitem[\protect\citeauthoryear{{Stacy}, {Bromm} \& {Loeb}}{{Stacy}
  et~al.}{2011}]{Stacy2011}
{Stacy} A.,  {Bromm} V.,    {Loeb} A.,  2011, \apjl, 730, L1

\bibitem[\protect\citeauthoryear{{Stacy}, {Greif}, {Klessen}, {Bromm} \&
  {Loeb}}{{Stacy} et~al.}{2013}]{Stacy2013}
{Stacy} A.,  {Greif} T.~H.,  {Klessen} R.~S.,  {Bromm} V.,    {Loeb} A.,  2013,
  \mnras, 431, 1470

\bibitem[\protect\citeauthoryear{{Stark}, {Schenker}, {Ellis}, {Robertson},
  {McLure} \& {Dunlop}}{{Stark} et~al.}{2013}]{Stark2013}
{Stark} D.~P.,  {Schenker} M.~A.,  {Ellis} R.,  {Robertson} B.,  {McLure} R.,
   {Dunlop} J.,  2013, \apj, 763, 129

\bibitem[\protect\citeauthoryear{{Toomre}}{{Toomre}}{1977}]{Toomre1977}
{Toomre} A.,  1977, \araa, 15, 437

\bibitem[\protect\citeauthoryear{{Tornatore}, {Borgani}, {Viel} \&
  {Springel}}{{Tornatore} et~al.}{2010}]{Tornatore2010}
{Tornatore} L.,  {Borgani} S.,  {Viel} M.,    {Springel} V.,  2010, \mnras,
  402, 1911

\bibitem[\protect\citeauthoryear{{Tornatore}, {Ferrara} \&
  {Schneider}}{{Tornatore} et~al.}{2007}]{Tornatore2007}
{Tornatore} L.,  {Ferrara} A.,    {Schneider} R.,  2007, \mnras, 382, 945

\bibitem[\protect\citeauthoryear{{Totani}, {Yoshii}, {Maihara}, {Iwamuro} \&
  {Motohara}}{{Totani} et~al.}{2001}]{Totani2001}
{Totani} T.,  {Yoshii} Y.,  {Maihara} T.,  {Iwamuro} F.,    {Motohara} K.,
  2001, \apj, 559, 592

\bibitem[\protect\citeauthoryear{{Tremonti}, {Heckman}, {Kauffmann},
  {Brinchmann}, {Charlot}, {White}, {Seibert}, {Peng}, {Schlegel}, {Uomoto},
  {Fukugita} \& {Brinkmann}}{{Tremonti} et~al.}{2004}]{Tremonti2004}
{Tremonti} C.~A.,  {Heckman} T.~M.,  {Kauffmann} G.,  {Brinchmann} J.,
  {Charlot} S.,  {White} S.~D.~M.,  {Seibert} M.,  {Peng} E.~W.,  {Schlegel}
  D.~J.,  {Uomoto} A.,  {Fukugita} M.,    {Brinkmann} J.,  2004, \apj, 613, 898

\bibitem[\protect\citeauthoryear{{Tseliakhovich} \& {Hirata}}{{Tseliakhovich}
  \& {Hirata}}{2010}]{TH2010}
{Tseliakhovich} D.,  {Hirata} C.,  2010, \prd, 82, 083520

\bibitem[\protect\citeauthoryear{{Vazza}, {Dolag}, {Ryu}, {Brunetti},
  {Gheller}, {Kang} \& {Pfrommer}}{{Vazza} et~al.}{2011}]{vazza2011}
{Vazza} F.,  {Dolag} K.,  {Ryu} D.,  {Brunetti} G.,  {Gheller} C.,  {Kang} H.,
    {Pfrommer} C.,  2011, \mnras, 418, 960

\bibitem[\protect\citeauthoryear{{Vogelsberger}, {Genel}, {Sijacki}, {Torrey},
  {Springel} \& {Hernquist}}{{Vogelsberger} et~al.}{2013}]{Vogelsberger2013}
{Vogelsberger} M.,  {Genel} S.,  {Sijacki} D.,  {Torrey} P.,  {Springel} V.,
  {Hernquist} L.,  2013, ArXiv e-prints

\bibitem[\protect\citeauthoryear{{Whalen}, {O'Shea}, {Smidt} \&
  {Norman}}{{Whalen} et~al.}{2008}]{Whalen2008}
{Whalen} D.,  {O'Shea} B.~W.,  {Smidt} J.,    {Norman} M.~L.,  2008, \apj, 679,
  925

\bibitem[\protect\citeauthoryear{{White} \& {Rees}}{{White} \&
  {Rees}}{1978}]{WhiteRees1978}
{White} S.~D.~M.,  {Rees} M.~J.,  1978, \mnras, 183, 341

\bibitem[\protect\citeauthoryear{{Wise} \& {Abel}}{{Wise} \&
  {Abel}}{2007}]{WiseAbel2007}
{Wise} J.~H.,  {Abel} T.,  2007, \apj, 665, 899

\bibitem[\protect\citeauthoryear{{Wise}, {Turk}, {Norman} \& {Abel}}{{Wise}
  et~al.}{2012}]{Wise2012}
{Wise} J.~H.,  {Turk} M.~J.,  {Norman} M.~L.,    {Abel} T.,  2012, \apj, 745,
  50

\bibitem[\protect\citeauthoryear{{Xu}, {Wise} \& {Norman}}{{Xu}
  et~al.}{2013}]{Xu2013arXiv}
{Xu} H.,  {Wise} J.~H.,    {Norman} M.~L.,  2013, ArXiv e-prints

\bibitem[\protect\citeauthoryear{{Yoshida}, {Abel}, {Hernquist} \&
  {Sugiyama}}{{Yoshida} et~al.}{2003}]{Yoshida2003}
{Yoshida} N.,  {Abel} T.,  {Hernquist} L.,    {Sugiyama} N.,  2003, \apj, 592,
  645

\bibitem[\protect\citeauthoryear{{Zheng}, {Postman}, {Zitrin} \& {33
  co-authros}}{{Zheng} et~al.}{2012}]{Zheng2012}
{Zheng} W.,  {Postman} M.,  {Zitrin} A.,    {33 co-authros} 2012, \nat, 489,
  406

\end{thebibliography}
%
%
\begin{figure*}
\centering
\includegraphics[width=0.33\textwidth]{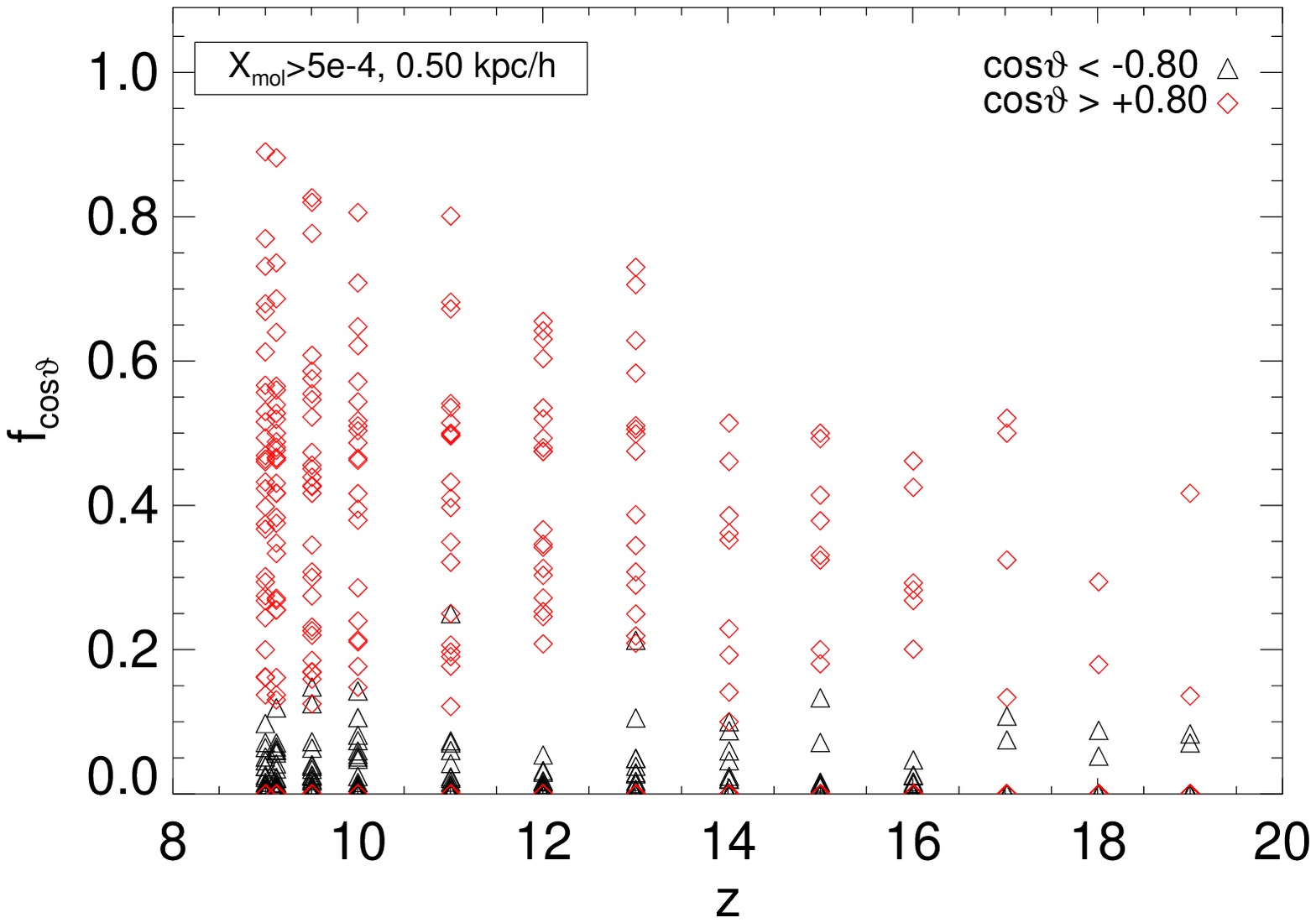}
\includegraphics[width=0.33\textwidth]{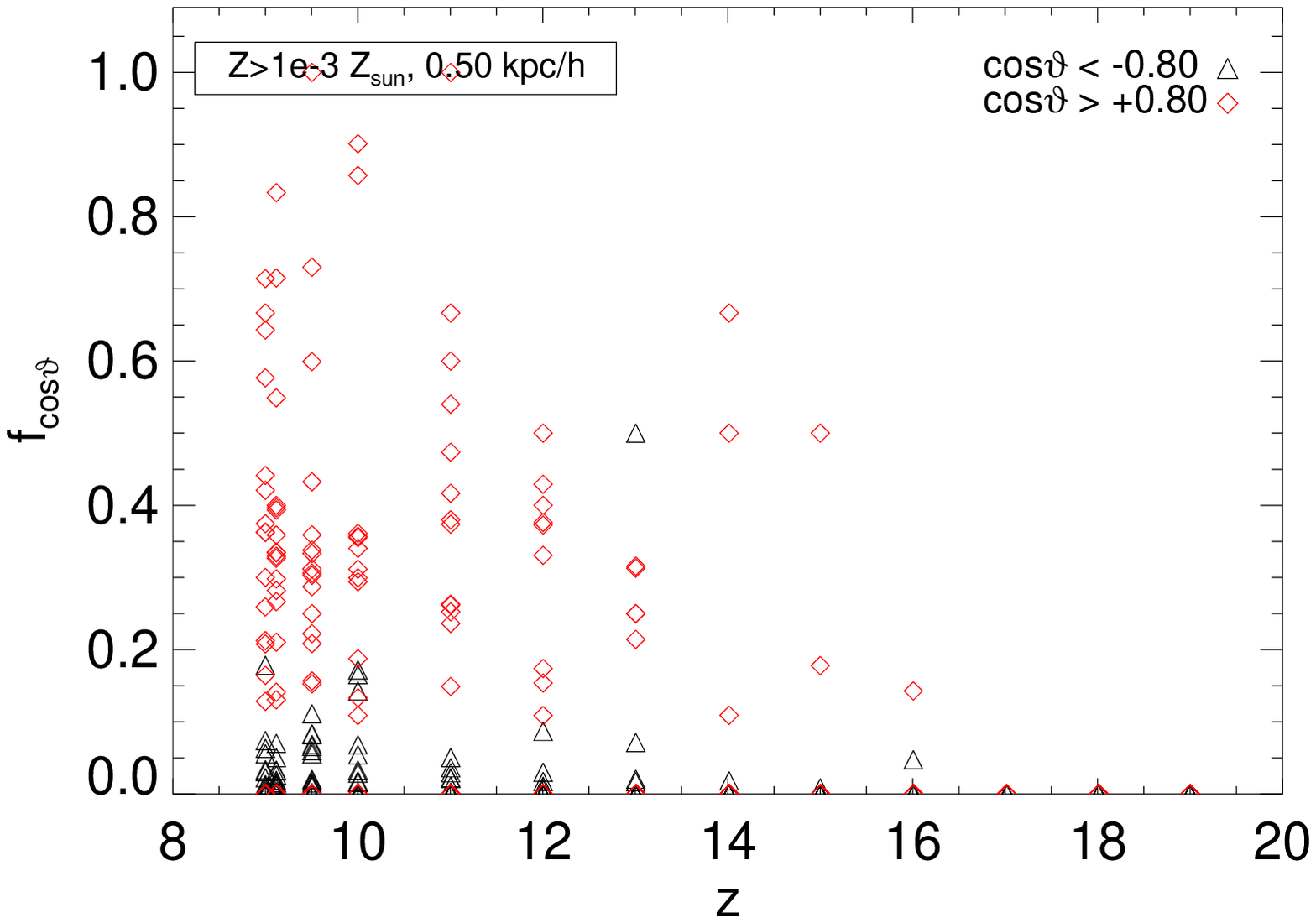}
\includegraphics[width=0.33\textwidth]{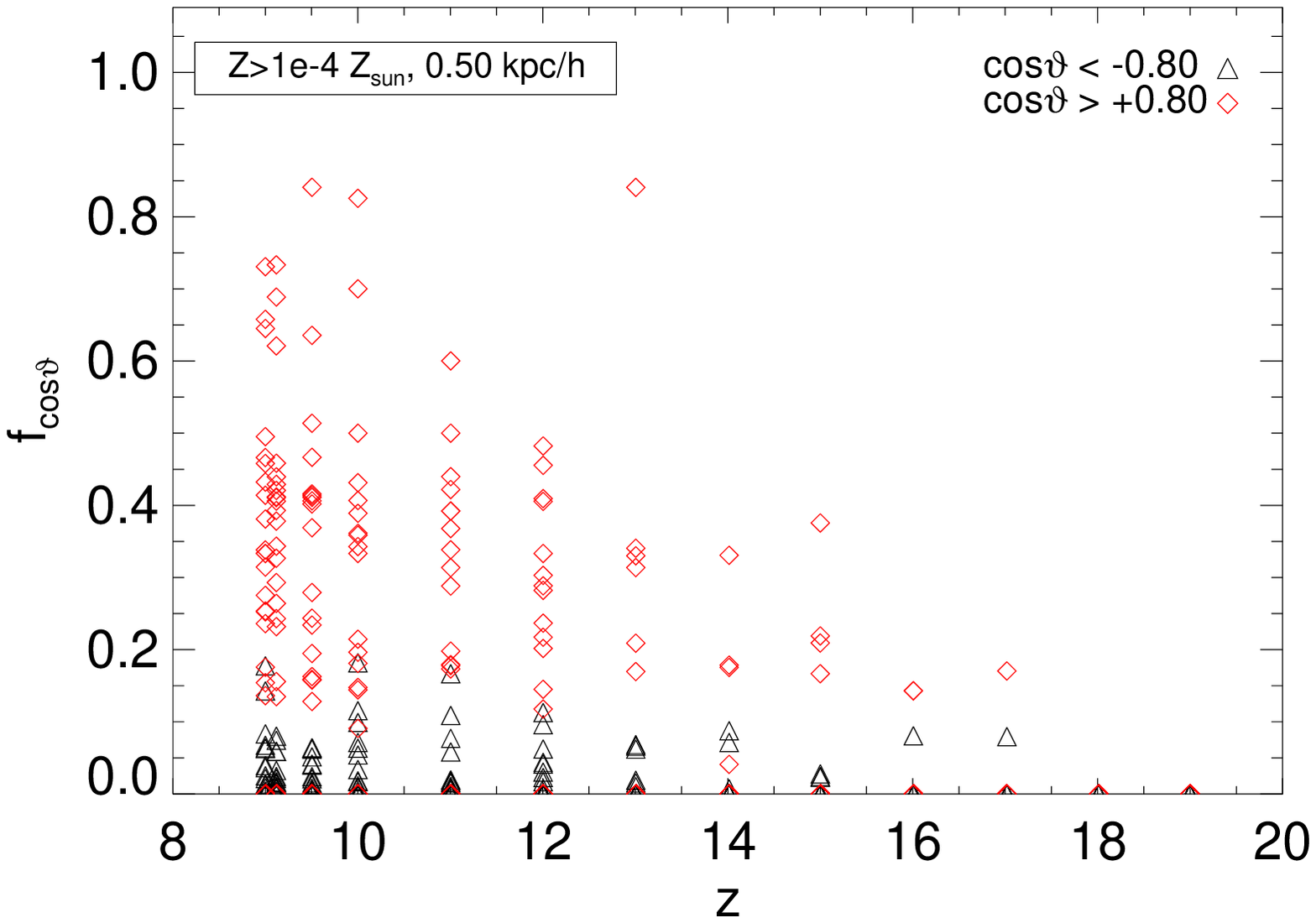}\\
\includegraphics[width=0.33\textwidth]{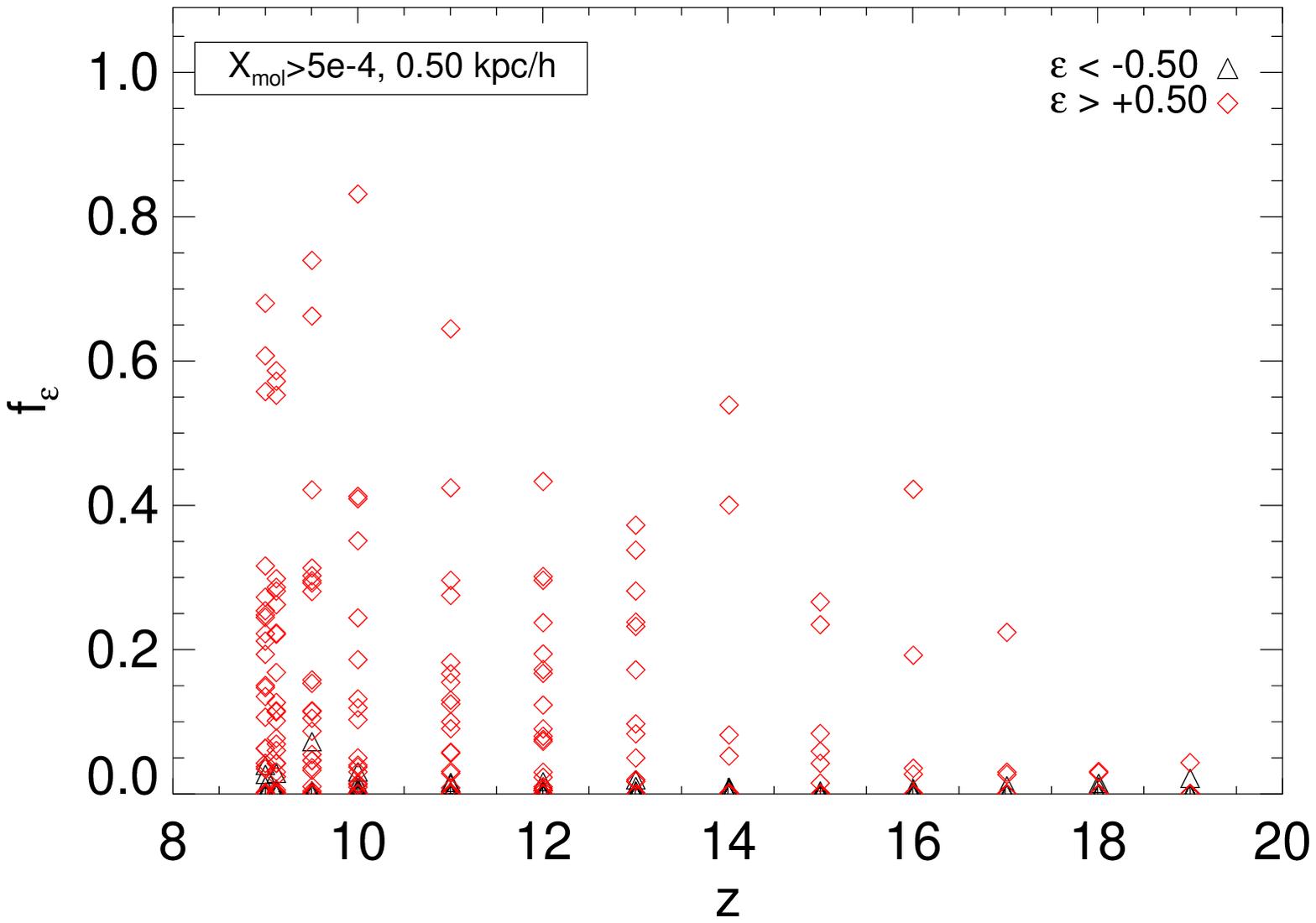}
\includegraphics[width=0.33\textwidth]{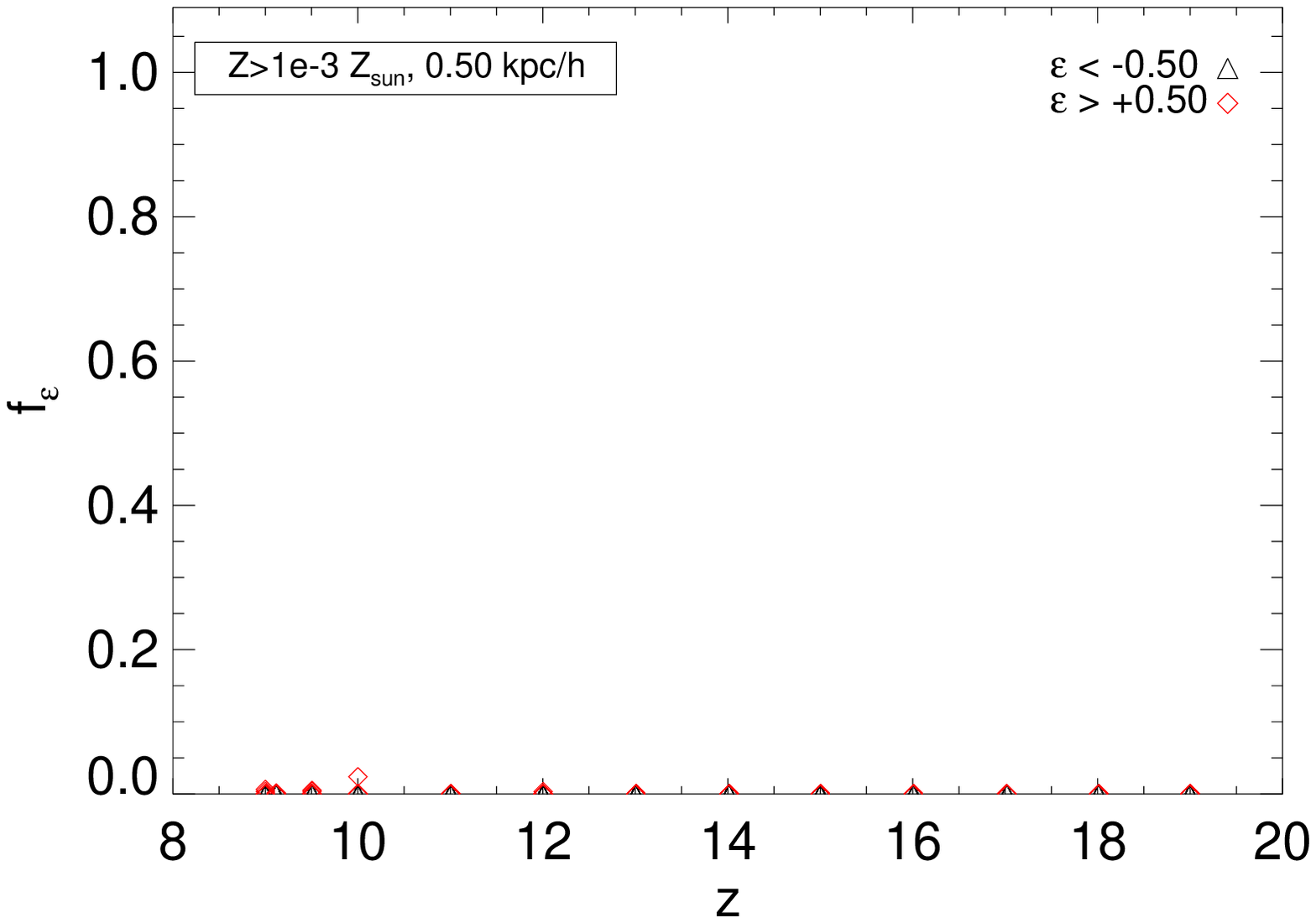}
\includegraphics[width=0.33\textwidth]{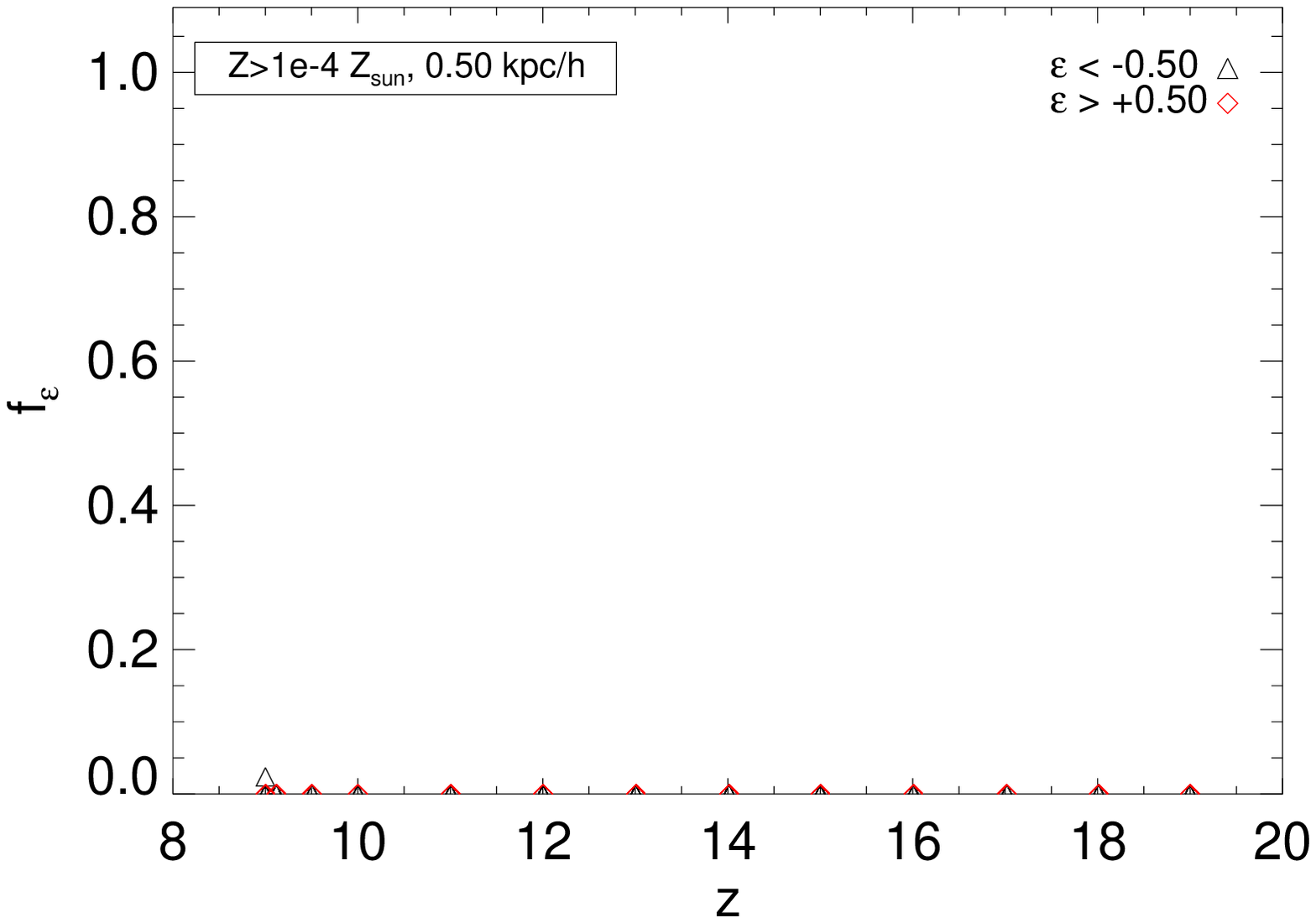}
\caption{Fractional distributions $ f_{cos\vartheta} $ (upper row) and $
  f_{\varepsilon} $ (lower row) as a function of redshift.   
  Columns refer to the gas phase considered, namely: molecular-rich gas
  with $x_{mol} >5\times 10^{-4}$ (left); metal-rich gas with
  $Z>10^{-3}\Zsun$ (middle); metal-rich gas with $Z>10^{-4}\Zsun$.  
  In every panel, each halo is marked with two points, representing the
  fraction of co-rotating (red diamonds) or contra-rotating (black
  triangles) gas (see legend and text). 
}
\label{fig:param_examples}
\end{figure*}
\appendix
\section[]{Parameter study}\label{appA}
We present here some demonstration cases of the parameter study done
to ascertain the reliability of the conclusions drawn, independently
of the assumptions made for the calculations.
More precisely, we have studied parameter dependences for the
fractional distributions $f_{\cos\vartheta}$ and $f_{\varepsilon}$,
i.e. the fractions of gas with $\cos\vartheta > |\cos\vartheta_{\rm  th}|$
and $\varepsilon > |\varepsilon_{\rm th}|$, respectively.
\\
First, we find that the trends are not significantly sensitive to the
precise sampling of the central regions, for 1.00, 0.50, and
0.25~kpc/{\it h} comoving radii.
In fact, since we aim at studying rotational patterns in the molecular-rich and
metal-rich gas, the selected gaseous components mainly reside, in any
case, in the very central sites of the halo, so that no striking
difference is introduced by the exact core size. 
\\
Moreover, the results on rotational patterns 
do not show any strong dependence on the
threshold angle $\vartheta_{\rm th}$, 
used to point out the fraction of aligned, potentially rotating gas, 
as long as $\cos\vartheta_{\rm th}\gtrsim 0.7$.
To ascertain this, the values explored as minimum threshold for the
inclination angle are
\begin{itemize}
\item $\cos\vartheta_{\rm th} = 0.7$, corresponding to $\theta_{\rm th}\simeq 45^o$,
\item $\cos\vartheta_{\rm th} = 0.8$, corresponding to $\theta_{\rm th}\simeq 37^o$,
\item $\cos\vartheta_{\rm th} = 0.9$, corresponding to $\theta_{\rm th}\simeq 26^o$.
\end{itemize}
Given the strong similarity of the results with varying parameters, 
in \fig\ref{fig:param_examples} we only show the reference cases (used
in the computations for the previous sections), for which the size of the core region
is assumed to be 0.5~kpc/{\it h} and the minimum threshold angle is
$\theta_{\rm th}\simeq 37^o$.
The figure
displays the fractional distributions $f_{\varepsilon}$ and
$f_{cos\vartheta}$ as a function of redshift, for molecular gas
($x_{mol}>5\times 10^{-4}$; left column) and
metal-rich gas 
with $\Zcrit=10^{-3}\Zsun$ (middle column) 
and  $\Zcrit=10^{-4}\Zsun$ (right column).
In the plots, each halo is marked with two points, representing the
fraction $f_{cos\vartheta}$ of gas for which $\cos\vartheta$ is
$> \cos\vartheta_{\rm th}$ (red diamond; co-rotating fraction) and $<
-\cos\vartheta_{\rm th}$ (black triangle; contra-rotating fraction),
respectively (upper-row panels of
\fig\ref{fig:param_examples}).
The same is done for $f_{\varepsilon}$ (lower-row panels).
\\
As already highlighted by the representative haloes discussed
in \sec\ref{sub:cases} and \sec\ref{sub:evol},
even this statistical analysis confirms that for metal-rich gas, no
significant alignment is found between gas particle angular momentum
and the circular reference value, $j_{circ}$.
Moreover, this result does not depend on the core size, since the
$f_{\varepsilon}$ distribution is always found to be flat around zero
at all redshifts, as in the central and right panels of the lower row
of \fig\ref{fig:param_examples}. 
Basically, this indicates that on average there is no evidence of
ordered, rotational patterns in the metal-rich gas component. 
This parameter study also reasserts that a different case is
represented by the molecular-rich gas, for which a significant number
of haloes, increasing with cosmic time, shows a large gas
fraction with co-rotating features, as indicated not only from the $\cos\vartheta$
diagnostics (upper row panels) but also via the more robust
$\varepsilon$ indicator (lower row panels).



\label{lastpage}
\end{document}